\documentclass[paperpaper,twocolumn,english,superscriptaddress,aps,prx,floatfix,amsfonts,amssymb]{revtex4-1}

\usepackage{epsfig}% Include figure files
\usepackage{graphicx}% Include figure files
\usepackage{dcolumn}% Align table columns on decimal point
\usepackage{bm}% bold math

\newcommand{\EQ}{\begin{equation}}
\newcommand{\EN}{\end{equation}}
\newcommand{\ea}{\end{eqnarray}}
\newcommand{\ba}{\begin{eqnarray}}

\newcommand{\bear}{\begin{eqnarray}}
\newcommand{\ear}{\end{eqnarray}}

\begin{document}
\title{Finite-temperature charge and spin transport in the one-dimensional Hubbard model accounting for its
global $[SU (2)\times SU(2)\times U(1)]/Z_2^2$ symmetry}

%\date{}

\author{J. M. P. Carmelo}
\affiliation{Center of Physics of University of Minho and University of Porto, LaPMET, P-4169-007 Oporto, Portugal}
\affiliation{CeFEMA, Instituto Superior T\'ecnico, Universidade de Lisboa, LaPMET, Av. Rovisco Pais, P-1049-001 Lisboa, Portugal}

\author{J. E. C. Carmelo}
\affiliation{CFTC, Universidade de Lisboa,  Fac. Ci\^encias, P-1749-016 Lisboa, Portugal}
\affiliation{CeFEMA, Instituto Superior T\'ecnico, Universidade de Lisboa, LaPMET, Av. Rovisco Pais, P-1049-001 Lisboa, Portugal}

%\date{\today}

%%%%%%%%%%%%%%%%%%%%%%%%%%%%%%%%%%%%%%%%%%%%%%%%%%%%%%%%%%%%%%%%%%%%%%%%%%
%                              abstract                                  %
%%%%%%%%%%%%%%%%%%%%%%%%%%%%%%%%%%%%%%%%%%%%%%%%%%%%%%%%%%%%%%%%%%%%%%%%%%

\begin{abstract}
Most studies on the one-dimensional (1D) Hubbard model with transfer integral $t$ and on-site repulsion $U$ have 
assumed that for $u=U/4t>0$ and $h=\mu =0$, where $h$ denotes the magnetic field and $\mu$ the chemical potential,
its global symmetry is $SO (4) = [SU (2)\times SU(2)]/Z_2$. However, both in 1D and for the $U>0$ Hubbard model
on any {\it bipartite lattice} it is actually larger and given by $[SU (2)\times SU(2)\times U(1)]/Z_2^2 = [SO (4)\times U(1)]/Z_2$. 
The studies of this paper account for the irreducible representations of that symmetry to identify both the spin and charge 
carriers that populate each of the energy eigenstates that span that model {\it full} Hilbert space. The global $\tau$-translational 
$U(1)$ symmetry beyond $SO (4)$ is found to describe the relative translational degrees of freedom of such spin and charge carriers, 
respectively. We show that important finite-temperature transport quantities as the $h>0$ spin and $\mu >0$ charge stiffnesses 
and the $h = 0$ spin and $\mu = 0$ charge diffusion constants are controlled by microscopic processes associated
with the spin and charge elementary currents carried by these spin and charge carriers, respectively. We describe the 
model finite-temperature transport properties in terms of the microscopic processes associated with the 
spin and charge elementary currents carried by the corresponding carriers. We expect that the general $\tau$- and physical 
$\alpha$-spin representation used in our study is that suitable to address the open problems on finite-temperature transport in the 
1D Hubbard model also discussed in this paper.
\end{abstract}
\maketitle
%%%%%%%%%%%%%%%%%%%%%%%%%%%%%%%%%%%%%%%%%%%%%%%%%%%%%%%%%%%%%%%%%%%%%%%%%%
%                              body of paper                             
%%%%%%%%%%%%%%%%%%%%%%%%%%%%%%%%%%%%%%%%%%%%%%%%%%%%%%%%%%%%%%%%%%%%%%%%%%
%%%%%%%%%%%%%%%%%%%%%%%%%%%%%%%%%%%%%%%%%%%%%%%%%%%%%%%%%%%%%%%%%%%%%%%%%%

\section{Introduction}
\label{SECI}

Over the years the Hubbard model \cite{Gutzwiller_63,Hubbard_63} with transfer integral $t$ and on-site repulsion $U$
has become more important, for it plays an essential role in several topics in condensed-matter physics and related quantum 
problems. For instance, the one-dimensional (1D) Hubbard model defined in a lattice of length $L$ and with $N_a$ sites \cite{Lieb_68,Lieb_03,Takahashi_72,Martins_97,Martins_98} is the paradigmatic quantum system for low-dimensional strongly correlated electron 
systems and materials.

It is well known that at zero magnetic field and zero chemical potential the 1D Hubbard model has a global 
$SO (4) = [SU (2)\times SU(2)]/Z_2$ symmetry whose two global $SU(2)$ symmetries refer
to spin and $\eta$-spin, respectively \cite{Essler_05}. On the other hand, the local $SU (2)\times SU(2)\times U(1)$ gauge symmetry 
of both the 1D Hubbard model and that model on any bipartite lattice for $u=U/4t\rightarrow\infty$ \cite{Ostlund_91}, has been lifted to a global 
$[SU (2)\times SU(2)\times U(1)]/Z_2^2$ symmetry at finite values of
$u=U/4t>0$ \cite{Carmelo_10,Carmelo_24}. The irreducible representations
of the global $\tau$-translational $U(1)$ symmetry beyond $SO (4)$
can be shown to ensure the independence of the global spin $SU (2)$ and $\eta$-spin
$SU (2)$ symmetries, respectively, in $[SU (2)\times SU(2)\times U(1)]/Z_2^2$,
which we often denote by $\alpha$-spin $SU(2)$ 
symmetries where $\alpha = s$ for spin and $\alpha = \eta$ for $\eta$-spin. 

However, that the global symmetry of the Hubbard model is for $u=U/4t>0$ larger than $SO (4)$
and specifically given by $[SO (4) \times U(1)]/Z_2 = [SU (2)\times SU(2)\times U(1)]/Z_2^2$ has been ignored by
most studies, which assume it only being the global $SU (2)\times SU(2)$ symmetry in 
$SO (4) = [SU (2)\times SU(2)]/Z_2$ \cite{Essler_05}. And this applies to recent studies on spin and charge transport 
relying on hydrodynamic theory and Kardar-Parisi-Zhang (KPZ) scaling
\cite{Ilievski_18,Fava_20,Moca_23}. More recently it was shown that such studies prediction of finite-temperature
$T>0$ charge transport at $\mu =0$ being anomalous superdiffusive is not true.
And this is due to they not accounting for the
important role played by the $\tau$-translational $U(1)$ symmetry beyond $SO (4)$ \cite{Carmelo_24}.

While the studies of Ref. \onlinecite{Carmelo_24} focus on the $T>0$ charge transport in the $\mu =0$ 
half-filled 1D Hubbard model, in this paper we study both $T>0$ spin and charge transport in 1D Hubbard model
for all values of the magnetic field $h$ and chemical potential $\mu$. Due to symmetry, this can be
achieved by only considering $h \geq 0$ and $\mu \geq 0$ values. 

Both the spin and charge carriers that populate each of the energy eigenstates that span that 
model {\it full} Hilbert space for $U>0$ are identified. One 
of our goals is to show that important finite-temperature transport quantities such as the $h>0$ spin and $\mu >0$ 
charge stiffnesses and the $h = 0$ spin and $\mu = 0$ charge diffusion constants are controlled by microscopic 
processes described in terms of the spin and charge elementary currents carried by these spin and charge carriers, 
respectively. To reach that goal, we account for the relation of the model global symmetry to charge and spin transport.
For instance, the irreducible representations of the global $\tau$-translational $U(1)$ symmetry beyond $SO (4)$ 
are found to describe the relative translational degrees of freedom of charge and spin carriers.

The study of spin and charge transport at any finite temperature involves the 1D Hubbard model in its full Hilbert space
spanned by $4^{N_a}$ energy eigenstates. Rather than the usual spinon and holon representations \cite{Essler_05,Pereira_12,Nocera_18},
which are valid only in some subspaces, the studies of this paper rely on a $\tau$- and $\alpha =s,\eta$ physical $\alpha$-spin representation 
that in the thermodynamic limit is for $U>0$ defined in the model's full Hilbert space. 

As justified in this paper, that representation's {\it physical spins $1/2$} refer indeed to electronic
physical spins and its {\it physical $\eta$-spins $1/2$} correspond to well-defined electronic configurations
of the electrons associated with the creation and annihilation operators in the expression of
the model's Hamiltonian given below in Eq. (\ref{H}).
That representation naturally emerges from the complete set of $4^{N_a}$ energy and momentum eigenstates 
referring for $u=U/4t>0$ and all values of magnetic field $h$ and chemical potential $\mu$ to 
a corresponding set of $4^{N_a}$ irreducible representations of the 1D Hubbard model's 
global $[SU (2)\times SU(2)\times U(1)]/Z_2^2$ symmetry at $h=0$ and $\mu =0$. 

We denote by $S_s$ and $S_{\eta}$ the energy eigenstate's spin and $\eta$-spin, respectively. 
The interest for transport of spin and charge of the $\tau$- and physical $\alpha$-spin representation 
is that it naturally identifies for all $4^{N_a}$ energy eigenstates the set of unpaired physical spins $1/2$ 
in spin-multiplet configurations whose number is $2S_s$ and the set of unpaired physical $\eta$-spins $1/2$ 
in $\eta$-spin-multiplet configurations whose number is number of $2S_{\eta}$ as the spin and charge carriers,
respectively. Actually, out of the $4^{N_a}$ energy eigenstates, only the $S_s>0$ and/or $S_{\eta}>0$ such states 
have finite spin and/or charge current expectation values, respectively.

It is well established that in the thermodynamic limit the spin and charge stiffnesses
of the 1D Hubbard model exactly vanish at $h=0$ and $\mu =0$, respectively,
for all temperatures $T>0$ \cite{Ilievski_17,Carmelo_18}. 
Hydrodynamic theory and KPZ scaling have then been used to study dynamical scaling properties of 
the corresponding nonballistic spin and charge transport in the 1D Hubbard model at $h=0$ and $\mu =0$, respectively
\cite{Ilievski_18,Fava_20,Moca_23}. 

The KPZ scaling refers to a universality class that involves a broad range of classical stochastic 
growth models that exhibit similar scaling behavior to the original KPZ equation \cite{Kardar_86,Krug_97,Kriecherbauer_10,Corwin_12}. 
It has also been shown to describe high-temperature $T\rightarrow\infty$ and thus classical 
dynamics of certain many-body systems near equilibrium \cite{Ljubotina_19,Ljubotina_17}. On the other hand, 
it was found that KPZ scaling prevails for all finite temperatures $T>0$ in the case of the dynamical spin 
structure factor of the $SU(2)$ symmetrical spin-$1/2$ $XXX$ chain,
which was found to be exactly described by the KPZ correlation function. Consistently, that spin chain exhibits 
anomalous superdiffusive behavior with dynamical scaling exponent $z=3/2$. 

This also thus applies to the gapless spin degrees of freedom of the 1D Hubbard model at $h=0$ \cite{Ilievski_18,Fava_20,Moca_23}.
As mentioned above, it was argued by studies relying on hydrodynamic theory and KPZ scaling that the 
$\mu =0$ finite-temperature charge transport is also anomalous
superdiffusive for $T>0$ \cite{Ilievski_18,Fava_20,Moca_23}. 
However, recent results have revealed that the $\mu =0$ charge diffusion constant is finite in the $k_B T/\Delta_{\eta}\ll 1$ regime
where $\Delta_{\eta}$ is the Mott-Hubbard gap \cite{Carmelo_24}. 

This brings about the interesting unsolved issue of how charge transport evolves 
upon increasing $T$ for all finite temperatures $T>0$.
The $\tau$- and physical $\alpha$-spin representation and the corresponding microscopic processes 
associated with the spin and charge elementary currents carried by the transport carriers is the suitable framework 
for the further study elsewhere of that interesting scientific issue.

As a first needed step, in this paper we clarify the role of the 1D Hubbard model's 
global $[SU (2)\times SU(2)\times U(1)]/Z_2^2$  symmetry in the finite-temperature transport properties 
in terms of such microscopic processes. For $U>0$ that model describes correlations associated with the on-site repulsion of 
$N = N_{\uparrow} + N_{\downarrow}$ electrons of spin projection $\sigma = \uparrow,\downarrow$.
We use natural units in which the Planck constant, electronic charge, and lattice spacing
are equal to one, with $N_a=L$ being an even integer number. We then consider 
in the thermodynamic limit, $L\rightarrow\infty$, the Hamiltonian of the 1D Hubbard model 
in a magnetic field magnetic field $h \geq 0$ 
and chemical potential $\mu \geq 0$ being under periodic boundary conditions given by,
\begin{equation}
\hat{H} = -t\sum_{\sigma, j}\left[c_{j,\sigma}^{\dag}\,c_{j+1,\sigma} + 
{\rm h.c.}\right] + U\sum_{j}\hat{\rho}_{j,\uparrow}\hat{\rho}_{j,\downarrow} -
\sum_{\alpha}\mu_{\alpha} {\hat{S}}_{\alpha}^{z} \, .
\label{H}
\end{equation}

Here the sum $\sum_{\alpha}$ is over $\alpha = s,\eta$ with $\mu_{s} = 2\mu_B h$
and $\mu_{\eta} = 2\mu$ where $\mu_B$ is the Bohr magneton, $c_{j,\sigma}^{\dag}$ 
creates one electron of spin projection $\sigma$ at site $j$,
$\hat{\rho}_{j,\sigma}= (\hat{n}_{j,\sigma}-1/2)$, $\hat{n}_{j,\sigma}=c_{j,\sigma}^{\dag}\,c_{j,\sigma}$,
and the two $\alpha = s,\eta$ diagonal generators  ${\hat{S}}_{\alpha}^{z}$
of the $\alpha$-spin $SU(2)$ symmetries are given below in Eq. (\ref{SSSS}) and 
have eigenvalues $S_s^z = {1\over 2}(N_{\uparrow} -  N_{\downarrow})$
and $S_{\eta}^z = {1\over 2}(L - N_{\uparrow} -  N_{\downarrow})$, respectively.
Our $h \geq 0$ and $\mu \geq 0$ values correspond to densities $m_{\eta z} = 2S_{\eta}^z/L = (N_a-N)/L \in [0,1]$ and 
$m_{s z} = 2S_{s}^z/L = (N_{\uparrow} - N_{\downarrow})/L\in [0, (1-m_{\eta z})]$.

The paper is organized as follows. The 1D Hubbard model's global $[SU (2)\times SU(2)\times U(1)]/Z_2^2$ symmetry
and the related $\tau$- and physical $\alpha$-spin representation used in the studies of this
paper are the subjects addressed in Sec. \ref{SECII}. In Sec. \ref{SECIII} the spin and charge carriers are 
identified and their $\alpha =s,\eta$ $\alpha$-spin elementary currents
introduced. The microscopic processes associated with such spin and charge elementary currents 
that control the $h>0$ spin and $\mu >0$ charge conducting-phases spin and charge Drude weights 
and the $h=0$ spin and $\mu =0$ charge diffusion constants, respectively,
are the issues studied in Sec. \ref{SECIV}. Finally, the conclusions and discussion 
of the results are presented in Sec. \ref{SECV}. Some complementary side results needed for 
the studies of this paper are presented in the Appendix.

\section{Symmetry and the $\tau$- and physical $\alpha$-spin representation}
\label{SECII}

Our study of the finite-temperature spin and charge transport in the 1D Hubbard model 
accounting for its global $[SU (2)\times SU(2)\times U(1)]/Z_2^2$ symmetry
requires the description of the $\tau$- and physical $\alpha$-spin representation in some detail.
That is the issue addressed in this section.

\subsection{The $[SU (2)\times SU(2)\times U(1)]/Z_2^2$ symmetry}

\label{SECIIA}

Since the mid-1980s, representations in terms of spinons and holons have been widely used to successfully describe the 
static and dynamical properties of both integrable 1D models and the physics of the materials they describe \cite{Essler_05,Pereira_12,Nocera_18}. 
Hence such representations became the paradigm of the 1D physics. 

However, the study of spin and charge transport in the 1D Hubbard model at arbitrary finite temperature $T>0$ requires
the identification of the spin and charge carriers that populate all the $S_s>0$ and $S_{\eta}>0$ energy and momentum eigenstates.
The fulfillment of such a requirement implies the use of a general representation beyond the holon-spinon paradigm 
that is defined in the model's full Hilbert space. 

At $U=0$ the global symmetry of the 1D Hubbard model, Eq. (\ref{H}), is $O(4)/Z_2 = SO(4) \times Z_2$ 
at $h=\mu=0$ \cite{Carmelo_10}. Here the factor $Z_2$ refers to the particle-hole transformation 
on a single spin under which the interacting term of the Hamiltonian, Eq. (\ref{H}), is not invariant \cite{Ostlund_91}.
The requirement of commutability of the global symmetry generators with the $u = U/4t\neq 0$ interacting Hamiltonian 
replaces such a $U=0$ global symmetry by $[SU (2)\times SU(2)\times U(1)]/Z_2^2 = [SO(4) \times U(1)]/Z_2$ 
rather than by only $SO(4)$. The global $\tau$-translational $U(1)$ symmetry beyond $SO(4)$ in $[SO(4) \times U(1)]/Z_2$ was 
preliminarily identified with the charge degrees of freedom \cite{Carmelo_10}. In this paper we confirm it 
rather is a translational symmetry that describes the relative translational degrees of freedom of charge 
and spin carriers \cite{Carmelo_24}.
 
In the global $[SU (2)\times SU(2)\times U(1)]/Z_2^2$ 
symmetry for $u=U/4t>0$ the factor $1/Z_2^2$ imposes that both $(S_{s} + S_{\tau})$ 
and $(S_{\eta} + S_{\tau})$ are integers. 
Here $S_{\tau} = {1\over 2}L_{\eta} = {1\over 2}(L-L_{s})$ is the 
eigenvalue of the generator of the $\tau$-translational $U(1)$ symmetry
and $L_{\eta}$ and $L_{s}$ such that $L_{\eta}+L_{s}=L$ are, as justified below, the numbers 
of physical $\eta$-spins $1/2$ and physical spins $1/2$, respectively.

The global $[SU (2)\times SU(2)\times U(1)]/Z_2^2=SO (3) \times SO (3) \times U (1)$ symmetry of the 
Hubbard model at $\mu =0$ and $h=0$ refers for $U>0$ to it on any bipartite lattice \cite{Carmelo_10}.
It plays an important role in the present case of the 1D Hubbard model's finite-temperature transport 
properties \cite{Carmelo_24}, as confirmed in this paper.

For 1D the diagonal and off-diagonal generators of the spin and $\eta$-spin $SU(2)$ symmetries
in $[SU (2)\times SU(2)\times U(1)]/Z_2^2$ are given by,
\begin{eqnarray}
{\hat{S}}^{z}_{s} & = & {1\over 2}\sum_{j=1}^{L}\Bigl({\hat{n}}_{j,\uparrow} - {\hat{n}}_{j,\downarrow}\Bigr) \, ; \,
{\hat{S}}^{+}_{s} = \left({\hat{S}}^{-}_{s}\right)^{\dag} = \sum_{j=1}^{L}c_{j,\downarrow}^{\dag}\,c_{j,\uparrow} 
\nonumber \\
{\hat{S}}^{z}_{\eta} & = & {1\over 2}\sum_{j=1}^{L}{\hat{n}}^h_{j} \, ; \,
{\hat{S}}^{+}_{\eta} = \left({\hat{S}}^{-}_{\eta}\right)^{\dag} = \sum_{j=1}^{L}(-1)^j\,c_{j,\downarrow}^{\dag}\,c_{j,\uparrow}^{\dag} \, ,
\label{SSSS}
\end{eqnarray}
where ${\hat{n}}^h_{j} = 1 - {\hat{n}}_{j}$ and $\hat{n}_{j} = \sum_{\sigma}\hat{n}_{j,\sigma}$.

The $u=U/4t >0$ global $\tau$-translational $U(1)$ symmetry beyond $SO(4)$ in $[SO(4) \times U(1)]/Z_2$ 
remained hidden because the expression of its generator involves a unitary operator ${\hat{V}}_u$ that
in this paper we call $u$-unitary operator. It is such that we can generate from any of the $4^L$ finite-$u$ energy
eigenstates $\vert\Phi,u\rangle$ of the 1D Hubbard model a corresponding energy eigenstate 
$\vert\Phi,\infty\rangle$ for $u\rightarrow\infty$ as,
\begin{equation}
\vert\Phi,\infty\rangle = {\hat{V}}_u\vert\Phi,u\rangle \hspace{0.20cm}{\rm and}\hspace{0.20cm}
\vert\Phi,u\rangle = {\hat{V}}_u^{\dag}\vert\Phi,\infty\rangle \, .
\label{Phistates}
\end{equation} 
The $u$-unitary operator ${\hat{V}}_u$ is uniquely defined in Eq. (11) of Ref. \onlinecite{Carmelo_17A} by its matrix 
elements between the model's $4^L$ finite-$u$ energy and momentum eigenstates.

Due to the $u\rightarrow\infty$ highly degenerated spin and $\eta$-spin configurations, there are many choices
for a set of $4^L$ $u\rightarrow\infty$ energy eigenstates of the 1D Hubbard model. A uniquely defined choice
for such $4^L$ $u\rightarrow\infty$ energy eigenstates refers to those generated from the $4^L$ finite-$u$ energy eigenstates
as given in Eq. (\ref{Phistates}).

As for all other choices of $u\rightarrow\infty$ energy eigenstates, the lattice electronic occupancy configurations of the 
specific states $\vert\Phi,\infty\rangle$, Eq. (\ref{Phistates}), are such that the numbers of $\sigma = \uparrow$ 
and $\sigma = \downarrow$ singly occupied sites, unoccupied sites, and doubly occupied sites by electrons are good quantum numbers.
Although such numbers are not good quantum numbers at finite $u=U/4t$ values, the construction of the
exact many-electron energy eigenstates $\vert\Phi,u\rangle$ involves at each finite value of $u=U/4t$ the transformation 
performed by the $u$-unitary operator ${\hat{V}}_u$. 

As discussed below in Secs. \ref{SECIIC} and \ref{SECIID}, 
the corresponding Bethe-ansatz $u$-independent quantum numbers that define a finite-$u$ energy
eigenstate $\vert\Phi,u\rangle$ are directly related to the above four good 
quantum numbers of the corresponding state $\vert\Phi,\infty\rangle = {\hat{V}}_u\vert\Phi,u\rangle$,
Eq. (\ref{Phistates}), with which $\vert\Phi,u\rangle$ has a one-to-one relationship.

On the other hand, the information for finite values of $u=U/4t$
on the lattice electronic occupancy configurations of any
energy eigenstate $\vert\Phi,u\rangle$ for which the numbers of $\sigma = \uparrow$ 
and $\sigma = \downarrow$ singly occupied sites, unoccupied sites, and doubly occupied sites 
by electrons {\it are not} good quantum numbers is for each such a finite value of $u$ stored in its involved
wave function amplitudes $\langle x_1,...,x_{N_{\sigma}},\sigma_1,...,\sigma_{N_{\sigma}}\vert\Phi,u\rangle$,
as discussed in Sec. \ref{SECIID}.

The generator of the hidden $\tau$-translational $U(1)$ symmetry whose eigenvalues read
$S_{\tau} = {1\over 2}L_{\eta} = {1\over 2}(L - L_{s})$ is given by,
\begin{equation}
{\hat{S}}_{\tau} = {1\over 2}\sum_{j=1}^L\Bigl(1 - \sum_{\sigma}{\tilde{n}}_{j,\sigma}{\tilde{n}}^h_{j,-\sigma}\Bigr) \, ,
\label{hatStau}
\end{equation}
where,
\begin{equation}
{\tilde{n}}_{j,\sigma} = {\tilde{c}}_{j,\sigma}^{\dag}\,{\tilde{c}}_{j,\sigma} 
\hspace{0.20cm}{\rm and}\hspace{0.20cm}
{\tilde{n}}^h_{j,-\sigma} = 1 - {\tilde{n}}_{j,-\sigma} \, ,
\label{nn}
\end{equation}
and the operators read,
\begin{equation}
{\tilde{c}}_{j,\sigma}^{\dag} =
{\hat{V}}_u^{\dag}\,c_{j,\sigma}^{\dag}\,{\hat{V}}_u 
\hspace{0.20cm}{\rm and}\hspace{0.20cm}
{\tilde{c}}_{j,\sigma} =
{\hat{V}}_u^{\dag}\,c_{j,\sigma}\,{\hat{V}}_u \, .
\label{rotated-operators}
\end{equation} 
Here $c_{j,\sigma}^{\dag}$ and $c_{j,\sigma}$ are the usual creation and annihilation operators
of one electron of spin projection $\sigma$ at the lattice site $j$.

All seven generators of the global $[SU (2)\times SU(2)\times U(1)]/Z_2^2$ symmetry
given in Eqs. (\ref{SSSS}) and (\ref{hatStau}) commute with the Hamiltonian, Eq. (\ref{H}),
at $h=0$ and $\mu =0$ \cite{Carmelo_10}. Within the $\tau$- and physical $\alpha$-spin representation 
valid for $u>0$, the complete set of $4^L$ energy eigenstates that span the
full Hilbert space for {\it all} magnetic field $h$ and chemical potential $\mu$ values in 
the Hamiltonian, Eq. (\ref{H}), are generated by the corresponding $4^L$ irreducible representations of the global 
$[SU (2)\times SU(2)\times U(1)]/Z_2^2$ symmetry. 

As discussed below in Secs. \ref{SECIID} and \ref{SECIIIA}, this is confirmed by Eqs. (\ref{FullDimension})-(\ref{NNsinglet}) of 
the Appendix where the numbers $M_{\alpha n}$ are for $\alpha = s,\eta$ and $n=1,...,\infty$ 
good quantum numbers defined below in Sec. \ref{SECIIB}. Also the spin $S_s$ and its projection 
$S_s^z ={1\over 2}(N_{\uparrow} - N_{\downarrow})$,
the $\eta$-spin $S_{\eta}$ and its projection $S_{\eta}^z= {1\over 2}(L-N_{\uparrow} - N_{\downarrow})$, and 
$S_{\tau} = {1\over 2}L_{\eta} = {1\over 2}(L-L_{s})$ are good quantum numbers. 

\subsection{The $\tau$- and physical $\alpha$-spin representation's  
squeezed effective lattices}
\label{SECIIB}

There are two equivalent representations that account for the transformation performed by the
$u$-unitary operator ${\hat{V}}_u$. The first associates the operators ${\tilde{c}}_{j,\sigma}^{\dag}$ and ${\tilde{c}}_{j,\sigma}$,
Eq. (\ref{rotated-operators}), with rotated electrons of spin projection $\sigma$ acting onto the original lattice site $j$, as in Refs. \onlinecite{Carmelo_10,Carmelo_18,Carmelo_17A}. On acting onto that lattice
to generate the involved finite-$u$ lattice occupancy configurations that describe an energy eigenstate $\vert\Phi,u\rangle$, 
the rotated-electron operators ${\tilde{c}}_{j,\sigma}^{\dag}$ and ${\tilde{c}}_{j,\sigma}$ are inherently constructed to
rather generate the occupancy configurations of the corresponding state 
$\vert\Phi,\infty\rangle = {\hat{V}}_u\vert\Phi,u\rangle$, Eq. (\ref{Phistates}). For the latter
the numbers of $\sigma = \uparrow$ and $\sigma = \downarrow$ singly occupied sites, unoccupied sites, and doubly 
occupied sites are good quantum numbers.

Equivalently and alternatively, we can consider that for any finite-$u$ energy eigenstate $\vert\Phi,u\rangle$
the original lattice with $j=1,...,L$ sites is mapped by the $u$-unitary operator ${\hat{V}}_u$ onto 
a {\it main effective lattice} also with $j=1,...,L$ sites that is inherently constructed 
to its occupancy configurations being those of $\vert\Phi,\infty\rangle = {\hat{V}}_u\vert\Phi,u\rangle$.
Hence the usual electron operators $c_{j,\sigma}^{\dag}$ and $c_{j,\sigma}$ on acting onto that lattice
generate occupancy configurations for a $u>0$ energy eigenstate $\vert\Phi,u\rangle$ that
are those of the corresponding state $\vert\Phi,\infty\rangle = {\hat{V}}_u\vert\Phi,u\rangle$, Eq. (\ref{Phistates}), 
in the original lattice. Hence in the main effective lattice the energy eigenstates $\vert\Phi,u\rangle$ 
have numbers of $\sigma = \uparrow$ and $\sigma = \downarrow$ singly occupied sites, unoccupied sites, and doubly 
occupied sites by electrons that are good quantum numbers for $u>0$.

Importantly, the six generators of the two $\alpha = s,\eta$ global $SU (2)$ symmetries given in Eq. (\ref{SSSS}) commute with the
$u$-unitary operator ${\hat{V}}_u$. For the above first representation, this implies that such six generators have exactly 
the same expressions in terms of the usual operators $c_{j,\sigma}^{\dag}$ and $c_{j,\sigma}$ and of the
corresponding rotated-electron operators ${\tilde{c}}_{j,\sigma}^{\dag}$ and ${\tilde{c}}_{j,\sigma}$, Eq. (\ref{rotated-operators}),
respectively. 

The $\tau$- and physical $\alpha$-spin representation used in the studies of this paper rather refers to the 
alternative main effective lattice representation. For it, the six generators of the two global $SU (2)$ symmetries 
commuting with the $u$-unitary operator ${\hat{V}}_u$
implies that on acting onto either the original lattice with $j=1,...,L$ sites
or onto the main effective lattice also with $j=1,...,L$ sites, they generate {\it exactly} the same processes 
in terms of electron occupancies. 

However, in contrast to the six generators of the two $\alpha = s,\eta$ global $SU (2)$ symmetries, that of
the $\tau$-translational $U(1)$ symmetry does not commute with the $u$-unitary operator ${\hat{V}}_u$. 
A property of crucial importance of the $\tau$- and physical $\alpha$-spin representation 
is thus that on acting onto the main effective lattice, the generator of that $U(1)$ symmetry, Eq. (\ref{hatStau}), 
has the following expression in terms of the usual electron creation and annihilation operators,
\begin{equation}
{\hat{S}}_{\tau} = {1\over 2}\sum_{j=1}^L\Bigl(1 - \sum_{\sigma}n_{j,\sigma}n^h_{j,-\sigma}\Bigr) \, ,
\label{hatStauMEL}
\end{equation}
where $n^h_{j,-\sigma} = 1 - n_{j,-\sigma}$. This is $1/2$ times the number operator of unoccupied plus
doubly occupied sites.

Hence on acting onto the main effective lattice the seven generators 
of the global $[SU (2)\times SU(2)\times U(1)]/Z_2^2$ symmetry have in terms of usual electron creation and annihilation 
operators the expressions given in Eqs. (\ref{SSSS}) and (\ref{hatStauMEL}).

The $\tau$- and physical $\alpha$-spin representation involves the concept of a squeezed effective lattice 
that is well known in 1D correlated systems \cite{Ogata_90,Penc_97,Kruis_04}. As confirmed below in Sec. \ref{SECIIC}, 
{\it all} squeezed-space effects considered in the following are also accounted for by the Bethe-ansatz
solution of the 1D Hubbard model, Eq. (\ref{H}).

\subsubsection{The main effective lattice degrees of freedom separation}

Consistently with Eq. (\ref{FullDimension}) of the Appendix, within the $\tau$- and physical 
$\alpha$-spin representation of the exact energy eigenstates in terms of 
irreducible representations of the global $[SU (2)\times SU(2)\times U(1)]/Z_2^2$ symmetry,
the degrees of freedom of the main effective lattice's electronic occupancies separate 
in the thermodynamic limit into those of three effective lattices: A $\tau$ effective lattice with $L$ sites and length $L$ whose occupancy 
configurations generate irreducible representations of the global $\tau$-translational $U (1)$ symmetry;
A spin-squeezed effective lattice with $L_s = L - 2S_{\tau}$ sites and length $L$ 
whose occupancy configurations generate irreducible representations of the global $SU (2)$ spin symmetry; 
A $\eta$-spin-squeezed effective lattice with $L_{\eta} = 2S_{\tau}$ sites and length $L$ whose occupancy 
configurations generate irreducible representations of the global $SU (2)$ $\eta$-spin symmetry.

The separation of the main effective lattice's electronic occupancies into their degrees of freedom 
described by the $\tau$ effective lattice, spin-squeezed effective lattice, 
and $\eta$-spin-squeezed effective lattice occupancies applies to subspaces for which  
$S_{\tau} = (L-L_s)/2 =L_{\eta}/2$ is finite and fixed and such that $0<S_{\tau}<L/2$. Indeed, for $S_{\tau}=L_{\eta}/2 = 0$ 
there is no $\eta$-spin-squeezed effective lattice and for $S_{\tau}=L/2$ and thus $L_{s} = 0$ there is no spin-squeezed effective lattice.
In these cases the separation of the main effective lattice's electronic occupancies is only into their degrees of freedom 
described by the $\tau$ effective lattice and one of the $\alpha = s,\eta$ $\alpha$-spin-squeezed effective lattices occupancies.
In the following we focus our attention mainly onto the general case for which $0<S_{\tau}<L/2$.

The {\it $\tau$ effective lattice} (i) $N_{\tau} = L - 2S_{\tau}$ occupied sites ($\tau$-particles) and (ii) $N_{\tau}^h = 2S_{\tau}$ 
unoccupied sites ($\tau$-holes) refer to the (i) $\sigma = \uparrow$ plus $\sigma = \downarrow$ singly occupied sites 
and (ii) unoccupied plus doubly occupied sites, respectively, of the main effective lattice. Since the occupancy configurations of the 
$\tau$ effective lattice generate irreducible representations of the $\tau$-translational $U(1)$ symmetry, they
do not distinguish main-effective-lattice $\sigma = \uparrow$ from $\sigma = \downarrow$ singly occupied sites and 
unoccupied from doubly occupied sites.

The two $\alpha =s,\eta$ {\it $\alpha$-spin-squeezed effective lattices} emerge in the thermodynamic limit 
from squeezed-space effects on the physical $\alpha$-spins moving in the main effective lattice.
A $L_s = L - 2S_{\tau} >0$ energy eigenstate is populated by a number $L_{s,\pm 1/2} = L_{s}/2 \pm S_{s}^z$ of physical 
spins of projection $\pm 1/2$.  They are indeed the {\it physical spins} of the electrons with spin projection $\pm 1/2$ 
that singly occupy the sites of the main effective lattice. 
A $L_{\eta} = 2S_{\tau} >0$ energy eigenstate is populated by a number $L_{\eta,\pm 1/2} = L_{\eta}/2 \pm S_{\eta}^z$ 
of physical $\eta$-spins of projections $+1/2$ and $-1/2$. They describe the $\eta$-spin degrees of freedom of the 
unoccupied and doubly occupied sites, respectively, of the main effective lattice. 

Squeezed-space effects are behind on moving in the main effective lattice
the $L_s = N_{\tau}$ physical spins only ``seeing'' its set of $N_{\tau} = L - 2S_{\tau}$ singly occupied sites 
and the $L_{\eta} = N_{\tau}^h$ physical $\eta$-spins only ``seeing'' its complementary set of $N_{\tau}^h = 2S_{\tau}$ 
unoccupied plus doubly occupied sites.

The length $L$ of the two $\alpha = s,\eta$ emerging $\alpha$-spin-squeezed effective lattices is the same as for the main effective lattice 
whereas their number $j = 1,...,L_{\alpha}$ of sites is such that $L_{\alpha}\leq L$. Therefore, for $0<S_{\tau}<L/2$ their spacing is 
larger than that of the main effective lattice. In the thermodynamic limit it corresponds to the average distance between their $j = 1,...,L_{\alpha}$ sites,
\begin{equation}
a_{\alpha} = {L\over L_{\alpha}}\,a \geq a \hspace{0.20cm}{\rm where}\hspace{0.20cm}
\alpha = s, \eta \, .
\label{aalpha}
\end{equation}
Here $a$ denotes the electronic lattice spacing, which in the natural units of this paper is one.

The $\tau$ effective lattice occupancy configurations generate irreducible representations of the $\tau$-translational
$U(1)$ symmetry in $[SU (2)\times SU(2)\times U(1)]/Z_2^2$ that store information on the relative positions in the main 
effective lattice of the $L_s = N_{\tau}$ sites of the spin effective lattice and $L_{\eta} = N_{\tau}^h$ sites of the 
$\eta$-spin-squeezed effective lattice, respectively. This ensures the independence of the occupancies of the spin-squeezed effective lattice 
and $\eta$-spin-squeezed effective lattice, which
do not ``see'' each other and generate independent irreducible representations of 
the spin $SU (2)$ symmetry and $\eta$-spin $SU (2)$ symmetry, respectively,
in $[SU (2)\times SU(2)\times U(1)]/Z_2^2$.

Accounting for the irreducible representations of the two $\alpha = s,\eta$ $\alpha$-spin $SU(2)$ symmetries,
we find, as for the spin-$1/2$ $XXX$ chain \cite{Carmelo_15,Carmelo_17,Carmelo_20},
that all finite-$S_{\alpha}$ energy eigenstates are populated by physical 
$\alpha$-spins in two types of configurations:\\

(i) A number $N_{\alpha} = N_{\alpha,+1/2} + N_{\alpha,-1/2} = 2S_{\alpha}$ of {\it unpaired physical $\alpha$-spins} that participate in a 
$\alpha$-spin multiplet configuration. The number $N_{\alpha,\pm 1/2 }$ of such unpaired physical $\alpha$-spins of projection $\pm 1/2$ of
an energy eigenstate is solely determined by its values of $S_{\alpha}$ and $S^z_{\alpha}$,
as it reads $N_{\alpha,\pm 1/2} = S_{\alpha} \pm S_{\alpha}^z$, so that,
\begin{equation}
2S_{\alpha}^z = N_{\alpha,+1/2} - N_{\alpha,-1/2} \hspace{0.20cm}{\rm and}\hspace{0.20cm}2S_{\alpha} = N_{\alpha,+1/2} + N_{\alpha,-1/2} 
\label{2Sz2Sq}
\end{equation}
for $\alpha = s,\eta$.\\

(ii) A complementary even number ${\cal{N}}_{\alpha} = {\cal{N}}_{\alpha,+1/2} + {\cal{N}}_{\alpha,-1/2}$
of {\it paired physical $\alpha$-spins} that participate in $\alpha$-spin singlet configurations where
${\cal{N}}_{\alpha,+1/2} = {\cal{N}}_{\alpha,-1/2} = L_{\alpha}/2-S_{\alpha}$ for $\alpha = s,\eta$.
Such $\alpha$-spin singlet configurations of the
${\cal{N}}_{\alpha} = {\cal{N}}_{\alpha,+1/2} + {\cal{N}}_{\alpha,-1/2} = 
L_{\alpha}-2S_{\alpha}$ paired physical $\alpha$-spins of an energy eigenstate
involve a number ${\cal{N}}_{\alpha}/2 = {\cal{N}}_{\alpha,\pm 1/2} = L_{\alpha}/2-S_{\alpha}$ of
$\alpha$-spin singlet pairs.\\

That this holds for {\it all} $\alpha = s,\eta$ finite-$S_{\alpha}$ energy eigenstates 
of the Hamiltonian, Eq. (\ref{H}), much simplifies our study. 
In the thermodynamic limit, the number ${\cal{N}}_{\alpha}/2 = L_{\alpha}/2-S_{\alpha}$ of
$\alpha$-spin singlet pairs of the $\alpha$-spin singlet configurations of an energy eigenstate
are distributed over a set of configurations that we call {\it $\alpha n$-pairs}.
Our general designation $\alpha n$-pairs where $\alpha =s,\eta$ 
refers both to {\it $\alpha 1$-pairs} and {\it $\alpha n$-string-pairs} for $n>1$:\\

(i) The internal degrees of freedom of a $\alpha 1$-pair correspond to one unbound $\alpha$-spin singlet
pair of physical $\alpha$-spins. It is described by a single $n=1$ spin ($\alpha=s)$ 
or charge ($\alpha=\eta)$ real Bethe rapidity. 
Due to both one $\alpha 1$-pair occupying two sites of the $\alpha$-spin-squeezed effective
lattice and further squeezed-space effects described below, the $\alpha 1$-pairs of
an energy eigenstate whose number we denote by $M_{\alpha 1}$ 
move in a $\alpha 1$-squeezed effective lattice with $j = 1,...,L_{\alpha 1}$
sites and length $L$ where the number $L_{\alpha 1}$ is given below.\\

(ii) The internal degrees of freedom of a $\alpha n$-string-pair refer to a number $n>1$ of bound 
$\alpha$-spin singlet pairs of physical $\alpha$-spins. They are bound within a configuration described 
below in Eq. (\ref{LambdaIm}) by a corresponding complex Bethe $\alpha n$-string.
Again due both one $\alpha n$-pair occupying $2n$ sites of the $\alpha$-spin-squeezed effective
lattice and further squeezed-space effects described below, the $\alpha n$-pairs of
an energy eigenstate whose number we call $M_{\alpha n}$ move in a $\alpha n$-squeezed effective lattice with $j = 1,...,L_{\alpha n}$
sites and length $L$ where the number $L_{\alpha n}$ is defined below.\\

It follows that the numbers $N_{\alpha,\pm 1/2}$ and ${\cal{N}}_{\alpha,\pm 1/2}$ of unpaired and paired
physical $\alpha$-spins of projection $\pm 1/2$, respectively, of an energy eigenstate can 
in the thermodynamic limit be exactly expressed as,
\begin{eqnarray} 
&& N_{\alpha,\pm 1/2} = S_{\alpha} \pm S_{\alpha}^z = L_{\alpha}/2 - \sum_{n=1}^{\infty}n\,M_{\alpha n} 
\pm S_{\alpha}^z \hspace{0.20cm}{\rm and}
\nonumber \\
&& {\cal{N}}_{\alpha,\pm 1/2} = L_{\alpha}/2-S_{\alpha} = \sum_{n=1}^{\infty}n\,M_{\alpha n}\hspace{0.20cm}{\rm with} 
\nonumber \\
&& L_{\alpha,\pm 1/2} = N_{\alpha,\pm 1/2} + {\cal{N}}_{\alpha,\pm 1/2}
\nonumber \\
&& L_{\alpha} = L_{\alpha,+1/2} + L_{\alpha,-1/2} \, ,
\label{MM}
\end{eqnarray}
where $L_{\alpha,\pm 1/2}$ is the total number of physical $\alpha$-spins of projection $\pm 1/2$.
The total number $L_s$ of physical spins and $L_{\eta}$ of physical $\eta$-spins that populate
an energy eigenstate can then be expressed as,
\begin{eqnarray}
L_s & = & N_{\tau} = L - 2S_{\tau} = 2S_s + \sum_{n=1}^{\infty}2n\,M_{sn}\hspace{0.20cm}{\rm and}
\nonumber \\
L_{\eta} & = & N_{\tau}^h = 2S_{\tau} = 2S_{\eta} + \sum_{n=1}^{\infty}2n\,M_{\eta n} \, ,
\label{LL}
\end{eqnarray}
respectively. Here $2S_{\alpha} = N_{\alpha} = N_{\alpha,+1/2} + N_{\alpha,-1/2}$ is the number of unpaired physical $\alpha$-spins and 
$\sum_{n=1}^{\infty}2n M_{\alpha n} = {\cal{N}}_{\alpha} = {\cal{N}}_{\alpha,+1/2} + {\cal{N}}_{\alpha,-1/2}$
that of paired physical $\alpha$-spins.

\subsubsection{The $\alpha$-spin-squeezed effective lattices separation
into $\alpha n$-squeezed effective lattices}

As mentioned above, there are further squeezed-space effects on the $\alpha n$-pairs moving in the $\alpha$-spin-squeezed effective lattice  
that are behind the emergence of a $\alpha n$-squeezed effective lattice for each branch of $\alpha n$-pairs. 
This applies to subspaces for which the number $M_{\alpha n}$ of $\alpha n$-pairs is finite and fixed. 

Each $\alpha n$-pair occupies a number $2n$ of $\alpha =s,\eta$ $\alpha$-spin-squeezed effective lattice's sites
where $n$ can have the values $n = 1,...,\infty$. Each such a set of $2n$ sites refers to a single occupied site of the
emerging $\alpha n$-squeezed effective lattice. The corresponding number $M_{\alpha n}$
of $\alpha n$-pairs then move in the $\alpha n$-squeezed effective lattice whose number of sites, 
$j=1,...,L_{\alpha n}$, is given in the following.

The emergence of the $\alpha n$-squeezed effective lattice also involves
squeezed-space effects on the $\alpha n$-pairs moving in the $\alpha$-spin-squeezed effective lattice
that involve their following interplay with: 1- All unpaired physical $\alpha$-spins whose number is $2S_{\alpha}$;
2- The $2n'$ paired physical $\alpha$-spins of each of the $M_{\alpha n'}$ $\alpha n'$-pairs for which $n'<n$;
3- A number $2 (n'-n)$ of paired physical $\alpha$-spins out of the $2n'$
paired physical $\alpha$-spins contained in each of the $M_{\alpha n'}$
$\alpha n'$-pairs for which $n'>n$:\\

(i) - {\it Unpaired physical $\alpha$-spins} - On moving in $\alpha$-spin-squeezed effective lattice, the $\alpha n$-pairs interchange 
position with the unpaired physical $\alpha$-spins in a number $N_{\alpha,+1/2} + N_{\alpha,-1/2}=N_{\alpha} = 2S_{\alpha}$ of sites
of that lattice, which play the role of unoccupied sites of their own $\alpha n$-squeezed effective lattice. In a very simplified schematic representation,
we describe a $\alpha$-singlet pair by $(\uparrow\downarrow)$ and a $\alpha 2$-pair by $[(\uparrow\downarrow)(\uparrow\downarrow)]$.
The latter interchanges position with one unpaired physical $\alpha$-spin of projection $+1/2$ 
described by $\uparrow$ as follows:
\begin{eqnarray}
[(\uparrow\downarrow)(\uparrow\downarrow)]\,\uparrow \hspace{0.20cm}\longrightarrow \hspace{0.20cm}\uparrow\,[(\uparrow\downarrow)(\uparrow\downarrow)]
\nonumber
\end{eqnarray}
The $\alpha 2$-pair sees the unpaired physical $\alpha$-spins as unoccupied sites of its $\alpha 2$-squeezed effective lattice.\\

(ii) - {\it $\alpha n'$-pairs for which $n'<n$} - Such $\alpha n'$-pairs ``see'' and interchange position with $2 (n-n')$
paired physical $\alpha$-spins of each $\alpha n$-pair. They use as unoccupied sites of their own $\alpha n'$-squeezed effective lattice
a corresponding number $2 (n-n')$ of sites out of the $2n$ sites of the $\alpha$-spin-squeezed effective lattice occupied
by each $\alpha n$-pair. However, due to squeezed-space effects the $\alpha n$-pairs do not ``see'' such $\alpha n'$-pairs for which $n'<n$,
which do not directly contribute to their own $\alpha n$-squeezed effective lattice. Again in a very simplified schematic representation,
one $\alpha 2$-pair $[(\uparrow\downarrow)(\uparrow\downarrow)]$ interchanges position with four of the eight spins of one $\alpha 4$-pair described by
$[(\uparrow\downarrow)(\uparrow\downarrow)(\uparrow\downarrow)(\uparrow\downarrow)]$. Due to squeezed-space effects,
the $\alpha 2$-pair ``sees'' the $\alpha 4$-pair merely as,
\begin{eqnarray}
\uparrow\downarrow\uparrow\downarrow
\nonumber
\end{eqnarray}
It thus interchanges position with the $\alpha 4$-pair as follows,
\begin{eqnarray}
[(\uparrow\downarrow)(\uparrow\downarrow)]\,\uparrow\downarrow\uparrow\downarrow\hspace{0.20cm}\longrightarrow \hspace{0.20cm}
\uparrow[(\uparrow\downarrow)(\uparrow\downarrow)]\downarrow\uparrow\downarrow
\nonumber
\end{eqnarray}
\begin{eqnarray}
\longrightarrow \hspace{0.20cm}\uparrow\downarrow[(\uparrow\downarrow)(\uparrow\downarrow)]\uparrow\downarrow
\nonumber
\end{eqnarray}
\begin{eqnarray}
\longrightarrow \hspace{0.20cm}\uparrow\downarrow\uparrow[(\uparrow\downarrow)(\uparrow\downarrow)]\downarrow
\nonumber
\end{eqnarray}
\begin{eqnarray}
\longrightarrow \hspace{0.20cm}\uparrow\downarrow\uparrow\downarrow\,[(\uparrow\downarrow)(\uparrow\downarrow)]
\nonumber
\end{eqnarray}
Indeed, the $\alpha 2$-pair only ``sees'' four spins of the $\alpha 4$-pair 
$[(\uparrow\downarrow)(\uparrow\downarrow)(\uparrow\downarrow)(\uparrow\downarrow)]$
as unoccupied sites of its $\alpha 2$-squeezed effective lattice.\\

(iii) - {\it $2 (n'-n)$ paired physical $\alpha$-spins of $\alpha n'$-pairs for which $n'>n$} - 
As reported above, on moving in the $\alpha$-spin-squeezed effective lattice, the $\alpha n$-pairs
``see'' and interchange position with $2 (n'-n)$ paired physical $\alpha$-spins of $\alpha n'$-pairs for which $n'>n$.
They use as unoccupied sites of their own $\alpha n$-squeezed effective lattice
a corresponding number $2 (n'-n)$ of sites out of the $2n'$ sites of the $\alpha$-spin-squeezed effective lattice occupied
by each $\alpha n'$-pair. Due to squeezed-space effects, they though do not ``see'' the remaining $2n$ paired physical $\alpha$-spins of 
$\alpha n'$-pairs for which $n'>n$ whose number is $2n'$.
On the other hand and also due to squeezed-space effects, such $\alpha n'$-pairs for which $n'>n$ do not ``see'' the $\alpha n$-pairs.
In the above very simplified schematic representation, one $\alpha 4$-pair 
$[(\uparrow\downarrow)(\uparrow\downarrow)(\uparrow\downarrow)(\uparrow\downarrow)]$
does not ``see'' the four $\alpha$ spins of one $\alpha 2$-pair $[(\uparrow\downarrow)(\uparrow\downarrow)]$
because they do not exist in the $\alpha 4$-effective squeezed lattice.

In both (ii) and (iii), one $\alpha n$-pair uses $2 (n'-n)$ sites out of the $2n'$ sites of 
the $\alpha$-spin-squeezed effective lattice occupied by each $\alpha n'$-pair for which $n'>n$ as unoccupied 
sites of its own $\alpha n$-squeezed effective lattice. 
In (ii) the numbers $n'$ and $n$ are interchanged yet the processes are the same and for simplicity 
let us consider that $n'>n$. Why only $2 (n'-n)$ sites rather than all $2n'$ sites of each $\alpha n'$-pair for which $n'>n$ are
used by $\alpha n$-pairs as unoccupied sites of their own $\alpha n$-squeezed effective lattice? 

The clarification of this issue involves 1D linear site order. On moving to the left or two the right
in the $\alpha$-spin-squeezed effective lattice, the 
$\alpha n$-pair interchanges position with the first $2(n'-n)$ paired physical
$\alpha$-spins of each $\alpha n'$-pair for which $n'>n$. The point is that the remaining $2n$ paired physical $\alpha$-spins
of the $\alpha n'$-pair become those of the $\alpha n$-pair after it has interchanged position with
its first $2(n'-n)$ paired physical $\alpha$-spins. 
This process involves some degree of binding breaking and
rebinding of the $n'$ and $n$ singlet pairs of physical $\alpha$-spins of the involved $\alpha n'$-pair and $\alpha n$-pair.

The net result is indeed that the $\alpha n$-pair only 
interchanges position with the first set of $2(n'-n)$ paired 
physical $\alpha$-spins of the $\alpha n'$-pair for which $n'>n$.
It follows that out of the $2n'$ sites the $\alpha$-spin-squeezed effective lattice occupied by
the $2n'$ unpaired physical $\alpha$-spins of each $\alpha n'$-pair, a $\alpha n$-pair 
only uses $2(n'-n)$ sites as unoccupied sites of its own $\alpha n$-squeezed effective lattice, as it
only interchanges position with a corresponding number $2(n'-n)$ of paired physical $\alpha$-spins.

The result of the above combined squeezed-space effects is that out of the 
${\cal{N}}_{\alpha,\pm 1/2} = \sum_{n=1}^{\infty}n\,M_{\alpha n}$ paired physical $\alpha$-spins of projection $\pm 1/2$
that populate an energy eigenstate, Eq. (\ref{MM}), the $\alpha n$-pairs only ``see'' and interchange position with a number 
${\cal{N}}_{\alpha n,\pm 1/2}<{\cal{N}}_{\alpha,\pm 1/2}$ of such paired physical $\alpha$-spins given by,
\begin{eqnarray}
&& {\cal{N}}_{\alpha n,\pm 1/2} = \sum_{n'=n+1}^{\infty}(n'-n)M_{\alpha n'}\hspace{0.20cm}{\rm so}\hspace{0.20cm}{\rm that}
\nonumber \\
&& N_{\alpha n,\pm 1/2} = N_{\alpha,\pm 1/2} + {\cal{N}}_{\alpha n,\pm 1/2}
\nonumber \\
&& M_{\alpha n}^h = N_{\alpha n,+1/2} + N_{\alpha n,-1/2} \, .
\nonumber \\
&& \hspace{0.8cm} =  2S_{\alpha} + \sum_{n'=n+1}^{\infty}2(n'-n)M_{\alpha n'}\hspace{0.20cm}{\rm and}
\nonumber \\
&& L_{\alpha n} = M_{\alpha n} + M_{\alpha n}^h \, .
\label{Nnh}
\end{eqnarray}
Here $N_{\alpha n,\pm 1/2}$ is the total number of physical $\alpha$-spins of projection $\pm 1/2$ which the $\alpha n$-pairs ``see'' 
and interchange position with.

The above processes then determine the number of $j = 1,...,L_{\alpha n}$ sites of each $\alpha n$-squeezed effective, 
$L_{\alpha n} = M_{\alpha n} + M_{\alpha n}^h$, Eq. (\ref{Nnh}). The corresponding 
numbers of occupied and unoccupied sites read $M_{\alpha n}$ and 
$M_{\alpha n}^h = 2S_{\alpha} + \sum_{n'=n+1}^{\infty}2(n'-n)M_{\alpha n'}$, respectively.

The $\alpha n$-squeezed effective lattices have the same length $L$ as both the main effective lattice
and the $\alpha$-spin-squeezed effective lattice and a number of sites 
$L_{\alpha n} =  2S_{\alpha} + \sum_{n'=n+1}^{\infty}2(n'-n)M_{\alpha n'} + M_{\alpha n}$ smaller
than both $L = \sum_{\alpha} L_{\alpha}$ and $L_{\alpha} = 2S_{\alpha} + \sum_{n=1}^{\infty}2n M_{\alpha n}$.
In the thermodynamic limit its spacing corresponds to the average distance between 
its $L_{\alpha n} = M_{\alpha n} + M_{\alpha n}^h$ sites,
\begin{equation}
a_{\alpha n} = {L\over L_{\alpha n}}\,a > a \, .
\label{an}
\end{equation}
This spacing is larger than that of the corresponding $\alpha$-spin-squeezed effective lattice,
Eq. (\ref{aalpha}).

\subsection{Relation to the quantum numbers of the Bethe ansatz and beyond it}
\label{SECIIC}

\subsubsection{Relation to the quantum numbers of the Bethe ansatz}
\label{SECIIC1}

Under the construction of the exact many-electron energy and momentum eigenstates,
the Bethe ansatz itself implicitly accounts for the transformation performed by the $u$-unitary
operator ${\hat{V}}_u$ and related irreducible representations of the global 
$[SU (2)\times SU(2)\times U(1)]/Z_2^2$ symmetry.
For that ansatz, the position space associated with the $\tau$ effective lattice with $j = 1,...,L$ sites and length $L$
and the position spaces associated with the $\alpha =s,\eta$ and $n=1,...,\infty$ 
$\alpha n$-squeezed effective lattices with 
$j = 1,...,L_{\alpha n}$ sites and length $L$ are associated with momentum spaces that correspond to the $\tau$-band with 
$j = 1,...,L$ discrete momentum values $q_j$ and the set of 
$\alpha n$-bands with $j = 1,...,L_{\alpha n}$ discrete momentum
values $q_j$, respectively. 

Actually, in the thermodynamic limit the Bethe-ansatz quantum numbers \cite{Lieb_68,Takahashi_72} 
$I_j^{\tau}$ and $I_j^{\alpha n}$ are the discrete $\tau$-band and $\alpha n$-band momentum values, 
\begin{eqnarray}
q_j & = & {2\pi\over L} I_j^{\tau}\hspace{0.20cm}{\rm for}\hspace{0.20cm}j = 1,..,L \hspace{0.20cm}{\rm and}
\nonumber \\
q_j & = & {2\pi\over L} I_j^{\alpha n}\hspace{0.20cm}{\rm for}\hspace{0.20cm}j = 1,..,L_{\alpha n} \, ,
\label{qq}
\end{eqnarray}
respectively, in units of $2\pi/L$, where,
\begin{eqnarray}
I_j^{\tau} & = & 0,\pm 1, \pm 2,...\hspace{0.20cm}{\rm for}\sum_{\alpha =s,\eta}M_{\alpha n}\hspace{0.20cm}{\rm even}
\nonumber \\
& = & \mp 1/2,\pm 3/2, \pm 5/2,...\hspace{0.20cm}{\rm for}\hspace{0.20cm}\sum_{\alpha =s,\eta}M_{\alpha n}\hspace{0.20cm}{\rm odd} \, ,
\label{Itau}
\end{eqnarray}
and
\begin{eqnarray}
&& I_j^{\alpha n} = 0,\pm 1, \pm 2,...\hspace{0.20cm}{\rm for}\hspace{0.20cm}2S_{\alpha} + M_{\alpha n}\hspace{0.20cm}{\rm odd}
\nonumber \\
&& \hspace{0.30cm} = \mp 1/2,\pm 3/2, \pm 5/2,...\hspace{0.20cm}{\rm for}\hspace{0.20cm}2S_{\alpha} + M_{\alpha n}\hspace{0.20cm}{\rm even} \, .
\label{Ian}
\end{eqnarray}

That the lengths of the $\tau$ effective lattice and all $\alpha\nu$-squeezed effective lattices is $L$ is consistent
with the $\tau$-band and $\alpha\nu$-bands discrete momentum values, Eqs. (\ref{qq})-(\ref{Ian}),
having separation $q_{j+1}-q_j = 2\pi/L$.
The $\tau$-band $j=1,...,L$ and $\alpha n$-bands $j=1,...,L_{\alpha n}$ discrete set of $q_j$'s,
Eq. (\ref{qq}), have values in the intervals $q_j \in [q_{\tau}^-,q_{\tau}^+]$ and $q_j \in [q_{\alpha n}^-,q_{\alpha n}^+]$, 
respectively, where,
\begin{eqnarray}
q_{\tau}^{\pm} & = & q_{\tau 0}^{\pm}  + q^{\Delta} 
\nonumber \\
q_{\tau 0}^{\pm} & = & \pm{\pi\over L} (L-1) 
\hspace{0.20cm}{\rm for}\hspace{0.20cm}\sum_{\alpha =s,\eta}\sum_{n=1}^{\infty}M_{\alpha n}\hspace{0.20cm}{\rm odd}
\nonumber \\
q_{\tau 0}^+ & = & + \pi \hspace{0.20cm}{\rm and}\hspace{0.20cm}q_{\tau 0}^- = - {\pi\over L} (L - 2)
\hspace{0.20cm}{\rm for}\sum_{\alpha =s,\eta}\sum_{n=1}^{\infty}M_{\alpha n}\hspace{0.20cm}{\rm even}
\nonumber \\
q_{\alpha n}^{\pm} & = & \pm {\pi\over L}(L_{\alpha n} -1) 
\hspace{0.20cm}{\rm for}\hspace{0.20cm}\alpha = s,\eta\hspace{0.20cm}{\rm and}\hspace{0.20cm}n = 1,....\infty \, ,
\label{qqq}
\end{eqnarray}
and the shift $\tau$-band momentum $q^{\Delta}$ is defined in Eq. (\ref{Bat}) of the Appendix.
For ground states and for excited states generated from them by a finite number of processes,
$q^{\Delta}=0$ in the thermodynamic limit.

Hence we can define the $N_{\tau}$ $\tau$-particles, $N_{\tau}^h$ $\tau$-holes, $M_{\alpha n}$ $\alpha n$-pairs,
and $M_{\alpha n}^h$ $\alpha n$-holes both (i) in the $\tau$ effective lattice and $\alpha n$-squeezed effective lattices
in position space and (ii) in the $\tau$-band and $\alpha n$-bands in momentum space.
The corresponding sets of quantum numbers $\{I_j^{\tau}\}$ and $\{I_j^{\alpha n}\}$, Eqs. (\ref{Itau})-(\ref{Ian}), and thus the 
corresponding sets $\{q_j\}$, Eq. (\ref{qq}), of discrete $\tau$-band and discrete $\alpha n$-band momentum values,
respectively, have Pauli-like occupancies: The $\tau$-band and $\alpha n$-band momentum distributions read 
(i) $N_{\tau} (q_j) = 1$ and $M_{\alpha n} (q_j) = 1$ and (ii) $N_{\tau} (q_j) = 0$ and $M_{\alpha n} (q_j) = 0$
for (i) occupied and (ii) unoccupied $q_j$'s, respectively. We also define corresponding $\tau$-hole
and $\alpha n$-hole distributions $N_{\tau}^h (q_j) = 1 - N_{\tau} (q_j)$ and
$M_{\alpha n}^h (q_j) = 1 - M_{\alpha n} (q_j)$, respectively.

Each of the energy eigenstates that span the full Hilbert space is specified by 
$S_{\tau} = {1\over 2}N_{\tau}^h = {1\over 2}L_{\eta} = {1\over 2}(L - L_s)$, numbers
$N_{\alpha,\pm 1/2} = S_{\alpha} \pm S_{\alpha}^z$ of unpaired $\alpha$-physical spins of projection $\pm 1/2$ 
for $\alpha = s,\eta$, and a set of distributions $N_{\tau} (q_j)$ and
$\{M_{\alpha n} (q_j)\}$ for $\alpha = s,\eta$ and $n=1,...,\infty$. 

Only the ${\alpha n}$-bands for which $M_{\alpha n}>0$ contribute to the
physical quantities associated with the corresponding energy eigenstate.
The eigenvalues and expectation values of different physical operators can be written in functional form in terms
of the distributions $N_{\tau} (q_j)$ and $M_{\alpha n} (q_j)$ or 
$N_{\tau}^h (q_j)$ and $M_{\alpha n}^h (q_j)$. For example, the momentum eigenvalues can 
in the thermodynamic limit be written as,
\begin{eqnarray}
&& P = P_{\tau} + \sum_{n=1}^{\infty}\sum_{j=1}^{L_{sn}}q_j\,M_{sn} (q_j)
\nonumber \\
&& + \sum_{n=1}^{\infty}\sum_{j=1}^{L_{\eta n}}\Bigl(\pi (n+1) - q_j\Bigr)\,M_{\eta n} (q_j) +\,\pi\,N_{\eta,-1/2}
\nonumber \\
&& {\rm where}\hspace{0.20cm}P_{\tau} = \sum_{j=1}^{L}q_j\,N_{\tau} (q_j) = - \sum_{j=1}^{L}q_j\,N_{\tau}^h (q_j) \, .
\label{PP}
\end{eqnarray}

Some physical quantities such as the energy eigenvalues and the $\alpha =s,\eta$ $\alpha$-spin current
expectation values whose general expressions are given below in other sections depend
on the occupancies of the $\tau$-band and $\alpha =s,\eta$ $\alpha n$-bands 
specified by the set of distributions $N_{\tau} (q_j)$ and $\{M_{\alpha n} (q_j)\}$
through corresponding dependences on a $\tau$-band momentum rapidity function 
$k (q_j)$ and $\alpha n$-band rapidity functions $\Lambda_{\alpha n} (q_j)$.

In the thermodynamic limit, the set of $\tau$-band and $\alpha n$-band discrete momentum values
$q_j$ such that $q_{j+1}-q_j = 2\pi/L$ can be described by continuum $\tau$-band and $\alpha n$-band continuum
momentum variables $q \in [q_{\tau}^{-},q_{\tau}^{+}]$ and $q \in [q_{\alpha n}^{-},q_{\alpha n}^{+}]$, respectively,
whose limiting momentum values are given in Eq. (\ref{qqq}). 
The corresponding $\tau$-band momentum rapidity function $k (q)$ and $\alpha n$-band rapidity functions $\Lambda_{\alpha n} (q)$
can be defined in terms their inverse functions: The $\tau$-band momentum function 
$q_{\tau}= q_{\tau} (k)$ whose continuum $\tau$-band momentum rapidity variable has
values in the interval $k \in [-\pi,\pi]$ and $\alpha n$-band momentum functions 
$q_{\alpha n} = q_{\alpha n} (\Lambda)$ whose continuum $\alpha n$-band rapidity variable has
values in the interval  $\Lambda\in [-\infty,\infty]$, respectively. They contain the same information as their inverse functions,
the $\tau$-band momentum rapidity function $k (q)$ and $\alpha n$-band rapidity functions $\Lambda_{\alpha n} (q)$,
respectively. For each energy eigenstate, the former functions are solutions of the coupled 
Bethe-ansatz equations given in functional form in Eqs. (\ref{Bat})-(\ref{BAetan}) of the Appendix.

Also the corresponding $\tau$-band momentum-rapidity-variable distribution ${\bar{N}}_{\tau} (k)$ and $\alpha n$-band 
rapidity-variable distributions ${\bar{M}}_{\alpha n} (\Lambda)$ store the same information
as the $\tau$-band momentum distribution $N_{\tau} (q)$ and set of
$\alpha n$-band momentum distributions $M_{\alpha n} (q)$, respectively. The former distributions
and the corresponding hole distributions obey the following relations,
\begin{eqnarray}
&& {\bar{N}}_{\tau} (k) = 1 -  {\bar{N}}^h_{\tau} (k)
\hspace{0.20cm} {\rm for}\hspace{0.20cm} k\in [-\pi,\pi]
\hspace{0.20cm} {\rm and}
\nonumber \\
&& {\bar{M}}_{\alpha n} (\Lambda) = 1 -  {\bar{M}}^h_{\alpha n} (\Lambda)
\hspace{0.20cm} {\rm for}\hspace{0.20cm} \Lambda\in [-\infty,\infty]
\hspace{0.20cm} {\rm where}
\nonumber \\
&& {\bar{N}}_{\tau} (k (q)) = N_{\tau} (q)\hspace{0.20cm} {\rm and}\hspace{0.20cm}{\bar{M}}_{\alpha n} (\Lambda_{\alpha n} (q)) 
= M_{\alpha n} (q) 
\nonumber \\
&& \hspace{0.80cm}{\rm for}\hspace{0.20cm}\alpha = s,\eta
\hspace{0.20cm} {\rm and}\hspace{0.20cm}n=1,...,\infty \, .
\label{NNrela}
\end{eqnarray}

The $N_{\tau} = L_s$ $\tau$-particles and $N_{\tau}^h = L_{\eta}$ $\tau$-holes refer to the $\tau$-translational $U(1)$ symmetry 
and thus have no internal degrees of freedom. Their translational degrees of freedom are associated with the $\tau$-band 
momentum values $q_j \in [q_{\tau}^-,q_{\tau}^+]$. 

The translational degrees of freedom of the $\alpha n$-pairs are associated with the $\alpha n$-band 
momentum values $q_j \in [q_{\alpha n}^-,q_{\alpha n}^+]$.
Their internal degrees of freedom correspond to a number $n=1,...,\infty$ of
$\alpha$-spin singlet pairs of physical $\alpha$-spins that are described by a real $(n=1)$ or a $(n>1)$ complex Bethe rapidity.
The corresponding $\alpha n$-string rapidity's structure depends on the system size. In the thermodynamic limit 
associated with the $\tau$- and $\alpha$-spin representation in which that structure simplifies, 
the rapidities $\Lambda_{\alpha n,l,j} = \Lambda_{\alpha n,l} (q_j) = \Lambda_{\alpha n,l} (q)$ can both for $n=1$ and $n>1$ be expressed as
\cite{Takahashi_72},
\begin{equation}
\Lambda_{\alpha n,l} (q) = \Lambda_{\alpha n} (q) + i(n + 1 -2l)\,u \hspace{0.20cm}{\rm where}\hspace{0.20cm}l=1,...,n \, .
\label{LambdaIm}
\end{equation}
Here the index $l=1,...,n$ labels each of the $n=1,...,\infty$ singlet $\alpha$-spin pairs contained in the $\alpha n$-pair.
The rapidity $\Lambda_{\alpha n,l} (q)$ is real for $n=1$ , $\Lambda_{\alpha 1,1} (q)=\Lambda_{\alpha 1} (q)$, and its real part refers 
to the $\alpha n$-band rapidity function $\Lambda_{\alpha n} (q)$ for $n>1$.

For $n>1$ the imaginary part, $i(n + 1 -2l)\,u$, of $\Lambda_{\alpha n,l} (q)$ is in general finite and describes 
the binding of $l=1,...,n$ $\alpha$-spin singlet pairs of physical $\alpha$-spins within each $\alpha n$-pair. 
The $\alpha n$ rapidities, Eq. (\ref{LambdaIm}), thus describe the internal degrees of freedom of each of the 
$\alpha n$-pairs of an energy eigenstate that contain its ${\cal{N}}_{\alpha} = L_{\alpha} - 2S_{\alpha} = \sum_{n=1}^{\infty}2n\,M_{\alpha n}$ 
paired physical $\alpha$-spins.

On the other hand, the $\alpha$-spin internal degrees of freedom of the $N_{\alpha} = 2S_{\alpha}$ 
unpaired physical $\alpha$-spins in the $\alpha$-spin multiplet configuration of
a finite-$S_{\alpha}$ energy eigenstate is an issue beyond the Bethe ansatz.
Here $N_{\alpha} = N_{\alpha,+1/2} + N_{\alpha,-1/2}$ where $N_{\alpha,\pm 1/2} = S_{\alpha} \pm S_{\alpha}^z$, Eq. (\ref{MM}).

The highest weight states (HWSs) and the lowest weight states (LWSs) of the $\alpha = s,\eta$ $\alpha$-spin $SU(2)$ 
symmetries are two important types of energy eigenstates. They are such that $S_{\alpha}^z = S_{\alpha}$ and 
$S_{\alpha}^z = - S_{\alpha}$, respectively. It is well known that the Bethe ansatz refers only to subspaces spanned either by the 
HWSs or the LWSs of the $\alpha = s,\eta$ $\alpha$-spin $SU(2)$ symmetries \cite{Essler_91}. 
For such states, all the $N_{\alpha} = 2S_{\alpha}$ unpaired physical $\alpha$-spins
have the same projection $+1/2$ or $-1/2$, respectively. 

\subsubsection{$\tau$- and $\alpha$-spin representation beyond the Bethe ansatz}
\label{SECIIC12}

Beyond the Bethe-ansatz subspace, the $\tau$- and physical $\alpha$-spin representation applies to the whole Hilbert space and
thus accounts for the internal degrees of freedom of the $N_{\alpha,\pm 1/2} = S_{\alpha} \pm S_{\alpha}^z$ 
unpaired physical $\alpha$-spins in the $\alpha$-spin multiplet configuration of
a finite-$S_{\alpha}$ energy eigenstate.

It is useful for our studies to denote the energy eigenstates $\vert\Phi,u\rangle$, Eq. (\ref{Phistates}),
by $\left\vert l_{\rm r}^{u},S_{\tau},S_{s},S_{s}^z,S_{\eta},S_{\eta}^z\right\rangle$.
Here $l_{\rm r}^{u}$ refers both to a fixed value of $u=U/4t$ and
the set of $u$-independent quantum numbers other than $S_{\tau}$, $S_{s}$, $S_{s}^z$, $S_{\eta}$, and
$S_{\eta}^z$ needed to specify the state. At fixed values of $S_{\tau}$
and $S_{\alpha}$ there are for $\alpha =s,\eta$ several sets $\{M_{\alpha n}\}$ of $\alpha n$-pair numbers that satisfy the
exact sum rule $L_{\alpha} - 2S_{\alpha} = \sum_{n = 1}^{\infty}2n M_{\alpha n}$. 
$l_{\rm r}^{u}$ includes one of such sets of fixed finite values $\{M_{\alpha n}\}$ 
both for $\alpha =s$ and $\alpha = \eta$, respectively. In addition,
it includes full information on the distributions $N_{\tau} (q_j)$ and 
$\{M_{\alpha n} (q_j)\}$ for $\alpha = s$, $\alpha = \eta$, and $n=1,...,\infty$ that
specify which discrete momentum values $q_j$'s of the $\tau$-band and each $\alpha n$-band 
are occupied and unoccupied. 

Consider a $\alpha$-spin HWS $\left\vert l_{\rm r}^{u},S_{\tau},S_{s},S_{s},S_{\eta},S_{\eta}\right\rangle$ both
for spin $(\alpha = s)$ and $\eta$-spin $(\alpha = \eta)$ described by the Bethe ansatz.
A number $2S_{\alpha}$ of $\alpha$-spin $SU (2)$ symmetry non-HWSs outside the Bethe-ansatz solution, which refer to different 
$\alpha$-spin multiplet configurations of the $N_{\alpha} = N_{\alpha,+1/2} + N_{\alpha,-1/2} = 2S_{\alpha}$ 
unpaired physical $\alpha$-spins, are generated from that HWS both for $\alpha = s$ and 
$\alpha = \eta$ as \cite{Carmelo_18,Carmelo_17A,Essler_91},
\begin{eqnarray} 
&& \left\vert l_{\rm r}^{u},S_{\tau},S_{s},S_{s}^z,S_{\eta},S_{\eta}^z\right\rangle =
\nonumber \\
&& \prod_{\alpha = s,\eta}{1\over \sqrt{{\cal{C}}_{\alpha}}}({\hat{S}}^{+}_{\alpha})^{n_{\alpha}^z}
\left\vert l_{\rm r}^{u},S_{\tau},S_{s},S_{s},S_{\eta},S_{\eta}\right\rangle 
\nonumber \\
&& {\rm where}\hspace{0.20cm}{\cal{C}}_{\eta} = (n_{\alpha}^z)!\prod_{\iota =1}^{n_{\alpha}^z}(S_{\alpha} +1 - \iota) \, .
\label{state}
\end{eqnarray} 
Here $n_{\alpha}^z = S_{\alpha} - S_{\alpha}^z = N_{\alpha,-1/2} =1,...,2S_{\alpha}$ so that
$S_{\alpha}^z = S_{\alpha} - n_{\alpha}^z$ and the off-diagonal generators ${\hat{S}}^{\pm}_{\alpha}$ 
of the $\alpha = s,\eta$ $\alpha$-spin $SU (2)$ symmetries are given in Eq. (\ref{SSSS}).

For the $\alpha$-spin non-HWSs, Eq. (\ref{state}), the two sets of $n_{\alpha}^z = S_{\alpha} - S_{\alpha}^z = N_{\alpha,-1/2}$ 
and $2S_{\alpha}-n_{\alpha}^z = S_{\alpha} + S_{\alpha}^z = N_{\alpha,+1/2}$ unpaired physical $\alpha$-spins have opposite 
projections $-1/2$ and $+1/2$, respectively. 

The $\alpha$-spin multiplet configurations associated
with the internal degrees of freedom of the $N_{\alpha,\pm 1/2} = S_{\alpha} \pm S_{\alpha}^z$  
unpaired physical $\alpha$-spins of projection $\pm 1/2$ are generated as given in Eq. (\ref{state}). 
Such configurations are associated with the factor $(2S_{\alpha} + 1)$ in the $\alpha$-spin dimension given in
Eq. (\ref{NNsinglet}) of the Appendix.

\subsection{Quantum numbers and wave function amplitudes}
\label{SECIID}

As mentioned in previous sections, the $\tau$- and physical $\alpha$-spin representation
naturally emerges for $U>0$ from the $4^L$ energy and momentum eigenstates that span the full Hilbert space referring
to irreducible representations of the model's global symmetry. The dimension of
that space can thus be expressed as the summation over the
integer values of $2S_{\tau}\geq 0$, $2S_s\geq 0$, and $2S_{\eta}\geq 0$ of the product of three numbers: Those of irreducible
representations of the $\tau$-translational $U(1)$ symmetry, spin $SU (2)$ symmetry, and $\eta$-spin $SU (2)$ symmetry,
respectively, in the model global $[SU (2)\times SU(2)\times U(1)]/Z_2^2$ symmetry, as given in 
Eqs. (\ref{FullDimension})-(\ref{NNsinglet}) of the Appendix.

Consistently with the Bethe ansatz accounting for the transformation performed by the $u$-unitary
operator ${\hat{V}}_u$, except for the value of $u=U/4t>0$ itself, all $u$-independent quantum numbers of
an energy eigenstate $\left\vert l_{\rm r}^{u},S_{\tau},S_{s},S_{s}^z,S_{\eta},S_{\eta}^z\right\rangle$
reported above in Sec. \ref{SECIIC} have exactly the same values for the corresponding state
$\left\vert l_{\rm r}^{\infty},S_{\tau},S_{s},S_{s}^z,S_{\eta},S_{\eta}^z\right\rangle$, which can
be generated from the former state as ${\hat{V}}_u\left\vert l_{\rm r}^{u},S_{\tau},S_{s},S_{s}^z,S_{\eta},S_{\eta}^z\right\rangle$,
Eq. (\ref{Phistates}).

On the other hand, the information on the electronic occupancy configurations in the original lattice
of a finite-$u$ energy eigenstate $\vert\Phi,u\rangle = \left\vert l_{\rm r}^{u},S_{\tau},S_{s},S_{s}^z,S_{\eta},S_{\eta}^z\right\rangle$ 
for which the numbers of $\sigma = \uparrow$ and $\sigma = \downarrow$ singly occupied sites, unoccupied sites, 
and doubly occupied sites by electrons {\it are not} good quantum numbers
is stored in its involved wave function amplitude $\langle x_1,...,x_{N_{\sigma}},\sigma_1,...,\sigma_{N_{\sigma}}\vert\Phi,u\rangle$. 
For a $\alpha$-spin HWS it is defined by Eqs. (2.5)-(2.10) of Ref. \onlinecite{Woynarovich_82},
its extension to a $\alpha$-spin non-HWS, Eq. (\ref{state}), being trivial. 

Only in the $u\rightarrow\infty$ limit, when the numbers of $\sigma = \uparrow$ and $\sigma = \downarrow$ 
singly occupied sites, unoccupied sites, and doubly occupied sites by electrons become good quantum numbers, 
the separation of the independent global $\eta$-spin $SU(2)$ symmetry, global spin $SU(2)$ symmetry,
and global $\tau$-translational $U(1)$ symmetry irreducible representations becomes visible in the form of the
wave function amplitude $\langle x_1,...,x_{N_{\sigma}},\sigma_1,...,\sigma_{N_{\sigma}}\vert\Phi,\infty\rangle$
of an energy eigenstate $\vert\Phi,\infty\rangle = {\hat{V}}_u\vert\Phi,u\rangle$. 

The point is that in the $u=U/4t\rightarrow\infty$ limit the Hubbard model in 1D or on any other bipartite lattice 
has at $h=\mu = 0$ a {\it local} $SU (2)\times SU(2)\times U(1)$ gauge symmetry \cite{Ostlund_91}.
That local symmetry is that directly captured by the dependence on the spatial coordinates of the wave function amplitudes, which 
in that limit can be expressed as $\langle x_1,...,x_{N_{\sigma}},\sigma_1,...,\sigma_{N_{\sigma}}\vert{\hat{V}}_u\vert\Phi,u\rangle$. 
Within the notations used in Ref. \onlinecite{Woynarovich_82A} and as given in its Eq. (2.23), they can be expressed as a 
product of the following three wave function amplitudes,
\begin{eqnarray}
&& (-1)^Q e^{i\sum_{l=1}n_l^d}\,\varphi_1 (y_1^d,...,y_L^d) \, ;\hspace{0.40cm}
\varphi_2 (y_1^s,...,y_{M-L}^s) 
\nonumber \\
&& {\rm and}\hspace{0.40cm} \varphi_3 = \sum_{P} (-1)^P e^{i\{\sum_{j=1}^{N-H-2L} k_{P_j} n^s_{Q_j}\}} \, ,
\label{3wf}
\end{eqnarray}
Here $(-1)^Q e^{i\sum_{l=1}n_l^d}\,\varphi_1$, $\varphi_2$, and $\varphi_3$
are indeed associated with the $\eta$-spin $SU(2)$ symmetry, spin $SU(2)$ symmetry, and $\tau$-translational $U(1)$ symmetry,
respectively. Consistently, $\varphi_1$ and $\varphi_2$ are essentially $SU (2)$ Heisenberg eigenfunctions \cite{Woynarovich_82A}
and $\varphi_3$ is an eigenfunction of noninteracting spin-less fermions, called $\tau$-particles within our representation. 

As reported in Ref. \onlinecite{Carmelo_10}, the $u=U/4t\rightarrow\infty$ local $SU (2)\times SU(2)\times U(1)$ gauge symmetry can be lifted to 
a global $[SU (2)\times SU(2)\times U(1)]/Z_2^2$ symmetry for the model with finite $u=U/4t>0$. Therefore, the global symmetry being 
$[SU (2)\times SU(2)\times U(1)]/Z_2^2 = [SO(4)\times U(1)]/Z_2$ rather than only $SO(4)$ has as starting point the 
$u=U/4t\rightarrow\infty$ local $SU (2)\times SU(2)\times U(1)$ gauge symmetry 
captured by wave function amplitudes being a product of three corresponding wave function amplitudes, Eq. (\ref{3wf}).

And this was already known by Woynarovich in 1982 \cite{Woynarovich_82A}, before the $u\rightarrow\infty$ 
local $SU (2)\times SU(2)\times U(1)$ gauge symmetry was formally identified \cite{Ostlund_91} and 
lifted to a global $[SU (2)\times SU(2)\times U(1)]/Z_2^2$ symmetry of the 1D Hubbard model for finite 
values of $u=U/4t$ \cite{Carmelo_10}.

On acting onto a $\alpha$-spin HWS as given in Eq. (\ref{state}), the off-diagonal generators ${\hat{S}}^{+}_{\alpha}$ of the 
two $\alpha = s,\eta$ global $SU (2)$ symmetries only perform $\alpha$-spin flips of unpaired physical $\alpha$-spins.
In each local occupancy configuration of an energy eigenstate, such unpaired physical $\alpha$-spins
occupy a number $2S_s + 2S_{\eta}$ of lattice sites. 

Importantly, {\it both} in the original lattice and the main effective lattice, such a set of $2S_s + 2S_{\eta}$ sites of
each local occupancy configuration refer for all finite-$S_s$ and finite-$S_{\eta}$ energy eigenstates
to a number $S_s + S_s^z$ of sites singly occupied by $\sigma = \uparrow$ electrons, a
number $S_s - S_s^z$ of sites singly occupied by $\sigma = \downarrow$ electrons,
a number $S_{\eta} + S_{\eta}^z$ of sites unoccupied by electrons, and a number
$S_{\eta} - S_{\eta}^z$ of sites doubly occupied by electrons. For $u>0$ the occupancy configurations 
of such $2S_s + 2S_{\eta}$ sites are for all finite-$S_s$ and finite-$S_{\eta}$ energy eigenstates independent of $u$.

That symmetry invariance follows from the off-diagonal generators ${\hat{S}}^{\pm}_{\alpha}$ of the two $\alpha = s,\eta$ global $SU (2)$ symmetries, 
Eq. (\ref{SSSS}), commuting with the $u$-unitary operator ${\hat{V}}_u$. This indeed implies that the  $\alpha$-spin multiplet occupancy configurations  
of a number $2S_s + 2S_{\eta}$ of sites in each of the local lattice occupancy configurations of
a given finite-$S_s$ and finite-$S_{\eta}$ energy eigenstate are {\it exactly the same} both in the 
original electronic lattice and in the main effective lattice, respectively.

That the numbers of $\sigma = \uparrow$ and $\sigma = \downarrow$ singly occupied sites, unoccupied sites, and doubly occupied sites 
by electrons are not good quantum numbers for finite values of $u=U/4t$ thus results only from the contributions of the occupancy
configurations of a number $L - 2S_s - 2S_{\eta} = L_s - 2S_s + L_{\eta} - 2S_{\eta}$ of sites. Those are occupied by 
a number $L_s - 2S_s$ of paired physical spins and a number $L_{\eta} - 2S_{\eta}$ of paired physical $\eta$-spins. 

Indeed, at finite values of $u$ only the occupancies of these $L - 2S_s - 2S_{\eta}$ sites are for 
finite-$S_s$ and finite-$S_{\eta}$ energy eigenstates different for the original lattice and the main effective lattice, respectively. 
In the cases of the former and latter lattice, the numbers of $\sigma = \uparrow$ and $\sigma = \downarrow$ singly occupied sites, unoccupied 
sites, and doubly occupied sites by electrons are not and are, respectively, good quantum numbers
for the occupancies of such $L - 2S_s - 2S_{\eta}$ sites.

\section{The spin and charge carriers and their $\alpha$-spin elementary currents}
\label{SECIII}

The Hamiltonian, Eq. (\ref{H}), in the presence of uniform $\sigma = \uparrow,\downarrow$ vector
potentials $\Phi_{\sigma}/L$ (twisted boundary conditions) remains solvable 
by the Bethe ansatz \cite{Carmelo_18,Shastry_90}. We then find that under such boundary conditions the
$\tau$-band, $s n$-band, and  $\eta n$-band discrete momentum values 
$q_j={2\pi\over L}I_j^{\tau}$, $q_j={2\pi\over L}I_j^{s n}$, and $q_j={2\pi\over L}I_j^{\eta n}$ in Eq. (\ref{qq}),
respectively, are shifted by the $\sigma = \uparrow,\downarrow$ vector
potentials $\Phi_{\sigma}/L$ as follows,
\begin{eqnarray}
q_j & \rightarrow & q_j + {\Phi_{\uparrow}\over L}\hspace{0.20cm}{\rm for}\hspace{0.20cm}j = 1,..,L 
\nonumber \\
q_j & \rightarrow & q_j - {n(\Phi_{\uparrow}+\Phi_{\downarrow})\over L}\hspace{0.20cm}{\rm for}\hspace{0.20cm}j = 1,..,L_{sn}
\nonumber \\
q_j & \rightarrow & q_j - {n(\Phi_{\uparrow}-\Phi_{\downarrow})\over L}\hspace{0.20cm}{\rm for}\hspace{0.20cm}j = 1,..,L_{\eta n} \, .
\label{qqPhi}
\end{eqnarray}
Here $\Phi_{\uparrow}=\Phi_{\downarrow}=\Phi$ for spin $\alpha = s$ and 
$\Phi_{\uparrow}=-\Phi_{\downarrow}=\Phi$ for charge/$\eta$-spin $\alpha = \eta$. 

Under the use in the thermodynamic limit of $\tau$-band momentum rapidity and $\alpha n$-band
rapidity variables, they are shifted by the $\sigma = \uparrow,\downarrow$ vector
potentials $\Phi_{\sigma}/L$ as follows,
\begin{eqnarray}
k & \rightarrow & k + {\Phi_{\uparrow}\over L\,2\pi\rho (k)}\hspace{0.20cm}{\rm for}\hspace{0.20cm}k\in [-\pi,\pi]
\nonumber \\
\Lambda & \rightarrow & \Lambda - {n(\Phi_{\uparrow}+\Phi_{\downarrow})\over L\,2\pi\sigma_{s n} (\Lambda )}
\hspace{0.20cm}{\rm for}\hspace{0.20cm} \Lambda \in [-\infty,\infty]
\nonumber \\
\Lambda & \rightarrow & \Lambda - {n(\Phi_{\uparrow}-\Phi_{\downarrow})\over L\,2\pi\sigma_{\eta n} (\Lambda )}
\hspace{0.20cm}{\rm for}\hspace{0.20cm} \Lambda \in [-\infty,\infty] \, ,
\label{kLambdaPhi}
\end{eqnarray}
where the functions $2\pi\rho (k)$, $2\pi\sigma_{s n} ( \Lambda )$, and $2\pi\sigma_{\eta n} (\Lambda )$
are defined by the coupled integral equations, Eqs. (\ref{Grho})-(\ref{Gsigmaetan}) of the Appendix.

\subsection{The role of the $\tau$-translational $U(1)$ symmetry in 
the identification of the $\alpha = s,\eta$ $\alpha$-spin carriers}
\label{SECIIIA}

Due to only accounting for the global $SO (4)$ symmetry in $[SO (4))\times U(1)]/Z_2 = [SU (2)\times SU(2)\times U(1)]/Z_2^2$, 
most previous studies have associated the $j=1,...,L$ quantum numbers $q_j = {2\pi\over L} I_j^{\tau}$, Eqs. (\ref{qq}) and (\ref{Itau}), 
$j=1,...,L$ momentum rapidities $k_j = k (q_j)$, and corresponding momentum-rapidity variable $k \in [-\pi,\pi]$
only with the charge degrees of freedom \cite{Essler_05,Ilievski_18,Fava_20,Moca_23}.
However, they are rather associated with the $\tau$-translational $U(1)$ 
symmetry in $[SU (2)\times SU(2)\times U(1)]/Z_2^2$, Eq. (\ref{qqPhi}) giving,
\begin{eqnarray}
q_j & \rightarrow & q_j + {\Phi\over L}\hspace{0.20cm}{\rm for}\hspace{0.20cm}j = 1,..,L 
\nonumber \\
q_j & \rightarrow & q_j - {2n\Phi\over L}\hspace{0.20cm}{\rm for}\hspace{0.20cm}j = 1,..,L_{sn}
\nonumber \\
q_j & \rightarrow & q_j \hspace{0.20cm}{\rm for}\hspace{0.20cm}j = 1,..,L_{\eta n} \, ,
\label{qqPhispin}
\end{eqnarray}
for spin and,
\begin{eqnarray}
q_j & \rightarrow & q_j + {\Phi\over L}\hspace{0.20cm}{\rm for}\hspace{0.20cm}j = 1,..,L 
\nonumber \\
q_j & \rightarrow & q_j\hspace{0.20cm}{\rm for}\hspace{0.20cm}j = 1,..,L_{sn}
\nonumber \\
q_j & \rightarrow & q_j - {2n\Phi\over L}\hspace{0.20cm}{\rm for}\hspace{0.20cm}j = 1,..,L_{\eta n} \, ,
\label{qqPhicharge}
\end{eqnarray}
for charge.

The first relation in Eqs. (\ref{qqPhispin}) and (\ref{qqPhicharge})
confirms that the $j=1,...,L$ quantum numbers $q_j = {2\pi\over L} I_j^{\tau}$, Eqs. (\ref{qq}) and (\ref{Itau}), couple
{\it both} to spin and charge. Indeed, the $\tau$-particle and $\tau$-hole occupancy configurations
that generate irreducible representations of the $\tau$-translational $U(1)$ symmetry 
whose number for each fixed-$S_{\tau}$ subspace reads $d_{\tau} (S_{\tau})$, Eq. (\ref{dtau}) of the Appendix,
are not related only to the charge degrees of freedom. 

The expressions in Eqs. (\ref{qqPhispin}) and (\ref{qqPhicharge}) also show that
the $j=1,...,L_{s n}$ quantum numbers $q_j = {2\pi\over L} I_j^{s n}$ and the
$j=1,...,L_{\eta n}$ quantum numbers $q_j = {2\pi\over L} I_j^{\eta n}$ in Eqs. (\ref{qq})-(\ref{Ian}) 
couple {\it only} to spin and charge, respectively. Consistently, the corresponding $\alpha=s,\eta$ $\alpha n$-band momentum
occupancy configurations of $\alpha$-spin singlet $\alpha n$-pairs whose number for each fixed-$S_{\tau}$ and 
fixed-$S_{\alpha}$ subspace is given by ${\cal{N}}_{\rm singlet} (S_{\alpha},L_{\alpha})$, Eq. (\ref{NNsinglet}) of the Appendix,
generate irreducible representations of the spin $SU(2)$ symmetry $(\alpha =s)$
and $\eta$-spin $SU(2)$ symmetry $(\alpha = \eta)$ in $[SU (2)\times SU(2)\times U(1)]/Z_2^2$.

The $\tau$-band whose $j = 1,...,L$ momentum values $q_j$ are such that $q_{j+1}-q_j=2\pi/L$
is for each energy eigenstate populated by $N_{\tau} = L - 2S_{\tau}$ $\tau$-particles and
$N_{\tau}^h = 2S_{\tau}$ $\tau$-holes. That band's particle-hole symmetry 
ensures that mathematically exactly the same contribution to the spin or charge current expectation values
is reached whether one uses the $\tau$-particles or the $\tau$-holes as corresponding transport carriers. 
The contributions to such expectation values from the $\tau$ branch are indeed generated by processes 
where, on moving in the $\tau$ effective lattice, the $N_{\tau} = L - 2S_{\tau}$ $\tau$-particles
interchange position with the $N_{\tau}^h = 2S_{\tau}$ $\tau$-holes and vice versa. 

For both spin and charge, the physical criterion for the choice of one of these two mathematical equivalent choices 
is univocally imposed by the $\tau$-translational $U(1)$ symmetry's irreducible representations: 
in case of general energy eigenstates for which $0<S_{\tau}<L/2$, 
the $\tau$-particles and the $\tau$-holes describe the
relative translational degrees of freedom of the $L_s = N_{\tau}$ physical spins and
$L_{\eta} = N_{\tau}^h$ physical $\eta$-spins, respectively. Specifically, that symmetry
associates the $N_{\tau}=L-2S_{\tau}$ $\tau$-particles and $N_{\tau}^h=2S_{\tau}$ $\tau$-holes 
with the $L_s = N_{\tau}=L-2S_{\tau}$ physical spins and $L_{\eta} = N_{\tau}^h=2S_{\tau}$ physical $\eta$-spins, respectively.
Therefore, the physically correct choice is to associate the $L_s = N_{\tau}$ $\tau$-particles with the spin carriers
whose translational degrees of freedom they describe and the $L_{\eta} = N_{\tau}^h$ $\tau$-holes with the charge 
carriers whose translational degrees of freedom they describe. 

As confirmed in the following, that is the {\it only} choice that allows the identification 
of the actual spin carriers and charge carriers of the 1D Hubbard model being for $U>0$ the $N_s = 2S_s$ 
unpaired physical spins in the spin multiplet configuration of a $S_s >0$ energy eigenstate
and the $N_{\eta} = 2S_{\eta}$ unpaired physical $\eta$-spins in the $\eta$-spin multiplet configuration 
of a $S_{\eta} >0$ energy eigenstate, respectively. 

On the other hand, consistently with Eq. (\ref{state})
the internal degrees of freedom of the $N_s = 2S_s$ 
unpaired physical spins and $N_{\eta} = 2S_{\eta}$ unpaired physical $\eta$-spins
are rather described by irreducible representations of the spin and $\eta$-spin $SU(2)$ symmetries, respectively.

As discussed in Sec. \ref{SECIIC}, the energy eigenstates that span the subspaces of the Bethe ansatz 
are $\alpha$-spin HWSs whose $N_{\alpha} = 2S_{\alpha}$ unpaired physical $\alpha$-spins have the same projection 
$+1/2$. For them, the momentum eigenvalues in the presence of a uniform vector potentials $\Phi/L$ that we 
denote by $P_{\Phi}$ are from the combined use of Eqs. (\ref{PP}) and (\ref{qqPhi})-(\ref{qqPhicharge}) 
found to be such that,
\begin{equation}
P_{\Phi} - P  = \iota_{\alpha}\left(L_{\alpha}-\sum_{n=1}^{\infty}2n\,M_{\alpha n}\right){\Phi\over L}
\hspace{0.20cm}{\rm for}\hspace{0.20cm}\alpha = s,\eta \, ,
\label{PeffUalpha}
\end{equation}
where $P$ denotes the corresponding momentum eigenvalues for $\Phi_{\uparrow}=\Phi_{\downarrow}=0$, Eq. (\ref{PP}), and,
\begin{equation}
\iota_{s} = 1\hspace{0.20cm}{\rm and}\hspace{0.20cm}\iota_{\eta} = -1 \, .
\label{cseta}
\end{equation}
As discussed in the following, this indeed shows that only the unpaired $\alpha$-spins whose number
reads $N_{\alpha} = L_{\alpha}-\sum_{n}2n\,M_{\alpha n} = 2S_{\alpha}$ couple to
$\alpha$-spin where $\alpha =s$ for spin and $\alpha = \eta$ for charge/$\eta$-spin.

As imposed by the $\tau$-translational $U(1)$ symmetry, to derive the term $\iota_{\alpha}L_{\alpha}\,{\Phi\over L}$ in the
general expression, Eq. (\ref{PeffUalpha}), we have used the alternative expressions
$\sum_{j=1}^{L}q_j\,N_{\tau} (q_j)$ in terms of $\tau$-particles
and $-\sum_{j=1}^{L}q_j\,N_{\tau}^h (q_j)$ in terms of $\tau$-holes given in Eq. (\ref{PP})
for the momentum eigenvalue term $P_{\tau}$ in the case of
spin ($\alpha =s$) and charge ($\alpha = \eta$), respectively. Under the replacement $q_j \rightarrow q_j + \Phi/L$, Eq. (\ref{qqPhi}),
and accounting for the sum rules, $L_s = \sum_{j=1}^{L}N_{\tau} (q_j)$ and $L_{\eta} = \sum_{j=1}^{L}N_{\tau}^h (q_j)$,
we have reached the $(P_{\Phi} - P)$'s expression, Eq. (\ref{PeffUalpha}), for spin $\alpha =s$ and charge $\alpha =\eta$,
respectively.

The other terms $-\iota_{\alpha}\sum_{n}2n\,M_{\alpha n}{\Phi\over L}$ in Eq. (\ref{PeffUalpha}) 
are straightforwardly derived by combining Eqs. (\ref{PP}) and (\ref{qqPhi}) for the $sn$ and $\eta n$ branches
with the sum rules $\sum_{j=1}^{L_{\alpha n}} M_{\alpha n} (q_j) = M_{\alpha n}$ for $\alpha =s,\eta$.

The term $\iota_{\alpha}L_{\alpha}\,{\Phi\over L}$ 
in $\iota_{\alpha}(L_{\alpha}-\sum_{n}2n\,M_{\alpha n}){\Phi\over L}$
refers to {\it all} $L_{\alpha}$ physical $\alpha$-spins coupling to the vector potential
in the absence of physical $\alpha$-spin singlet pairings. 

Consistently with for $\alpha = s$ such a term stemming from the coupling to spin of $\tau$-particles
whose occupancy configurations in the $\tau$ effective lattice generate irreducible representations of 
the $\tau$-translational $U(1)$ symmetry, such configurations do not distinguish
sites that in the main effective lattice are singly occupied 
by $\sigma = \uparrow$ electrons from those that are singly occupied by $\sigma = \downarrow$ electrons.

Similarly, for $\alpha = \eta$ the term $\iota_{\alpha}L_{\alpha}\,{\Phi\over L}$ stems from the coupling to 
$\eta$-spin of $\tau$-holes, the corresponding occupancy configurations in the $\tau$ effective lattice not distinguishing 
sites unoccupied from sites doubly occupied by electrons in the main effective lattice.

On the other hand, the coupling counterterms $-\iota_{\alpha}\sum_{n}2n\,M_{\alpha n}{\Phi\over L}$ 
in $\iota_{\alpha}(L_{\alpha}-\sum_{n}2n\,M_{\alpha n}){\Phi\over L}$ impose that the number  
${\cal{N}}_{\rm singlet} (S_{\alpha},L_{\alpha})$ of $\alpha$-spin singlet configurations is that 
given in Eq. (\ref{NNsinglet}) of the Appendix.
Such a coupling counterterms $-\iota_{\alpha}\sum_{n}2n\,M_{\alpha n}{\Phi\over L}$ 
refer to the number $2n$ of paired physical $\alpha$-spins in each $\alpha n$-pair. 
They {\it exactly cancel} the coupling in $\iota_{\alpha}L_{\alpha}\,{\Phi\over L}$ of the 
corresponding $2n$ paired $\alpha$-spins in each such an $\alpha n$-pair. 
It is as a result of such counterterms that only the $N_{\alpha} = L_{\alpha}-\sum_{n}2n\,M_{\alpha n} = 2S_{\alpha}$
unpaired physical $\alpha$-spins couple to the vector potential and thus carry spin $(\alpha =s)$ or charge 
$(\alpha = \eta)$ current.

Confirming what was above stated, such an identification of the $N_{\alpha} = 2S_{\alpha}$ unpaired 
physical $\alpha$-spins as the $\alpha$-spin carriers that for $u=U/4t>0$
populate finite-$S_{\alpha}$ energy eigenstates was possible because 
the requirement of the $\tau$-translational $U(1)$ symmetry that to obtain the $(P_{\Phi} - P)$'s expression, 
Eq. (\ref{PeffUalpha}), a representation in terms of the $N_{\tau} = L_s$ 
$\tau$-particles is used for the coupling to spin of the $L_s$ physical spins and a representation in terms of 
the $N_{\tau}^h = L_{\eta}$ $\tau$-holes is used for the coupling to charge of the $L_{\eta}$ physical $\eta$-spins
was accounted for.

In the case of $\alpha$-spin non-HWSs, the use of Eq. (\ref{state}) reveals that
the expression for $(P_{\Phi} - P)$, Eq. (\ref{PeffUalpha}), is transformed into
the following more general expression,
\begin{equation}
P_{\Phi} - P = \iota_{\alpha}(N_{\alpha,+1/2}-N_{\alpha,-1/2})\,{\Phi\over L}
\hspace{0.20cm}{\rm for}\hspace{0.20cm}\alpha = s,\eta \, .
\label{PeffUalphaNnonHWS}
\end{equation}

The {\it spin carriers} ($\alpha = s$) and {\it charge carriers} ($\alpha = \eta$) that populate {\it all} finite-$S_{\alpha}$ energy eigenstates
are thus the $N_{\alpha} = 2S_{\alpha}$ unpaired physical $\alpha$-spins. The general
expression, Eq. (\ref{PeffUalphaNnonHWS}), reveals in addition that, as expected, the coupling 
to $\alpha$-spin of unpaired physical $\alpha$-spins of projections $+1/2$ and $-1/2$ has opposite sign.

\subsection{Relation of the $\eta$-spin $SU(2)$ symmetry and charge carriers to the energy eigenvalues}
\label{SECIIIB}

The 1D Hubbard model as written in Eq. (\ref{H}) has at $h= \mu=0$ global 
$[SU (2)\times SU(2)\times U(1)]/Z_2^2$ symmetry \cite{Carmelo_10}. However, when
at $h= \mu=0$ it is written in its original and most usual form,
\begin{equation}
\hat{H} = -t\sum_{\sigma, j}\left[c_{j,\sigma}^{\dag}\,c_{j+1,\sigma} + 
{\rm h.c.}\right] + U\sum_{j}\hat{n}_{j,\uparrow}\hat{n}_{j,\downarrow} \, ,
\label{H0}
\end{equation}
its global $\eta$-spin $SU (2)$ symmetry is broken to $U (1)$. 

This can be shown by use and manipulation of symmetry operators algebras \cite{Essler_05},
yet it is easiest understood within our $\tau$- and physical $\alpha$-spin representation beyond
the Bethe ansatz. The physical reason why the latter alternative analysis of the problem 
by means of that representation is interesting for our study is that
it only involves the $2S_{\eta}$ unpaired physical $\eta$-spins in the $\eta$-spin multiplet
configuration of $S_{\eta}> 0$ energy eigenstates, which have been found in Sec. \ref{SECIIIA} 
to be the charge carriers.

When the Hamiltonian, Eq. (\ref{H0}), acts onto the full Hilbert space, its 
energy eigenvalues are in the thermodynamic limit given by,
\begin{eqnarray}
&& E (l_{\rm r}^{u},S_{\tau},S_s,S_{\eta},N_{\eta,-1/2}) = -2t\sum_{j=1}^L  N_{\tau} (q_j) \cos k (q_j) 
\nonumber \\
&& +\,4t\sum_{n=1}^{\infty}\sum_{j=1}^{L_{\eta n}}M_{\eta n} (q_j){\rm Re}\left\{\sqrt{1 - (\Lambda_{\eta n} (q_j) - i n u)^2}\right\} 
\nonumber \\
&& +\,U N_{\eta,-1/2} \, .
\label{ESS0}
\end{eqnarray}

While this energy eigenvalue expression is well known in the case of the Bethe-ansatz subspace \cite{Takahashi_72}, 
within the $\tau$- and physical $\alpha$-spin representation beyond that ansatz
its extension to the full Hilbert space involves the additional term $U N_{\eta,-1/2}$ where $N_{\eta,-1/2} = 0,1,...,2S_{\eta}$.

The form of the exact expression of that additional energy term 
obviously obeys the requirement of consistency with the limiting $\eta$-spin HWS and $\eta$-spin LWS Bethe-ansatz energy 
eigenvalue expressions. For the Bethe-ansatz $\eta$-spin HWS representation it vanishes \cite{Takahashi_72}, 
consistently with $N_{\eta,-1/2} = 0$ in $U N_{\eta,-1/2}$. For the 
Bethe-ansatz $\eta$-spin LWS representation it reads $U\,N_{\eta} = U\,2S_{\eta} = U\,2S_{\eta}^z$  and
thus $U\,(L - N_{\uparrow} - N_{\downarrow})$, consistently with
the relation given in Eq. (3) of Ref. \onlinecite{Lieb_03}, which can be rewritten as
$E (N_{\uparrow},N_{\downarrow}) + U\,(L - N_{\uparrow} - N_{\downarrow}) = 
E (L - N_{\uparrow},L - N_{\downarrow})$. Indeed, $N_{\eta,-1/2} = 2S_{\eta}$ in $U N_{\eta,-1/2}$
where $2S_{\eta}$ is given by $2S_{\eta} = (L - N_{\uparrow} - N_{\downarrow})$ for a $\eta$-spin LWS.

That in the general case of all $2S_{\eta} +1$ states of a $\eta$-spin tower,
such additional energy term reads $U N_{\eta,-1/2}$ where $N_{\eta,-1/2} = 0,1,...,2S_{\eta}$
is imposed both by the $\eta$-spin off-diagonal generator 
${\hat{S}}_{\eta}^+$ of the $\eta$-spin global $SU (2)$ symmetry given in Eq. (\ref{SSSS}) commuting 
with the $u$-unitary operator ${\hat{V}}_u$ and the symmetry invariance under the corresponding
$u$-rotated-unitary transformation of the occupancies of a number $2S_{\eta}$ of sites
of the original electronic lattice occupied by unpaired physical $\eta$-spins.
As discussed in Sec. \ref{SECIID}, for the occupancies of such $2S_{\eta}$ sites
the numbers $S_{\eta}+S_{\eta}^z$ of unoccupied sites and $S_{\eta}-S_{\eta}^z$ of doubly
occupied sites by electrons are good quantum numbers both in the main effective lattice
and the original lattice.

In addition, the form of the interaction term of the Hamiltonian, Eq. (\ref{H0}), 
is such that $U$ is the bare on-site energy for creation of one electron doubly occupied site.
Due to the above symmetry invariance, under each $\eta$-spin flip produced by the $\eta$-spin off-diagonal generator 
${\hat{S}}_{\eta}^+$ in Eq. (\ref{state}), one doubly occupied site
is created both in the main effective lattice and in the original electronic lattice. This is
why the term $U N_{\eta,-1/2}$ emerges from the extension to the full Hilbert space 
of the energy eigenvalue expression provided by the Bethe ansatz \cite{Takahashi_72}.
That extension directly accounts for the extra contributions from the 
$\eta$-spin non-HWSs and $\eta$-spin non-LWS states in Eq. (\ref{state}) for
$\alpha = \eta$.

Also the term $\pi\,N_{\eta,-1/2}$ in the momentum expression, Eq. (\ref{PP}), 
results from each $\eta$-spin flip produced by the $\eta$-spin off-diagonal 
generator ${\hat{S}}_{\eta}^+$ in Eq. (\ref{state}) involving an excitation momentum $\pi$.
In this case such a momentum $\pi$ stems from the form of that generator itself, Eq. (\ref{SSSS}),
whereas the energy $U$ associated with the term $U N_{\eta,-1/2}$ stems from the
form of the Hamiltonian's interaction term in Eq. (\ref{H0}).

The dependence on $N_{\eta,-1/2}$ of the energy term $U N_{\eta,-1/2}$ confirms 
that the 1D Hubbard model as written in Eq. (\ref{H0}) does not commute with the two off-diagonal generators
of the $\eta$-spin $SU (2)$ symmetry. Indeed, $\eta$-spin $SU (2)$ symmetry requires that at $\mu=0$ all
the $2S_{\eta} + 1$ energy eigenstates of the same $\eta$-spin tower must have the same energy. 
The energy term $U N_{\eta,-1/2}$ in Eq. (\ref{ESS0}) clearly violates that requirement.

The energy eigenvalues of the 1D Hubbard model as written in Eq. (\ref{H}) have 
relative to those of that given in Eq. (\ref{H0}) additional terms $-{U\over 4}L + U S_{\eta}^z$. 
Following Eq. (\ref{2Sz2Sq}), these extra terms 
can be written as $-{U\over 4}L + {U\over 2}(N_{\eta,+1/2}-N_{\eta,-1/2})$.
Addition of these two terms to $U N_{\eta,-1/2}$ then leads to $- {U\over 2}\left({L\over 2} - N_{\eta}\right) = - U \left({L\over 4} - S_{\eta}\right)$.
Therefore, the energy eigenvalues of the 1D Hubbard model, Eq. (\ref{H}), are at $h=\mu=0$ given by,
\begin{eqnarray}
&& E (l_{\rm r}^{u},S_{\tau},S_s,S_{\eta}) = 
\nonumber \\
&& -2t\sum_{j=1}^L  N_{\tau} (q_j) \cos k (q_j) - U \left({L\over 4} - S_{\eta}\right)
\nonumber \\
&& + 4t\sum_{n=1}^{\infty}\sum_{j=1}^{L_{\eta n}}M_{\eta n} (q_j){\rm Re}\left\{\sqrt{1 - (\Lambda_{\eta n} (q_j) - i n u)^2}\right\}  \, .
\nonumber \\
\label{ESS}
\end{eqnarray}
Such energy eigenvalues have the same value for all $2S_{\eta} + 1$ energy eigenstates of the same $\eta$-spin tower, 
consistently with the corresponding Hamiltonian, Eq. (\ref{H}) at $h=\mu=0$, commuting with the three generators of the
global $\eta$-spin $SU (2)$ symmetry, so that its global symmetry is $[SU (2)\times SU(2)\times U(1)]/Z_2^2$.

The energy eigenvalues of the 1D Hubbard model, Eq. (\ref{H}), are for finite values of $h$ and $\mu$ then given by,
\begin{equation}
E = E (l_{\rm r}^{u},S_{\tau},S_s,S_{\eta}) - 
{1\over 2}\sum_{\alpha =s,\eta}\mu_{\alpha}\left(N_{\alpha,+1/2} - N_{\alpha,-1/2}\right) \, ,
\label{E}
\end{equation}
where $\mu_s = 2\mu_B h$ and $\mu_{\eta} = 2\mu$, as in Eq. (\ref{H}).
The exact relation given in Eq. (\ref{2Sz2Sq}) was again used to obtain this expression. 

\subsection{The $\alpha$-spin elementary currents carried by the $\alpha$-spin carriers}
\label{SECIIIC}

The $z$ component of the spin and charge current operators are in units of spin $1/2$ and electronic
charge given by,
\begin{eqnarray}
\hat{J}_s^z & = & - it\sum_{\sigma}\sum_{j=1}^L(2\sigma)\left[c_{j,\sigma}^{\dag}\,c_{j+1,\sigma} -
c_{j+1,\sigma}^{\dag}\,c_{j,\sigma}\right]\hspace{0.20cm}{\rm and}
\nonumber \\
\hat{J}_{\eta}^z & = & - it\sum_{\sigma}\sum_{j=1}^L\left[c_{j,\sigma}^{\dag}\,c_{j+1,\sigma} -
c_{j+1,\sigma}^{\dag}\,c_{j,\sigma}\right] \, ,
\label{Jzseta}
\end{eqnarray}
respectively. In the factor $(2\sigma)$ of the spin current operator expression, $\sigma$ reads 
$+1/2$ for $\sigma = \uparrow$ and $-1/2$ for $\sigma = \downarrow$.

Only the term $E (l_{\rm r}^{u},S_{\tau},S_s,S_{\eta})$, Eq. (\ref{ESS}), of the energy eigenvalue expression, 
Eq. (\ref{E}), involves summations over the momentum values $q_j$ of the $\tau$-band and $\alpha n$-bands.
The $\alpha$-spin-current expectation value $\langle \hat{J}_{\alpha}^z (HWS)\rangle$ of
a $\alpha$-spin HWS can then be obtained from the energy eigenvalues of the Hamiltonian, Eq. (\ref{H}), in the presence 
of a vector potential $\Phi/L$ as follows,
\begin{equation}
\langle \hat{J}_{\alpha}^z (HWS)\rangle  = \lim_{\Phi/L\rightarrow 0}
{d E (l_{\rm r}^{u},S_{\tau},S_s,S_{\eta},\Phi/L)\over d (\Phi/L)} \, .
\label{currentHWS}
\end{equation}
Here $E (l_{\rm r}^{u},S_{\tau},S_s,S_{\eta},\Phi/L)$ is the energy eigenvalue term, Eq. (\ref{ESS}),
with the discrete momentum values $q_j$ of the $\tau$-band and $\alpha n$-bands
in its summations replaced by their expressions on the right-hand side of Eq. (\ref{qqPhispin}) for 
spin $\alpha = s$ and of Eq. (\ref{qqPhicharge}) for charge/$\eta$-spin $\alpha = \eta$. 

As in the case of the spin-$1/2$ $XXX$ chain \cite{Carmelo_15}, from manipulations involving the algebras
of the $\alpha$-spin $SU(2)$ symmetries, we find that the $\alpha$-spin-current expectation
values of the $\alpha$-spin non-HWSs, Eq. (\ref{state}), can be expressed in terms of the
$\alpha$-spin-current expectation value $\langle\hat{J}_{\alpha}^z (HWS)\rangle$, Eq. (\ref{currentHWS}),
of the corresponding $\alpha$-spin HWS as,
\begin{eqnarray}
\langle \hat{J}_{\alpha}^z \rangle & = &
{S_{\alpha}^z\over S_{\alpha}}\langle \hat{J}_{\alpha}^z (HWS)\rangle
\nonumber \\
& = & {(N_{\alpha,+1/2} - N_{\alpha,-1/2})\over (N_{\alpha,+1/2} + N_{\alpha,-1/2})}\,
\langle \hat{J}_{\alpha}^z (HWS)\rangle \, ,
\label{rel-currents-gen}
\end{eqnarray}
where Eq. (\ref{2Sz2Sq}) was used to reach the second expression. 
The calculations to reach Eq. (\ref{rel-currents-gen}) are relatively easy for non-HWSs whose 
generation from HWSs in Eq. (\ref{state}) involves small $n_{\alpha}^z = S_{\alpha} - S_{\alpha}^z$ values. 
The calculations become lengthy as the $n_{\alpha}^z$ value increases, but remain straightforward.

A general explicit expression for the spin ($\alpha = s$) and charge ($\alpha = \eta$)
current expectation values, Eq. (\ref{rel-currents-gen}), that accounts for
the corresponding expression of $\langle \hat{J}_{\alpha}^z (HWS)\rangle$ appearing 
in that equation and applies to all energy eigenstates is given in Eqs. (\ref{Jsz}) and (\ref{Jetaz}) 
of the Appendix, respectively.

Accounting for the generation of unpaired physical $\alpha$-spin flips described by Eq. (\ref{state}),
the exact relation, Eq. (\ref{rel-currents-gen}), can be expressed as,
\begin{equation}
\langle \hat{J}_{\alpha}^z \rangle = \sum_{\sigma =\pm 1/2}N_{\alpha,\sigma}\,j_{\alpha,\sigma} \, ,
\label{rel-currents}
\end{equation}
where,
\begin{equation}
j_{\alpha,\pm 1/2} = \pm {\langle \hat{J}_{\alpha}^z (HWS)\rangle\over 2S_{\alpha}} = 
\pm {\langle \hat{J}_{\alpha}^z (HWS)\rangle\over N_{\alpha}} \, ,
\label{elem-currents}
\end{equation}
and $N_{\alpha,\pm 1/2} = S_{\alpha} \pm S_{\alpha}^z$ is the number of unpaired physical $\alpha$-spins of projection $\pm 1/2$,
which are the $\alpha$-spin carriers.

It is straightforward to confirm that $j_{\alpha,\pm 1/2} $, Eq. (\ref{elem-currents}), is the 
$\alpha$-spin elementary current carried by the unpaired physical $\alpha$-spins of projection $\pm 1/2$
in the $\alpha$-spin multiplet configuration of any $S_{\alpha}>0$ energy eigenstate 
of $\alpha$-spin-current expectation value $\langle \hat{J}_{\alpha}^z \rangle =
{S_{\alpha}^z\over S_{\alpha}}\langle \hat{J}_{\alpha}^z (HWS)\rangle$. Here $S_{\alpha}^z$ can have $2S_{\alpha} +1$ values, 
$S_{\alpha}^z = S_{\alpha} - n_{\alpha}^z$ where $n_{\alpha}^z = 0,1,...,2S_{\alpha}$. 

Indeed, under each unpaired physical $\alpha$-spin flip generated by the operator 
${\hat{S}}^{+}_{\alpha}$, Eq. (\ref{SSSS}), in the expression given in Eq. (\ref{state}), the $\alpha$-spin-current expectation value 
$\langle \hat{J}_{\alpha}^z \rangle$, Eqs. (\ref{rel-currents-gen}) and (\ref{rel-currents}), exactly changes 
by $-2j_{\alpha,+1/2} = 2j_{\alpha,-1/2}$.

An important symmetry is that the $\alpha$-spin elementary currents $j_{\alpha,\pm 1/2}$, Eq. (\ref{elem-currents}), 
have the same value for all $2S_{\alpha} + 1$ states of the same $\alpha$-spin tower. That implies they do
not depend on $S_{\alpha}^z$ and thus on $m_{\alpha z}$, so that,
\begin{equation}
{\partial  j_{\alpha,\pm 1/2} \over \partial\,m_{\alpha z}} = 0 \hspace{0.20cm}{\rm for}\hspace{0.20cm}\alpha = s,\eta \, .
\label{deriv-elem-currents}
\end{equation}

These results reveal the deep physical meaning of the relation, Eqs. (\ref{rel-currents-gen}) and
(\ref{rel-currents}): It expresses the $\alpha$-spin-current expectation value 
$\langle \hat{J}_{\alpha}^z \rangle$ of a $S_{\alpha}>0$ energy eigenstate 
$\left\vert l_{\rm r}^{u},S_{\tau},S_{s},S_{s}^z,S_{\eta},S_{\eta}^z\right\rangle$ in terms of the $\alpha$-spin elementary currents
$j_{\alpha,\pm 1/2}$ carried by each of the $N_{\alpha,\pm 1/2} = S_{\alpha} \pm S_{\alpha}^z$ 
$\alpha$-spin carriers of projection $\pm 1/2$ in the $\alpha$-spin multiplet configuration of that state. 
This also shows that when $N_{\alpha} = N_{\alpha,+1/2} + N_{\alpha,-1/2} = 0$ the $\alpha$-spin-current expectation value 
$\langle \hat{J}_{\alpha}^z \rangle = \sum_{\sigma =\pm 1/2}N_{\alpha,\sigma}\,j_{\alpha,\sigma} $, Eq. (\ref{rel-currents}), 
vanishes, {\it i.e.} $S_{\alpha}=0$ energy eigenstates have vanishing $\alpha$-spin-current expectation value.

The combined use of Eqs. (\ref{currentHWS}), (\ref{rel-currents-gen}), and Eqs. (\ref{Jsz}), (\ref{Jetaz}) of
the Appendix in the expression given in Eq. (\ref{elem-currents}) leads to the following general expression for the 
spin and charge elementary currents carried by the corresponding transport carriers 
in the spin and $\eta$-spin multiplet configuration of finite-$S_s$ and finite-$S_{\eta}$ energy eigenstates,
\begin{eqnarray}
 && j_{s,\pm 1/2} = \pm {t\over\pi}{L\over N_{\tau}}\int_{-\pi}^{\pi}dk {\bar{N}}_{\tau} (k) \sin k
\pm {2t\over \pi^2 u}\sum_{n=1}^{\infty} {L\over M_{sn}^h}
 \nonumber \\
 && \times \int_{-\infty}^{\infty}d\Lambda {\bar{M}}_{sn}^h (\Lambda)\int_{-\pi}^{\pi}dk 
 {\bar{N}}_{\tau} (k) {2\pi\rho (k) \sin k\over 1 + \left({\Lambda - \sin k\over nu}\right)^2} \, ,
 \label{js}
 \end{eqnarray}
 and
 \begin{eqnarray}
 && j_{\eta,\pm 1/2} = \mp {t\over\pi}{L\over N_{\tau}^h}\int_{-\pi}^{\pi}dk {\bar{N}}_{\tau}^h (k) \sin k
 \pm {4t\over \pi}\sum_{n=1}^{\infty} {L\over M_{\eta n}^h}
 \nonumber \\
 && \times \int_{-\infty}^{\infty}d\Lambda
 {\bar{M}}_{\eta n}^h (\Lambda)\Bigl(n\,{\rm Re}\Bigl\{{(\Lambda - i n u)\over\sqrt{1 - (\Lambda - i n u)^2}}\Bigr\}
 \nonumber \\
 && + {1\over 2\pi u} \int_{-\pi}^{\pi}dk {\bar{N}}_{\tau} (k) {2\pi\rho (k) \sin k\over 1 + \left({\Lambda - \sin k\over nu}\right)^2}\Bigr) \, ,
 \label{jeta}
 \end{eqnarray}
respectively. The distributions appearing here are defined by Eq. (\ref{NNrela}) and 
Eq. (\ref{Grho}) of the Appendix.

In the $S_{\tau} = 0$ subspace for which $L_s = L$ and $L_{\eta} = 0$ in which the
1D Hubbard model is equivalent to the spin-$1/2$ $XXX$ chain when $u=U/4t$ is
large there are no physical $\eta$-spins. The 
expression of the spin elementary currents given in Eq. (\ref{js}) then simplifies to,
\begin{eqnarray}
 j_{s,\pm 1/2} & = & \pm {2t\over \pi^2 u}\sum_{n=1}^{\infty} {L\over M_{sn}^h}
 \int_{-\infty}^{\infty}d\Lambda {\bar{M}}_{sn}^h (\Lambda)
 \nonumber \\
 & \times & \int_{-\pi}^{\pi}dk {2\pi\rho (k) \sin k\over 1 + \left({\Lambda - \sin k\over nu}\right)^2} \, .
 \label{jsFu}
 \end{eqnarray}
This spin elementary current can be expressed only in terms of spin distributions by use in it 
of Eq. (\ref{Grho}) of the Appendix for the 
function $2\pi\rho (k)$ that for $S_{\tau} = 0$ and $L_s = L$ has no contributions from $\eta$-spin distributions. 

For large values of $u = U/4t$ the spin elementary current, Eq. (\ref{jsFu}), is up to first order in the energy 
$J = {4t^2\over U}$ the spin elementary current carried by the spin carriers of projection $\pm 1/2$ 
in the spin multiplet configuration of the spin-$1/2$ $XXX$ chain's finite-spin energy eigenstates.

In the subspace spanned by states for which $S_{\tau} = L/2$ and thus $L_s = 0$ and $L_{\eta} = L$ there are no physical spins. 
The expression of the $\eta$-spin elementary current given in Eq. (\ref{jeta}) then reads,
 \begin{eqnarray}
 && j_{\eta,\pm 1/2} = 
 \pm {4t\over \pi}\sum_{n=1}^{\infty} {L\over M_{\eta n}^h}\int_{-\infty}^{\infty}d\Lambda
 {\bar{M}}_{\eta n}^h (\Lambda)\,\Bigl(n \times 
 \nonumber \\
 && \,{\rm Re}\Bigl\{{(\Lambda - i n u)\over\sqrt{1 - (\Lambda - i n u)^2}}\Bigr\}
 + {1\over 2\pi u} \int_{-\pi}^{\pi}dk {2\pi\rho (k) \sin k\over 1 + \left({\Lambda - \sin k\over nu}\right)^2}\Bigr) \, .
 \nonumber \\
 \label{jetaFu}
 \end{eqnarray}
Again, this $\eta$-spin elementary current can be expressed solely in terms of $\eta$-spin distributions by use in it 
of Eq. (\ref{Grho}) of the Appendix for the 
function $2\pi\rho (k)$ that for $S_{\tau} = L/2$ and $L_{\eta} = L$ has no contributions from spin distributions.

It is confirmed in this paper that for $u=U/4t >0$ the 
$\alpha$-spin elementary currents $j_{\alpha,\pm 1/2}$ 
given in Eqs. (\ref{elem-currents}) and (\ref{js})-(\ref{jetaFu}) that are carried by each of the 
$N_{\alpha,\pm 1/2} = S_{\alpha} \pm S_{\alpha}^z$ unpaired physical $\alpha$-spins
of projection $\pm 1/2$ in the $\alpha$-spin multiplet configuration
of $u=U/4t>0$ finite-$S_{\alpha}$ energy eigenstates
play a key role in the finite-temperature spin and charge transport in 
the 1D Hubbard model, Eq. (\ref{H}). 

To discuss issues on the 1D Hubbard model's finite-temperature transport properties in terms of the 
microscopic processes associated with the 
spin and charge elementary currents $j_{s,+1/2}$ and $j_{\eta,+1/2}$,
Eqs. (\ref{js}) and (\ref{jeta}), respectively, below in Sec. \ref{SECV} they are expressed in terms of the 
$\tau$-particle distribution $N_{\tau} (q_j)$, $s n$-hole distributions $M_{s n}^h (q_j)$,
$\tau$-hole distribution $N_{\tau}^h (q_j)$, and $\eta n$-hole distributions $M_{\eta n}^h (q_j)$
that directly refer to the discrete momentum values $q_j = {2\pi\over L}I_j^{\tau}$ 
and $q_j = {2\pi\over L} I_j^{\alpha n}$ for $\alpha = s,\eta$, Eq. (\ref{qq}), and thus to the 
quantum numbers $I_j^{\tau}$ and $I_j^{\alpha n}$, Eqs. (\ref{Itau}) and (\ref{Ian}), respectively.

\section{Control of the $\alpha$-spin stiffness and diffusion constant by the 
$\alpha$-spin elementary currents}
\label{SECIV}  

The real part of the $\alpha =s,\eta$ $\alpha$-spin conductivity of the 1D Hubbard model, Eq. (\ref{H}), 
is for finite temperatures given by,
\begin{equation}
\sigma_{\alpha} (\omega,T) = 2\pi D_{\alpha}^z (T)\,\delta (\omega) + \sigma_{\alpha, {\rm reg}} (\omega,T) \, .
\label{sigma}
\end{equation}
When the spin $(\alpha =s)$ and charge $(\alpha = \eta)$ stiffness $D_{\alpha}^z (T)$ in its singular part is finite, 
the dominant $\alpha$-spin transport is ballistic. It is well established that in the thermodynamic limit the 
$\alpha$-spin stiffness is finite for finite magnetic field $h>0$ $(\alpha =s)$ and finite chemical $\mu >0$ $(\alpha = \eta)$
and exactly vanishes at $h=0$ and at $\mu = 0$, respectively, for all temperatures $T>0$ 
\cite{Fava_20,Moca_23,Ilievski_17,Carmelo_18}. 

The type of nonballistic $\alpha$-spin transport is determined by the value of the corresponding
diffusion constant $D_{\alpha} (T)$ associated with the regular
part of the $\alpha$-spin conductivity $\sigma_{\alpha, {\rm reg}} (\omega,T)$ in Eq. (\ref{sigma}):
It is normal diffusive when $D_{\alpha} (T)$ is finite and anomalous superdiffusive when
$D_{\alpha} (T) = \infty$ \cite{Ilievski_18,Fava_20,Moca_23}.

In the following we express $D_{\alpha}^z (T)$ and $D_{\alpha} (T)$ in terms of the 
$\alpha$-spin elementary currents carried by the $\alpha$-spin transport carriers in the $\alpha$-spin multiplet 
configuration of $S_{\alpha}>0$ energy eigenstates.
  
\subsection{$T>0$ spin and charge transport quantities}
\label{SECIVA}  

At finite temperature, $T>0$, the spin $(\alpha = s)$ and charge $(\alpha = \eta)$ stiffness $D_{\alpha}^z (T)$ 
in Eq. (\ref{sigma}) can be expressed as \cite{Mukerjee_08},
\begin{eqnarray}
D_{\alpha}^z (T) & = & {1\over 2 L\,k_B T}\sum_{S_{\tau}=0}^{L/2}
\sum_{S_{\alpha}=\vert S_{\alpha}^z\vert}^{L_{\alpha}/2}
\sum_{S_{\bar{\alpha}}=\vert S_{\bar{\alpha}}^z\vert}^{L_{\bar{\alpha}}/2=(L-L_{\alpha})/2}
\nonumber \\
& \times & \sum_{l_{\rm r}^{u}} 
p_{l_{\rm r}^{u},S_{\tau},S_{\alpha},S_{\alpha}^z,S_{\bar{\alpha}},S_{\bar{\alpha}}^z}\vert \langle \hat{J}_{\alpha}^z \rangle \vert^2 
\nonumber \\
& {\rm where} & \hspace{0.20cm}{\bar{s}} = \eta \hspace{0.20cm}{\rm and}\hspace{0.20cm}{\bar{\eta}} = s \, .
\label{D-all-T-simpA}
\end{eqnarray}
Here the summations run over {\it all} quantum-problem energy eigenstates 
$\left\vert l_{\rm r}^{u},S_{\tau},S_{\alpha},S_{\alpha}^z,S_{\bar{\alpha}},S_{\bar{\alpha}}^z\right\rangle$
with both fixed $S_{s}^z$ and $S_{\eta}^z$ values, $p_{l_{\rm r}^{u},S_{\tau},S_{\alpha},S_{\alpha}^z,S_{\bar{\alpha}},S_{\bar{\alpha}}^z}$ are 
the corresponding Boltzmann weights, and the summation $\sum_{S_{\tau}=0}^{L/2}$ where $S_{\tau} = L_{\eta}/2 = L - L_{s}/2$
is equivalent to both $\sum_{L_{\alpha}=0}^{L}$ and $\sum_{L_{\bar{\alpha}}=0}^{L}$ where
$L_{\alpha} + L_{\bar{\alpha}} = L$.

From the combined use of Eqs. (\ref{rel-currents-gen}), (\ref{rel-currents}), and (\ref{D-all-T-simpA}),
we find that $D_{\alpha}^z (T)$ can be expressed directly in terms of the $\alpha$-spin elementary currents 
carried by the $\alpha$-spin carriers, Eq. (\ref{elem-currents}), as,
\begin{eqnarray}
D_{\alpha}^z (T) & = & m_{\alpha z}^2\,{\Omega_{m_{\alpha z}} (T)\over 2k_B T} 
\nonumber \\
& = & m_{\alpha z}^2\,{L\over 2k_B T} \langle\vert j_{\alpha,\pm 1/2}\vert^2\rangle_{m_{\alpha z},m_{{\bar{\alpha}} z},T} \, ,
\label{D-all-T-simp-m}
\end{eqnarray}
where $\langle ...\rangle_{m_{\alpha z},m_{{\bar{\alpha}} z},T}$ denotes canonical equilibrium averages at temperature $T$
for fixed values of both $m_{\alpha z}$ and $m_{{\bar{\alpha}} z}$ and,
\begin{eqnarray}
&& \langle\vert j_{\alpha,\pm 1/2} \vert^2\rangle_{m_{\alpha z},m_{{\bar{\alpha}} z},T} = {\Omega_{m_{\alpha z}} (T)\over L}
\nonumber \\
&& = \sum_{S_{\tau}=0}^{L/2}
\sum_{S_{\alpha}=\vert S_{\alpha}^z\vert}^{L_{\alpha}/2}
\sum_{S_{\bar{\alpha}}=\vert S_{\bar{\alpha}}^z\vert}^{L_{\bar{\alpha}}/2} \sum_{l_{\rm r}^{u}} 
p_{l_{\rm r}^{u},S_{\tau},S_{\alpha},S_{\alpha}^z,S_{\bar{\alpha}},S_{\bar{\alpha}}^z}\vert j_{\alpha,\pm 1/2}\vert^2 \, ,
\nonumber \\
\label{jz2T}
\end{eqnarray}
so that $\langle\vert j_{\alpha,\pm 1/2} \vert^2\rangle_{m_{\alpha z},m_{{\bar{\alpha}} z},T} = \Omega_{m_{\alpha z}} (T)/L$ is the
thermal expectation value of the square of the absolute value $\vert j_{\alpha,\pm 1/2}\vert$ of the 
$\alpha$-spin elementary current carried by each $\alpha$-spin carrier of projection $\pm 1/2$, 
Eq. (\ref{elem-currents}), in the subspace spanned by states with fixed values of both $S_{s}^z$ and $S_{\eta}^z$. 

On the other hand, from manipulations of the Kubo formula and Einstein relation
under the use of the $\tau$- and $\alpha$-physical-spin representation and the $\alpha$-spin elementary currents 
$j_{\alpha, \pm 1/2}$ emerging from it, we find that for finite temperature $T>0$ and at $h=0$ for spin and at $\mu =0$ for charge the 
$\alpha$-spin diffusion constant $D_{\alpha} (T)$ associated with the regular
part of the $\alpha$-spin conductivity $\sigma_{\alpha, {\rm reg}} (\omega,T)$ in Eq. (\ref{sigma})
can be expressed as,
\begin{equation}
D_{\alpha} (T) = C_{\alpha} (T)\,\Pi_{\alpha} (T) = C_{\alpha} (T)\,L\,\langle\vert j_{\alpha,\pm 1/2}\vert^2\rangle_{m_{{\bar{\alpha}} z},T} \, ,
\label{DproptoOmega}
\end{equation}
where the coefficient $C_{\alpha} (T)$ is finite and,
\begin{eqnarray}
&& \langle\vert j_{\alpha,\pm 1/2}\vert^2\rangle_{m_{{\bar{\alpha}} z},T} = {\Pi_{\alpha} (T)\over L}
\nonumber \\
&& = \sum_{S_{\tau}=0}^{L/2}
\sum_{S_{\alpha}=1}^{L_{\alpha}/2}
\sum_{S_{\alpha}^z=-S_{\alpha}}^{S_{\alpha}}
\sum_{S_{\bar{\alpha}}=\vert S_{\bar{\alpha}}^z\vert}^{L_{\bar{\alpha}}/2} \sum_{l_{\rm r}^{u}} 
p_{l_{\rm r}^{u},S_{\tau},S_{\alpha},S_{\alpha}^z,S_{\bar{\alpha}},S_{\bar{\alpha}}^z}\vert j_{\alpha,\pm 1/2}\vert^2
\nonumber \\
&& = \sum_{S_{\tau}=0}^{L/2}
\sum_{S_{\alpha}=1}^{L_{\alpha}/2}
(2S_{\alpha}+1)
\sum_{S_{\bar{\alpha}}=\vert S_{\bar{\alpha}}^z\vert}^{L_{\bar{\alpha}}/2} \sum_{l_{\rm r}^{u}} 
p_{l_{\rm r}^{u},S_{\tau},S_{\alpha},0,S_{\bar{\alpha}},S_{\bar{\alpha}}^z}
\nonumber \\
&& \hspace{4.5cm}\times \vert j_{\alpha,\pm 1/2}\vert^2 \, .
\label{jz2TD}
\end{eqnarray}

Here $\langle\vert j_{\alpha,\pm 1/2}\vert^2\rangle_{m_{{\bar{\alpha}} z},T} = \Pi_{\alpha} (T)/L$ is thus the
thermal expectation value of the square of the absolute value $\vert j_{\alpha,\pm 1/2}\vert$ of the 
$\alpha$-spin elementary current carried by each $\alpha$-spin carrier of projection $\pm 1/2$, 
Eq. (\ref{elem-currents}), in the subspace spanned by states with fixed values of only $S_{\bar{\alpha}}^z$.
Indeed, that thermal expectation value involves summations over all finite-$S_{\alpha}$ energy eigenstates with fixed value of $S_{\bar{\alpha}}^z$,
thus including those with different $S_{\alpha}^z\in [-S_{\alpha},S_{\alpha}]$ values. 

Since the $\alpha$-spin diffusion constant, Eq. (\ref{DproptoOmega}), 
refers to $\mu_{\alpha} = 0$ in the energy eigenvalue, Eq. (\ref{E}), 
where $\mu_s = 2\mu_B h$ for $\alpha =s$ and $\mu_{\eta} = 2\mu$ for $\alpha =\eta$, the
$2S_{\alpha} + 1$ states of each $\alpha$-spin tower, Eq. (\ref{state}), have the same energy. In addition,
there is the symmetry, Eq. (\ref{deriv-elem-currents}), such that the 
$\alpha$-spin elementary currents $j_{\alpha,\pm 1/2}$, Eq. (\ref{elem-currents}), also have the same value for all 
the $2S_{\alpha} + 1$ states of each $\alpha$-spin tower, Eq. (\ref{state}),
and thus are independent of $S_{\alpha}^z$. 

That both the energies and $\alpha$-spin elementary currents have the same value for all 
such $2S_{\alpha} + 1$ states has thus allowed to perform the summation 
$\sum_{S_{\alpha}^z=-S_{\alpha}}^{S_{\alpha}}$ in Eq. (\ref{jz2TD}). For each $\alpha$-spin tower, a number $2S_{\alpha}+1$ of weights 
$\{p_{l_{\rm r}^{u},S_{\tau},S_{\alpha},S_{\alpha}^z,S_{\bar{\alpha}},S_{\bar{\alpha}}^z}\}$
where $S_{\alpha}^z = -S_{\alpha}, ..., S_{\alpha}$ have been replaced by a single weight
$p_{l_{\rm r}^{u},S_{\tau},S_{\alpha},0,S_{\bar{\alpha}},S_{\bar{\alpha}}^z}$
for which $S_{\alpha}^z = 0$ multiplied by the degenerescency factor $(2S_{\alpha}+1)$.

As a result of each $\alpha$-spin tower summation, $\sum_{S_{\alpha}^z=-S_{\alpha}}^{S_{\alpha}}$,
a set of $2S_{\alpha}+1$ states was thus replaced in the overall summations of Eq. (\ref{jz2TD})
by the tower state for which $S_{\alpha}^z = 0$. We could have chosen for instance the 
$\alpha$-spin HWS for which $S_{\alpha}^z = S_{\alpha}$. However, since both choices give the same 
value of $\Pi_{\alpha} (T) = L\,\langle\vert j_{\alpha,\pm 1/2}\vert^2\rangle_{m_{{\bar{\alpha}} z},T}$,
for the sake of comparison of the expression of $\Omega_{\alpha} (T) = \Omega_{m_{\alpha z}=0} (T)$ 
where $\Omega_{m_{\alpha z}} (T)$ is given in Eq. (\ref{jz2T}) and that of $\Pi_{\alpha} (T)$, Eq. (\ref{jz2TD}), we have chosen the 
tower state for which $S_{\alpha}^z = 0$.

It is known that the $h=0$ spin $(\alpha = s)$ and $\mu=0$ charge $(\alpha = \eta)$ diffusion constant
$D_{\alpha} (T)$, Eq. (\ref{DproptoOmega}), obeys the inequality \cite{Medenjak_17},
\begin{equation}
D_{\alpha} (T) \geq {k_B T\over 8 v_{\alpha {\rm LR}}\,\chi_{\alpha} (T)\,f_{\alpha} (T)}
{\partial^2 D_{\alpha}^z (T)\over\partial m_{\alpha z}^2}\Bigg|_{m_{\alpha z} = 0} \, .
\label{Dalphaineq}
\end{equation}
Here $v_{\alpha {\rm LR}}$ and $\chi_{\alpha} (T)$ are the spin ($\alpha = s$) and charge ($\alpha = \eta$) 
Lieb-Robinson velocity and static susceptibility, respectively, and $f_{\alpha} (T)$ is the second derivative of the 
free energy density with respect to $m_{\alpha z}$ at $m_{\alpha z} = 0$. The spin and charge 
Lieb-Robinson velocity are the maximal velocity with which the information can travel through the 
spin-squeezed and $\eta$-spin-squeezed effective lattices, respectively, both of length $L$.

From the use of Eqs. (\ref{D-all-T-simp-m}) and (\ref{jz2T}) we find that,
\begin{equation}
{\partial^2 D_{\alpha}^z (T)\over\partial m_{\alpha z}^2}\Bigg|_{m_{\alpha z} = 0} = 
{L\over k_B T} \langle\vert j_{\alpha,\pm 1/2}\vert^2\rangle_{0,m_{{\bar{\alpha}} z},T} = {\Omega_{\alpha} (T)\over k_B T} \, ,
\label{2derivD-all-T-simp-m}
\end{equation}
where $\Omega_{\alpha} (T) = \Omega_{m_{\alpha z}=0} (T)$, so that, 
\begin{eqnarray}
&& \langle\vert j_{\alpha,\pm 1/2}\vert^2\rangle_{0,m_{{\bar{\alpha}} z},T} =  {\Omega_{\alpha} (T)\over L} =
\nonumber \\
&& \sum_{S_{\tau}=0}^{L/2}
\sum_{S_{\alpha}=1}^{L_{\alpha}/2}
\sum_{S_{\bar{\alpha}}=\vert S_{\bar{\alpha}}^z\vert}^{L_{\bar{\alpha}}/2} \sum_{l_{\rm r}^{u}} 
\,p_{l_{\rm r}^{u},S_{\tau},S_{\alpha},0,S_{\bar{\alpha}},S_{\bar{\alpha}}^z}\vert j_{\alpha,\pm 1/2}\vert^2 \, ,
\label{jz2T0}
\end{eqnarray}
is the thermal expectation value $\langle\vert j_{\alpha,\pm 1/2} \vert^2\rangle_{m_{\alpha z},m_{{\bar{\alpha}} z},T}$
in Eq. (\ref{jz2T}) at $m_{\alpha z}=0$. Here we accounted for that $j_{\alpha,\pm 1/2}= 0$ at $S_{\alpha} = 0$ and that due to the form of the
energy eigenvalues, Eq. (\ref{E}), for $S_{\alpha}^z =0$ states the Boltzmann weights 
$p_{l_{\rm r}^{u},S_{\tau},S_{\alpha},0,S_{\bar{\alpha}},S_{\bar{\alpha}}^z}$ do not depend on $S_{\alpha}^z$.

The use of the expression, Eq. (\ref{2derivD-all-T-simp-m}), on the right-hand side of Eq. (\ref{Dalphaineq}) then leads to
the inequality,
\begin{equation}
D_{\alpha} (T) \geq {L\,\langle\vert j_{\alpha,\pm 1/2}\vert^2\rangle_{0,m_{{\bar{\alpha}} z},T}\over 8 v_{\alpha {\rm LR}}\,\chi_{\alpha} (T)\,f_{\alpha} (T)}
= {\Omega_{\alpha} (T)\over 8 v_{\alpha {\rm LR}}\,\chi_{\alpha} (T)\,f_{\alpha} (T)} \, .
\label{DalphaineqF}
\end{equation}

As justified in the following and consistent with
the lower bound of the inequality, Eq. (\ref{DalphaineqF}), we find strong evidence that the
following expression for the finite coefficient $C_{\alpha} (T)$ in the $\alpha$-spin-diffusion constant 
$D_{\alpha} (T)$, Eq. (\ref{Dalphaineq}), is either exact or a very good approximation, 
\begin{equation}
C_{\alpha} (T) = {1\over 8 v_{\alpha {\rm LR}}\,\chi_{\alpha} (T)\,f_{\alpha} (T)} \, .
\label{CalphaT}
\end{equation}
Its use in Eq. (\ref{DproptoOmega}) gives,
\begin{eqnarray}
D_{\alpha} (T) & = & {\Pi_{\alpha} (T)\over 8 v_{\alpha {\rm LR}}\,\chi_{\alpha} (T)\,f_{\alpha} (T)}
\hspace{0.20cm}{\rm such}\hspace{0.20cm}{\rm that}
\nonumber \\
D_{\alpha} (T) & > & {\Omega_{\alpha} (T)\over 8 v_{\alpha {\rm LR}}\,\chi_{\alpha} (T)\,f_{\alpha} (T)} \, ,
\label{Dalpha}
\end{eqnarray}
which indeed is consistent with the inequality, Eq. (\ref{DalphaineqF}).

There is some universality concerning some basic properties of the diffusion constants of the integrable model considered in 
Ref. \onlinecite{Medenjak_17} and the 1D Hubbard model. We have accounted for both that universality and the different 
notations and representations used in that reference and in this paper, respectively.

We have compared the state summations that contribute to the diffusion-constant expression given in Eq. (5) of 
Ref. \onlinecite{Medenjak_17} for $n\rightarrow\infty$ and those that contribute to its lower bound in the inequality provided in Eq. (6) of that reference.
For the 1D Hubbard model, that inequality has the form given in Eq. (\ref{Dalphaineq}). Within our representation in terms
of the elementary currents carried by the $\alpha$-spin carriers, it can be exactly written as given in Eq. (\ref{DalphaineqF}).

We have then accounted for the following properties:\\

(A) The different state summations that contribute to $\Pi_{\alpha} (T)$, Eq. (\ref{jz2TD}), and
$\Omega_{\alpha} (T)$, Eq. (\ref{jz2T0}), respectively, and the inequality $\Pi_{\alpha} (T)>\Omega_{\alpha} (T)$
that follows from such state summations.\\

(B) The different state summations that contribute to the diffusion-constant expression given in Eq. (5) of 
Ref. \onlinecite{Medenjak_17} for $n\rightarrow\infty$ and to its lower bound in the inequality provided in Eq. (6) of that reference,
respectively.\\

Taking into consideration the different notations and representations, that the two types of state summations in (A) correspond to 
the two types of state summations in (B), respectively, and the lower bound as written in Eq. (\ref{Dalpha})
and the expression of the $\alpha$-spin diffusion constant given in that equation have the same dimensions provides
indeed strong evidence that the latter expression and that of the finite coefficient $C_{\alpha} (T)$, Eq. (\ref{CalphaT}),
are either exact or a very good approximation.

At $h=0$ for spin and at $\mu =0$ for charge, the dominant $T>0$ transport is normal diffusive when
$D_{\alpha} (T)$ is finite and anomalous superdiffusive when $D_{\alpha} (T)$ diverges \cite{Ilievski_18,Fava_20,Moca_23}.
Independently of the form of the coefficient $C_{\alpha} (T)$, Eq. (\ref{CalphaT}), the point here is that it is finite. 
The relation, Eq. (\ref{DproptoOmega}), then shows that the thermal expectation value 
$\langle\vert j_{\alpha,\pm 1/2}\vert^2\rangle_{m_{{\bar{\alpha}} z},T}$, Eq. (\ref{jz2TD}), of 
the square of the absolute value $\vert j_{\alpha,\pm 1/2}\vert$ of the 
$\alpha$-spin elementary current carried by each $\alpha$-spin carrier of projection $\pm 1/2$, 
Eq. (\ref{elem-currents}), controls the type of $\alpha$-spin transport. 

Specifically, according to the following criteria involving the thermal expectation value
$\langle\vert j_{\alpha,\pm 1/2}\vert^2\rangle_{m_{{\bar{\alpha}} z},T} = \Pi_{\alpha} (T)/L$ one has that,
\begin{eqnarray}
&& \Pi_{\alpha} (T) = L\,\langle\vert j_{\alpha,\pm 1/2}\vert^2\rangle_{m_{{\bar{\alpha}} z},T} = 
\infty \Rightarrow \hspace{0.20cm}{\rm superdiffusion}
\nonumber \\
&& \Pi_{\alpha} (T) = L\,\langle\vert j_{\alpha,\pm 1/2}\vert^2\rangle_{m_{{\bar{\alpha}} z},T} 
\hspace{0.20cm}{\rm finite} \Rightarrow \hspace{0.20cm}{\rm diffusion} \, ,
\label{noramomD}
\end{eqnarray}
where $\alpha = s$ for spin at $h=0$ and $\alpha = \eta$ for charge at $\mu=0$.

\subsection{Effects of the $[SU (2)\times SU(2)\times U(1)]/Z_2^2$ symmetry on 
$T>0$ charge transport}
\label{SECIVB}  

The use of hydrodynamic theory and KPZ scaling
to study dynamical scaling properties of spin and charge transport
seems to indicate that the $\alpha$-spin diffusion constant $D_{\alpha} (T)$ 
at $h=0$ for spin ($\alpha =s$) and at $\mu =0$ for charge ($\alpha = \eta$) diverges at $T>0$, which would imply 
anomalous superdiffusive spin and charge transport \cite{Ilievski_18,Fava_20,Moca_23}.

As mentioned in Sec. \ref{SECI}, originally, the KPZ universality class refers actually to a broad range of classical stochastic 
growth models that exhibit similar scaling behavior to the original KPZ equation \cite{Kardar_86,Krug_97,Kriecherbauer_10,Corwin_12}.
It is thus not completely surprising that the KPZ scaling has also been shown to describe high-temperature 
$T\rightarrow\infty$ and thus classical dynamics of certain many-body systems near equilibrium
\cite{Ljubotina_19,Ljubotina_17}. 

What is somehow surprising is that quantum effects allow the
classical KPZ scaling to prevail for all finite temperatures $T>0$ in the case of the 
dynamical spin structure factor of the $SU(2)$ symmetrical spin-$1/2$ $XXX$ chain.
Indeed, it was found to be exactly described by the KPZ correlation function, and thus that chain exhibits 
anomalous superdiffusive behavior with dynamical scaling exponent $z=3/2$. 

And this also thus applies to the gapless spin degrees of freedom of the 1D Hubbard 
model at $h=0$ \cite{Ilievski_18,Fava_20,Moca_23}. This invariance under quantum effects of the classical
KPZ scaling is related to the gapless nature of spin excitations.
That $h=0$ spin transport is anomalous superdiffusive for the $U>0$ 1D Hubbard 
model implies that $\Pi_{s} (T) = \infty$ in Eq. (\ref{noramomD}) for spin $\alpha = s$ 
at finite temperature $T>0$.

On the other hand, the exact results of Ref. \onlinecite{Carmelo_24} show that
$\mu =0$ charge transport rather is normal diffusive in the $k_B T/\Delta_{\eta}\ll 1$ regime.
Here $\Delta_{\eta}$ denotes the Mott-Hubbard gap \cite{Lieb_68,Carmelo_24}.
The prediction of anomalous superdiffusive charge transport at any finite temperature by hydrodynamic theory (KPZ scaling) 
is based on the assumption that at half filling $\eta$-spin and spin $SU(2)$ symmetries 
are related by duality, which would imply that spin and charge transport have identical properties.
However, the physical reason why concerning charge transport that prediction fails is that for $k_B T/\Delta_{\eta}\ll 1$ 
and due to quantum effects accounted for by the $\tau$-translational $U(1)$ symmetry degrees of freedom 
the above duality relation is broken by the Mott-Hubbard gap \cite{Carmelo_24}.

At $\mu = h=0$ the 1D Hubbard Hamiltonian, Eq. (\ref{H}), commutes with the generators of global 
$[SU (2)\times SU(2)\times U(1)]/Z_2^2$ symmetry, Eqs. (\ref{SSSS}), (\ref{hatStau}), and (\ref{hatStauMEL}).
We have that $L_s = L$ and $L_{\eta} = 0$ for the corresponding absolute ground state that is not populated 
by physical $\eta$-spins. For $L$ even and $L/2$ odd its energy eigenvalue 
$E_{GS}$, Eq. (\ref{E}), and momentum eigenvalue, Eq. (\ref{PP}), read \cite{Lieb_68},
\begin{equation}
{E_{GS}\over L} = - \Bigl(4t\int_0^{\infty}d\omega {J_0 (\omega)J_1 (\omega)\over\omega (1 + e^{2\omega u})} 
+ {U\over 4}\Bigr) \hspace{0.20cm}{\rm and}\hspace{0.20cm} P_{GS} = 0 \, ,
\label{EAGS}
\end{equation}
respectively, where $J_0 (\omega)$ and $J_1 (\omega)$ are Bessel functions.
\begin{figure*}
\includegraphics[width=0.495\textwidth]{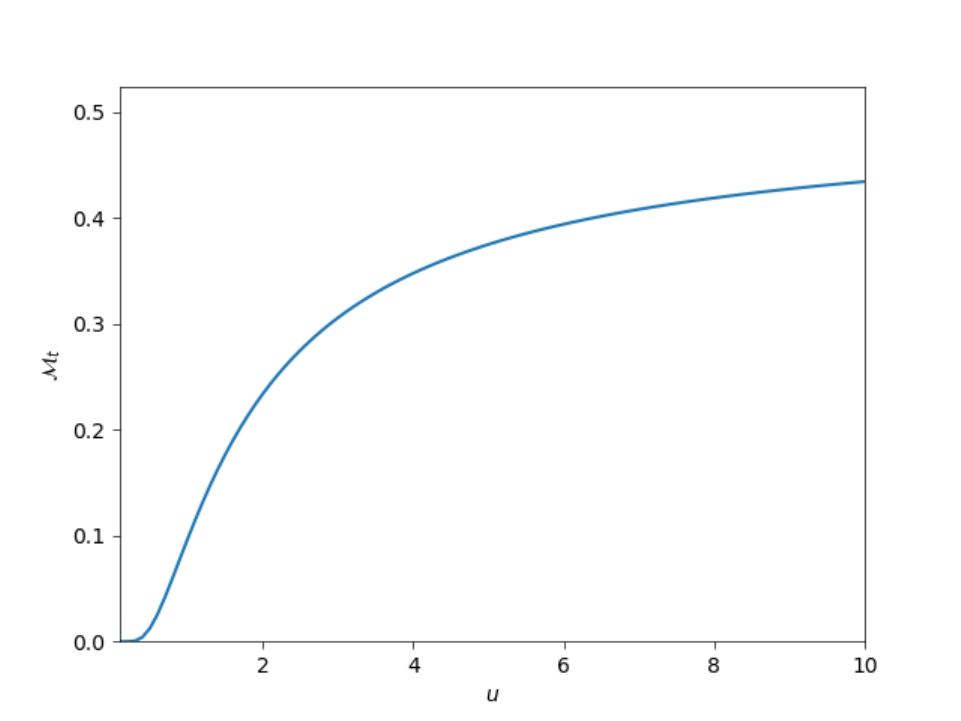}
\includegraphics[width=0.495\textwidth]{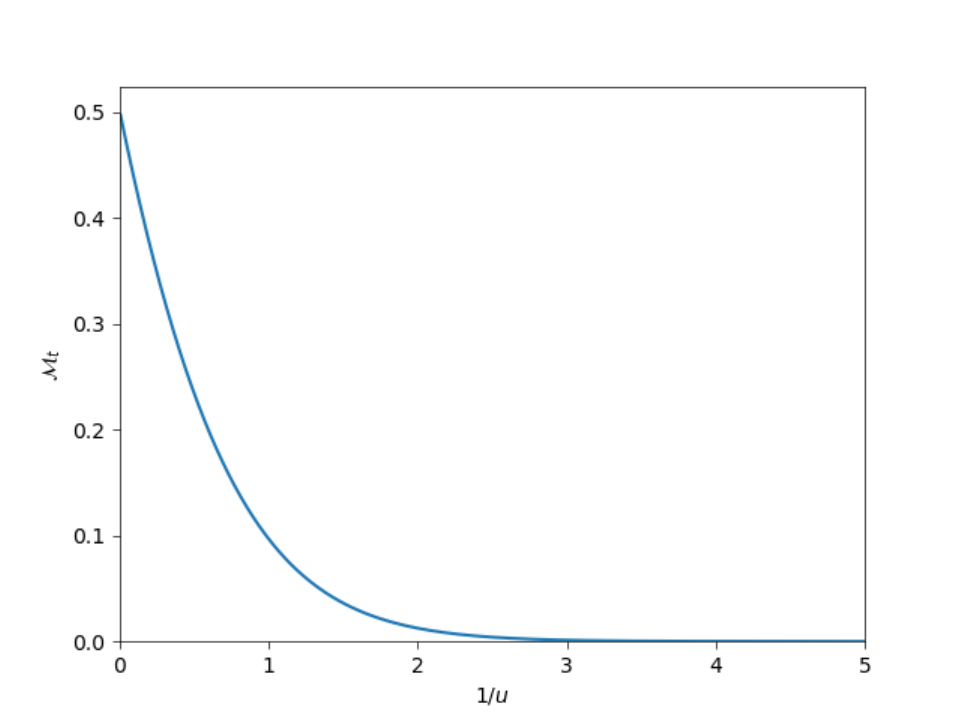}
\caption{The $\mu = h =0$ charge transport mass ${\cal{M}}_t= {\cal{M}}_{\tau}^h$ of the $\eta$-spin triplet pairs 
that are the charge carriers as a function of (a) $u=U/4t$ and (b) $1/u = 4t/U$. (Here
${\cal{M}}_{\tau}^h$ is the $\tau$-hole static mass, Eq. (\ref{invm0}).) This charge transport mass
appears in the charge-diffusion constant expression, Eq. (\ref{D}). That ${\cal{M}}_t=0$ at 
$u=0$ marks in the $k_B T/\Delta_{\eta}\ll 1$ regime the transition from ballistic charge transport at $U=0$ to 
normal diffusive transport for $U>0$. The $\mu = h =0$ charge transport mass ${\cal{M}}_t \in [0,1/2t]$ is enhanced 
on increasing $u=U/4t$, reaching its largest value $1/2t$ in the $u\rightarrow\infty$ limit.}
\label{figure1PRB_DD}
\end{figure*}

At $\mu = h=0$ charge transport is in the $k_B T/\Delta_{\eta}\ll 1$ regime controlled by excited states
populated by two unpaired physical $\eta$-spins with the same projection whose translational degrees
of freedom are described by two $\tau$-holes with $\tau$-band momentum values 
$-q_j \approx \mp\pi$ and $- q_{j'} = - q_j \pm 2\pi/L$ \cite{Carmelo_24}. Such states energy reads
$E (q_j,q_{j'}) = E_{GS} + \delta E (q_j,q_{j'})$ where the excitation energy $\delta E (q_j,q_{j'})$ and 
their momentum eigenvalue $P$ are given by,
\begin{equation} 
\delta E (q_j,q_{j'}) = - \varepsilon_{\tau} (q_j) - \varepsilon_{\tau} (q_{j'}) \hspace{0.20cm}{\rm and}\hspace{0.20cm}
P  = \pm \pi - q_j - q_{j'} \, .
\label{deltaE}
\end{equation}
Here,
\begin{equation}
\varepsilon_{\tau} (q_j) = - {\Delta_{\tau}\over 2} - {(\pm\pi -q_j)^2\over 2{\cal{M}}_{\tau}^h} \, ,
\label{varepsiloncpi}
\end{equation}
where at $\mu = h=0$ the energy $\Delta_{\tau}$ associated below with the $\tau$-translational $U(1)$ symmetry equals the Mott-Hubbard 
gap $\Delta_{\eta}$ and reads,
\begin{equation}
\Delta_{\tau} = \Delta_{\eta} = U - 4t + 8t\int_0^{\infty}d\omega {J_1 (\omega)\over\omega (1 + e^{2\omega u})} \, .
\label{gap0}
\end{equation}
The $\tau$-holes static mass also appearing in Eq. (\ref{varepsiloncpi}) is given by,
\begin{equation}
{\cal{M}}_{\tau}^h =
{1\over 2t}\,{\left(1 - 2\int_0^{\infty}d\omega {J_0 (\omega)\over 1 + e^{2\omega u}}\right)^2\over 
\vert 1 - 2\int_0^{\infty}d\omega {\omega J_1 (\omega)\over 1 + e^{2\omega u}}\vert} 
\in \left[0,{1\over 2t}\right] \, .
\label{invm0}
\end{equation}

While the $\tau$-band momentum values $q_j$ define the relative momentum-space translational 
degrees of freedom of the $L_s$ physical spins and $L_{\eta}$ $\eta$-physical spins, respectively, the 
corresponding $\tau$-band energy $\varepsilon_{\tau} (q_j)$, Eq. (\ref{varepsiloncpi}), defines their relative
energy scales. At $\mu = h =0$ the minimum energy of the gapped $\eta$-spin degrees of freedom relative to that of the
gapless spin degrees of freedom is $\Delta_{\tau} = -2\varepsilon_{\tau} (\pm\pi)$, Eq. (\ref{gap0}). 
That energy is associated with the creation of two holes in the $\tau$-band at exactly momentum values 
$q_j = \pm\pi$ and $q_j = \pm(\pi - 2\pi/L)$, respectively. 

Consistently with the relation $N_{\tau}^h = L_{\eta}$, two unpaired physical $\eta$-spins have also been created.
The conservation relation $L = L_{s} + L_{\eta}$ then implies that two physical spins have been
annihilated. This refers to the removal of one $s1$-pair at $s1$-band momentum $q_j =\pm\pi/2$ that
due to a $s1$-band overall momentum shift $\pm \pi/L$ is a zero-momentum process.
The minimum energy of such processes refers to the Mott-Hubbard gap that at $\mu = h =0$ 
when the quantum-problem global symmetry is 
$[SU (2)\times SU(2)\times U(1)]/Z_2^2$ and contains two $SU(2)$ symmetries
is {\it fully controlled} by the $\tau$-translational $U(1)$ symmetry, as it reads $\Delta_{\eta} = \Delta_{\tau}$, Eq. (\ref{gap0}).

The studies of Refs. \onlinecite{Ilievski_18,Fava_20,Moca_23} have only considered the $SU (2)\times SU(2)$
symmetries in  $[SU (2)\times SU(2)\times U(1)]/Z_2^2$. Therefore, they have not accounted for the key role
of the $\tau$-translational $U(1)$ symmetry: At the $\mu = h=0$ point its irreducible representations are behind the emergence of the Mott-Hubbard gap
$\Delta_{\eta} = \Delta_{\tau}$ that separates the gapped Mott-Hubbard insulator $\eta$-spin degrees of freedom
from the gapless spin degrees of freedom, respectively.
The duality relation of the $\eta$-spin $SU(2)$ symmetry and spin $SU(2)$ symmetry
assumed in these references is broken by the Mott-Hubbard gap brought about by the $\tau$-translational 
$U(1)$ symmetry in $[SU (2)\times SU(2)\times U(1)]/Z_2^2$ \cite{Carmelo_24}.

In the $k_B T/\Delta_{\eta} \ll 1$ regime the charge carriers are actually $\eta$-spin triplet pairs whose
translational degrees of freedom are described in the $\tau$-squeezed effective lattice by 
those of two adjacent unpaired physical $\eta$-spins with the same projection $+1/2$
or $-1/2$ \cite{Carmelo_24}. The corresponding two $\tau$-holes propagate with momenta that 
differ only by $2\pi/L\rightarrow 0$ in the thermodynamic limit. 
We call them {\it $\eta$-spin triplet charge carriers} to distinguish from their two unpaired physical $\eta$-spins. 
The spectrum of such $\eta$-spin triplet charge carriers is in the thermodynamic limit given by,
\begin{equation}
\epsilon_{\eta} (k) = \Delta_{\eta} + {(k\pm \pi)^2\over 2{\cal{M}}_{\eta}}
\hspace{0.20cm}{\rm where}\hspace{0.20cm} {\cal{M}}_{\eta} = 2{\cal{M}}_t = 2{\cal{M}}_{\tau}^h \, .
\label{epk}
\end{equation}
Here ${\cal{M}}_{\eta}$ and ${\cal{M}}_t$ are the static and transport masses 
of a $\eta$-spin triplet pair \cite{Carmelo_24}. Its energy spectrum $\epsilon_{\eta} (k)$ refers to the excitation energy and momentum 
in Eq. (\ref{deltaE}) with $k = P \approx \mp\pi$. 

On the other hand, in the case of the Hamiltonian, Eq. (\ref{H}), for $\mu =0$ and finite magnetic field $h\in [0,h_c]$, the 
spin $SU (2)$ symmetry in $[SU (2)\times SU(2)\times U(1)]/Z_2^2$ becomes a $U(1)$ symmetry.
Then the minimum energy of the gapped $\eta$-spin degrees of freedom relative to that of the
gapless spin degrees of freedom is $\Delta_{\eta} = \Delta_{\tau} + 2\mu_B h$. Here 
$\Delta_{\eta} = \epsilon_{\eta} (\pm\pi) = -2\varepsilon_{\tau} (\pm\pi) + 2\mu_B h$ remains being associated 
with the creation of two holes in the $\tau$-band at momentum values $q_j = \pm\pi$ and $q_j = \pm(\pi - 2\pi/L)$
that describe the translational degrees of freedom of the $\eta$-spin triplet pair and
two corresponding unpaired physical $\eta$-spins.

For $h\in ]0,h_c]$ both the $\tau$-translational $U(1)$ symmetry and the spin $U(1)$ symmetry contribute to the Mott-Hubbard gap.
The corresponding annihilation of two physical spins and of their $s1$-pair is a gapless process for the spin degrees 
of freedom that though creates an extra energy difference given by $2\mu_B h$ concerning the minimum value of the 
energy of the gapped $\eta$-spin degrees of freedom relative to that of the
gapless spin degrees of freedom. At $h=0$ also two physical spins are annihilated, yet the spin $SU(2)$ symmetry 
renders it a zero-energy process concerning the relative energy scales of the $\eta$-spin and 
spin degrees of freedom, respectively.

Since charge transport is normal diffusive in the $k_B T/\Delta_{\eta}\ll 1$ regime for $u>0$, $\mu =0$,
and $h\in [0,h_c]$, the corresponding charge diffusion constant $D_{\eta} (T)$,
Eq. (\ref{DproptoOmega}) for $\alpha = \eta$, is finite for very low finite temperatures. In
that regime it decreases on enhancing $T$ as it reads \cite{Carmelo_24},
\begin{equation}
D_{\eta} (T) = {e^{\Delta_{\eta}/k_B T} \over 4{\cal{M}}_t} = {e^{\Delta_{\eta}/k_B T} \over 2{\cal{M}}_{\eta}} \, .
\label{D}
\end{equation}
This diffusion-constant exponential behavior also occurs in
gapped 1D models in the quantum Sine-Gordon universality class \cite{Damle_05}.

The transport mass ${\cal{M}}_t = {\cal{M}}_{\tau}^h$ appearing in Eq. (\ref{D})
where ${\cal{M}}_{\tau}^h$ is given in Eq. (\ref{invm0}) for $\mu = h =0$ is plotted (a) as a function of $u$ and (b) as a
function of $1/u$ in Fig. \ref{figure1PRB_DD} for $\mu = h =0$.
Importantly, that ${\cal{M}}_t=0$ at $u=0$ marks in the $k_B T/\Delta_{\eta}\ll 1$ regime
the transition from ballistic charge transport at $U=0$ to normal diffusive transport for $U>0$.
That transport mass continuously increases from zero to $1/(2t)$ upon varying $u$ from $0$ to $\infty$.
The Mott-Hubbard gap $\Delta_{\eta}$ in Eq. (\ref{epk}) is also a continuously increasing function of $u$, vanishing
at $u=0$ and linearly depending on $U$, $\Delta_{\eta} = U$, and thus diverging as $u\rightarrow\infty$.

The $\eta$-spin triplet pair dispersion $\epsilon_{\eta} (k)$, Eq. (\ref{epk}), and charge-diffusion constant $D_{\eta} (T)$,
Eq. (\ref{D}), have for $h\in [0,h_c]$ the same general form as given in these equations with ${\cal{M}}_{\tau}^h={\cal{M}}_t$ in
${\cal{M}}_{\eta} = 2{\cal{M}}_{\tau}^h$ smoothly increasing from its value at $h=0$, Eq. (\ref{invm0}),
to ${\cal{M}}_{\tau}^h={\cal{M}}_t=1/(2t)$ at $h=h_c$, as shown in Fig. 1 (a) of Ref. \onlinecite{Carmelo_24}. 
The Mott-Hubbard gap $\Delta_{\eta} = \Delta_{\tau} + 2\mu_B h$ also appearing in Eqs. (\ref{epk}) and 
(\ref{D}) changes very little in the interval $h\in [0,h_c]$, as shown if Fig. 1 (a) of Ref. \onlinecite{Carmelo_21} where
it is denoted by $2\Delta_{MH}$.

Accounting for the spin transport being anomalous superdiffusive at $h=0$ for finite temperatures, we find that
the thermal expectation value $\langle\vert j_{\alpha,\pm 1/2}\vert^2\rangle_{m_{{\bar{\alpha}} z},T} = \Pi_{\alpha} (T)/L$,
Eq. (\ref{noramomD}), is for $\alpha = s$ at $h=0$ and for $\alpha = \eta$ at $\mu =0$ such that,
\begin{eqnarray}
\Pi_{s} (T) & = & L\,\langle\vert j_{s,\pm 1/2}\vert^2\rangle_{T} \hspace{0.20cm}{\rm at}\hspace{0.20cm}h = 0
\nonumber \\
& = & \infty \hspace{0.20cm}{\rm for}\hspace{0.20cm} T > 0
\nonumber \\
\Pi_{\eta} (T) & = & L\,\langle\vert j_{\eta,\pm 1/2}\vert^2\rangle_{T} \hspace{0.20cm}{\rm at}\hspace{0.20cm}\mu = 0
\nonumber \\
& = & {4v_{\eta {\rm LR}}\,f_{\eta} (T) \over \sqrt{\pi {\cal{M}}_t\,k_B T}}
\hspace{0.20cm}{\rm for}\hspace{0.20cm} T\ll \Delta_{\eta}/k_B \, .
\label{PietaTloq}
\end{eqnarray}

In the case of the $k_B T/\Delta_{\eta}\ll 1$ regime in which the charge static susceptibility reads
$\chi_{\eta} (T) = \sqrt{2{\cal{M}}_{\eta}/(\pi\,k_B T)}\,e^{-\Delta_{\eta}/k_B T}$ where
${\cal{M}}_{\eta} = 2{\cal{M}}_t$, we used in Eq. (\ref{PietaTloq}) the expression, Eq. (\ref{CalphaT}) for $\alpha = \eta$, of the coefficient $C_{\eta} (T)$ 
that appears in the diffusion-constant expression, Eq. (\ref{DproptoOmega}) for $\alpha = \eta$.

\section{Conclusions and discussion}
\label{SECV}

The studies of this paper have used a $\tau$- and physical $\alpha$-spin representation that naturally emerges 
from the complete set of energy and momentum eigenstates of the 1D Hubbard model, Eq. (\ref{H}), referring for 
$U>0$ and all values of magnetic field $h$ and chemical potential $\mu$ to irreducible representations of its global 
$[SU (2)\times SU(2)\times U(1)]/Z_2^2$ symmetry at $h=\mu =0$. The use of that representation
has allowed the identification of the spin and charge carriers in the $\alpha = s,\eta$ 
$\alpha$-spin multiplet configuration of all $S_s > 0$ and $S_{\eta} > 0$ 
energy eigenstates, respectively. Their spin elementary currents $j_{s,\pm 1/2}$ and 
charge elementary currents $j_{\eta,\pm 1/2}$ have been found in this paper
to control the transport properties of the model.

Specifically, the $\alpha =s,\eta$ $\alpha$-spin current expectation values $\langle \hat{J}_{\alpha}^z \rangle$,
$T>0$ and $\mu_{\alpha} >0$ $\alpha$-spin stiffness $D_{\alpha}^z (T)$, and $T>0$ and 
$\mu_{\alpha} = 0$ $\alpha$-spin-diffusion constant $D_{\alpha} (T)$ were found to read,
\begin{eqnarray}
\langle \hat{J}_{\alpha}^z \rangle & = & N_{\alpha,+1/2}\,\,j_{\alpha,+1/2} + N_{\alpha,-1/2}\,\,j_{\alpha,-1/2}
\nonumber \\
D_{\alpha}^z (T) & = & m_{\alpha z}^2\,{L\over 2k_B T} \langle\vert j_{\alpha,\pm 1/2}\vert^2\rangle_{m_{\alpha z},m_{{\bar{\alpha}} z},T}
\nonumber \\
D_{\alpha} (T) & = & {L\over
8 v_{\alpha {\rm LR}}\,\chi_{\alpha} (T)\,f_{\alpha} (T)}\langle\vert j_{\alpha,\pm 1/2}\vert^2\rangle_{m_{{\bar{\alpha}} z},T} \, ,
\label{3JDzD}
\end{eqnarray}
where $\mu_{s} = 2\mu_B h$ for spin and $\mu_{\eta} = 2\mu$ for charge and
all quantities and thermal averages in this equation have been defined in this paper.

In the case of finite magnetic field for spin and finite chemical potential for charge, $T>0$ transport
is ballistic, so that the form of the $\alpha = s,\eta$ $\alpha$-spin stiffness in Eqs. (\ref{jz2T}) and (\ref{3JDzD})
reveals that the thermal expectation value of the square of the spin and charge elementary currents
carried by the corresponding carriers is such that,
\begin{eqnarray}
&& \langle\vert j_{\alpha,\pm 1/2} \vert^2\rangle_{m_{\alpha z},m_{{\bar{\alpha}} z},T} = {\Omega_{m_{\alpha z}} (T)\over L} 
\nonumber \\
&& {\rm where}\hspace{0.20cm}\Omega_{m_{\alpha z}} (T)\hspace{0.20cm}
{\rm is}\hspace{0.20cm}{\rm finite}\hspace{0.20cm}{\rm for}\hspace{0.20cm}\mu_{\alpha} > 0 \, .
\label{jhmufinite}
\end{eqnarray}
This shows that for subspaces spanned by states with fixed values of $S_z^z$ and $S_{\eta}^z$ 
the average values of the square of the spin and charge elementary currents $j_{s,+1/2}$ and $j_{\eta,+1/2}$,
respectively, are of the order of $1/L$. 

In the case of $T>0$  and $\mu_{\alpha}=0$ nonballistic $\alpha$-spin transport, we have that,
\begin{equation}
\langle\vert j_{\alpha,\pm 1/2} \vert^2\rangle_{m_{{\bar{\alpha}} z},T} = {\Pi_{\alpha} (T)\over L} \, .
\label{jhmuzero}
\end{equation}

As given in Eq. (\ref{PietaTloq}), our results are consistent with $T>0$ $\alpha =s$ spin transport being 
anomalous superdiffusive, as predicted by hydrodynamic theory and KPZ scaling \cite{Ilievski_18,Fava_20,Moca_23}, 
so that $\Pi_{s} (T) = \infty$ for $T>0$. In the thermodynamic limit, $L\rightarrow\infty$, this is consistent with 
$\langle\vert j_{s,\pm 1/2} \vert^2\rangle_{m_{\eta z},T}$ being of order of one rather than of $1/L$. 
In contrast, $\langle\vert j_{s,\pm 1/2} \vert^2\rangle_{0,m_{\eta z},T} = \Omega_s (T)/L$,
which refers to Eq. (\ref{jhmufinite}) for $\alpha =s$ and $m_{s z} =0$, is of the latter order. 

$\Pi_{s} (T) = \infty$, Eq. (\ref{jz2TD}) for $\alpha = s$, and finite $\Omega_s (T)$, Eq. (\ref{jz2T0}) for 
$\alpha = s$, only differ in the summation $\sum_{S_{s}=1}^{L_{s}/2}$ that includes an extra factor
$(2S_{s}+1)$ in the case of $\Pi_{s} (T)$. That $L_s = L$ and $L_{\eta} = 0$ for the $h = \mu =0$ ground
state is consistent with many excited states having a number $L_s$ of physical spins of the order of $L$.
On reaching the largest values in the summation $\sum_{S_{s}=1}^{L_{s}/2}$, the extra factor $(2S_{s}+1)$ 
itself becomes of order of $L$. This is consistent with $\Pi_{s} (T) = \infty$ in spite of $\Omega_s (T)$ 
being finite for $T>0$.

In the case of $T>0$ $\alpha = \eta$ charge/$\eta$-spin transport, our analysis is consistent with 
the exact result \cite{Carmelo_24} that $\Pi_{\eta} (T)$ is finite in the $k_B T/\Delta_{\eta}\ll 1$ regime, 
see Eq. (\ref{PietaTloq}). This contradicts the prediction of hydrodynamic theory and KPZ scaling 
of anomalous superdiffusive for all finite temperatures \cite{Ilievski_18,Fava_20,Moca_23}.

As discussed in Sec. \ref{SECIVB}, that is due to quantum effects
associated with the Mott-Hubbard gap \cite{Carmelo_24}. They break the duality that is assumed by hydrodynamic theory and KPZ scaling 
to relate at $\mu = h = 0$ the $\eta$-spin and spin $SU(2)$ symmetries, respectively, which would imply 
that spin and charge transport have identical properties. Indeed, such effects prevent large-$S_{\eta}$ 
states to contribute to charge transport at low temperatures. 

A second reason is related to the $h = \mu =0$ ground state being populated by $L_s = L$ physical spins and no $L_{\eta} = 0$ physical
$\eta$-spins, so that to reach $L_{\eta}$ values of order of $L$ in the summation
$\sum_{S_{\eta}=1}^{L_{\eta}/2}$, Eq. (\ref{jz2TD}) for $\alpha = \eta$, requires
a huge number of microscopic processes associated with the generation
of states with very large energies.

The microscopic processes that control the spin and charge transport are described
by the expressions of the spin and charge elementary currents $j_{s,+1/2}$ and $j_{\eta,+1/2}$,
respectively, that appear in the transport quantities given in Eq. (\ref{3JDzD}). Such processes
become more explicit when such elementary currents, Eqs. (\ref{js}) and (\ref{jeta}), are expressed as,
\begin{eqnarray}
 && j_{s,\pm 1/2} = \pm {2t\over N_{\tau}}\sum_{j=1}^L {N_{\tau} (q_j)\over 2\pi\rho (k_j)} \sin k_j
\pm {(4t)^2\over U}{2\over L}\sum_{n=1}^{\infty} {1\over M_{sn}^h}
 \nonumber \\
 && \times \sum_{j=1}^{L_{sn}} {M_{sn}^h (q_j)\over 2\pi\sigma_{sn} (\Lambda_{sn,j})}\sum_{j'=1}^L
 {N_{\tau} (q_{j'}) \sin k_{j'}\over 1 + \left({\Lambda_{sn,j} - \sin k_{j'}\over nu}\right)^2} \, ,
 \label{jsqj}
 \end{eqnarray}
 and
 \begin{eqnarray}
 && j_{\eta,\pm 1/2} = \mp {2t\over N_{\tau}^h}\sum_{j=1}^L {N_{\tau}^h (q_j)\over 2\pi\rho (k_j)} \sin k_j
 \pm {8t\over L}\sum_{n=1}^{\infty} {1\over M_{\eta n}^h}
 \nonumber \\
 && \times \sum_{j=1}^{L_{\eta n}} {M_{\eta n}^h (q_j)\over
 2\pi\sigma_{\eta n}  (\Lambda_{\eta n,j})}\Bigl(n\,{\rm Re}\Bigl\{{(\Lambda_{\eta n,j} - i n u)\over\sqrt{1 - (\Lambda_{\eta n,j} - i n u)^2}}\Bigr\}
 \nonumber \\
 && + {4t\over U} \sum_{j'=1}^L {N_{\tau} (q_{j'}) \sin k_{j'}\over 1 + \left({\Lambda_{\eta n,j} - \sin k_{j'}\over nu}\right)^2}\Bigr) \, ,
 \label{jetaqj}
 \end{eqnarray}
respectively. The momentum distributions $N_{\tau} (q_j)$, $M_{s n}^h (q_j)$, $N_{\tau}^h (q_j)$, and 
$M_{\eta n}^h (q_j)$ in these expressions directly describe the different occupancy configurations of the 
discrete $\tau$-band and $\alpha =s,\eta$ $\alpha n$-band momentum values $q_j = {2\pi\over L}I_j^{\tau}$ 
and $q_j = {2\pi\over L} I_j^{\alpha n}$ given in Eq. (\ref{qq}) that generate the energy eigenstates.
They are such that $q_{j+1} - q_j = 2\pi/L$, as the corresponding quantum numbers $I_j^{\tau}$ and $I_j^{\alpha n}$ 
are integers or half-odd integers according to the boundary conditions, Eqs. (\ref{Itau}) and (\ref{Ian}), respectively.
 
In Eqs. (\ref{jsqj}) and (\ref{jetaqj}), $k_j = k (q_j)$ and $\Lambda_{\alpha n,j} = \Lambda_{\alpha n} (q_j)$ where $\alpha = s,\eta$
are rapidity functions whose inverse functions are solutions of the Bethe-ansatz equations given in functional
form in Eqs. (\ref{Bat})-(\ref{BAetan}) of the Appendix and the 
functions $2\pi\rho (k)$ and $2\pi\sigma_{\alpha n}  (\Lambda)$ are Jacobians associated with the transformation 
from $\tau$-band continuum momentum variables $q \in [q_{\tau}^-,q_{\tau}^+]$ and 
$\alpha n$-band continuum momentum variables $q \in [q_{\alpha n}^-,q_{\alpha n}^+]$ to momentum rapidity variables $k \in [-\pi,\pi]$
and rapidity variables $\Lambda \in [-\infty,\infty]$, respectively. Such 
functions are given by the derivatives in Eq. (\ref{rhosigma}) of the Appendix and obey Eqs. (\ref{Grho})-(\ref{Gsigmaetan}) of that Appendix.
  
The expressions of the $\alpha =s,\eta$ 
$\alpha$-spin elementary current in Eqs. (\ref{jsqj}) and (\ref{jetaqj}) 
refer to {\it all} $S_s>0$ and $S_{\eta}>0$ energy eigenstates, as they are functionals 
whose $\tau$-particle distribution $N_{\tau} (q_j)$, $s n$-hole distributions $M_{s n}^h (q_j)$,
$\tau$-hole distribution $N_{\tau}^h (q_j)$, and $\eta n$-hole distributions 
$M_{\eta n}^h (q_j)$ have Pauli-like occupancies, one or zero, whose different
occupancy configurations generate all such states.

The corresponding $N_{\tau} = 2S_s + {\cal{N}}_s$ $\tau$-particles
and $M_{s n}^h = 2S_{s} + {\cal{N}}_{s n}$ $s n$-holes describe translational degrees of freedom of
the $N_s = 2S_s$ unpaired physical spins and the $N_{\tau}^h = 2S_{\eta} + {\cal{N}}_{\eta}$ $\tau$-holes
and $M_{\eta n}^h = 2S_{\eta} + {\cal{N}}_{\eta n}$ $\eta n$-holes describe translational degrees of freedom of
the $N_{\eta} = 2S_{\eta}$ unpaired physical $\eta$-spins. Their occupancy configurations determine
the values of the elementary currents carried by such $N_s = 2S_s$ spin carriers and $N_{\eta} = 2S_{\eta}$ charge carriers,
respectively, which occurs through the functional, Eqs. (\ref{jsqj}) and (\ref{jetaqj}). 

The corresponding values of the $T>0$ and $\mu_{\alpha} >0$ $\alpha$-spin stiffness $D_{\alpha}^z (T)$
and $T>0$ and $\mu_{\alpha} = 0$ $\alpha$-spin-diffusion constant $D_{\alpha} (T)$ are controlled by those of the
corresponding two types of thermal averages of the square of such 
$\alpha$-spin elementary currents, as given 
in Eq. (\ref{3JDzD}). The general expressions of the $\alpha$-spin current expectation values $\langle \hat{J}_{\alpha}^z\rangle$ also
given in that equation are provided in Eqs. (\ref{Jsz}) and (\ref{Jetaz}) of the Appendix.
These expressions contain the factors ${L_{\alpha,+1/2} - L_{\alpha,-1/2}\over L_{\alpha,+1/2} + L_{\alpha,-1/2}}$ and
${N_{\alpha n,+1/2} - N_{\alpha n,-1/2}\over N_{\alpha n,+1/2} + N_{\alpha n,-1/2}}$ where $\alpha = s,\eta$.
They result from the coupling to $\alpha$-spin of the unpaired physical $\alpha$-spins. That coupling
has opposite sign in the case of unpaired physical $\alpha$-spins with opposite projection $\pm 1/2$.

The $u\rightarrow 0$ limit of the expressions given in Eqs. (\ref{jsqj}) and (\ref{jetaqj}) for the 
$\alpha$-spin elementary currents provide no information on $U=0$ spin ($\alpha =s$) and 
charge ($\alpha = \eta$) transport. Indeed, the spin and charge carriers are completely and qualitatively different at $U=0$. 
In addition, concerning the complementary number ${\cal{N}}_{\alpha, \pm 1/2} = L_{\alpha}/2 - S_{\alpha}$ of
paired physical $\alpha$-spins of projection $\pm 1/2$ that also populate a $u=U/4t>0$  finite-$S_{\alpha}$ energy eigenstate,
the binding of the $l=1,...,n$ $\alpha$-spin singlet pairs within each $\alpha n$-pair described by the 
imaginary part, $i(n + 1 -2l)\,u$, of the rapidities, Eq. (\ref{LambdaIm}), vanishes at $u=U/4t=0$. Therefore, 
the $u>0$ paired physical $\alpha$-spins become unpaired at $U=0$.

These qualitative differences are associated with rearrangements of the Hilbert space that are
consistent with the emergence of $h=0$ spin and $\mu =0$ charge ballistic transport at $U=0$ \cite{Fava_20,Moca_23} and the related 
replacement of the $U>0$ global symmetry $[SU (2)\times SU(2)\times U(1)]/Z_2^2 = [SO(4) \times U(1)]/Z_2$ 
by the $U=0$ global symmetry $O(4)/Z_2 = SO(4) \times Z_2$ whose irreducible representations are
different from those of the former symmetry.

The mechanisms that determine how charge transport at $\mu =0$ evolves for all finite temperatures $T>0$ upon increasing $T$
beyond the $k_B T/\Delta_{\alpha}\ll 1$ limit is a very complex quantum problem that will be 
further studied elsewhere. 

The $\tau$- and physical $\alpha$-spin representation used in the studies of this paper
that describes $T>0$ spin and charge transport in terms of the spin and charge elementary currents
carried by the spin and charge carriers in the spin and $\eta$-spin
multiplet configurations of all $S_s>0$ and $S_{\eta}>0$
energy eigenstates, respectively, of the 1D Hubbard model is the suitable framework to clarify the interesting
open problems associated with $T>0$ charge transport in that model.

Concerning both experimental \cite{Jin_19,Vianez_22}
and theoretical \cite{Tsyplyatyev_22} results that detected effects beyond the Luttinger liquid limit,
the use of our representation for the 1D Hubbard model \cite{Carmelo_17A} and extended versions of it  
\cite{Carmelo_19,Carmelo_19A} also described related effects beyond that limit.

The results of this paper are on the transport properties of the 1D Hubbard model. On the other hand, the global
$[SU (2)\times SU(2)\times U(1)]/Z_2^2 = [SO (4)\times U(1)]/Z_2$ symmetry refers for $U>0$ to the Hubbard model on 
any {\it bipartite lattice} \cite{Carmelo_10}. However, most studies have assumed it to be only $SO (4)$
\cite{Essler_05,Ilievski_18,Fava_20,Moca_23}.

Last but not least, the role of the $\tau$-translational $U(1)$ symmetry beyond $SO (4)$ 
in the properties of the model on higher dimensional bipartite lattices and of the materials and systems it 
describes \cite{Carmelo_10,Carmelo_12} is an interesting scientific issue that deserves further studies. A universal 
property of all such quantum problems is that the $\tau$ $U(1)$ symmetry describes the relative translational 
degrees of freedom of the spin $SU (2)$ and $\eta$-spin $SU (2)$ symmetries, respectively.

%%%%%%%%%%%%%%%%%%%%%%%%%%%%%%%%%%%%%%%%%%%%%%%%%%%%%%%%%%%%%%%%%%%%%%%%%%
\acknowledgements
We thank Iveta R. Pimentel, Toma\v{z} Prosen, Subir Sachdev, and Pedro D. Sacramento 
for illuminating discussions. J. M. P. C. acknowledges support from Funda\c{c}\~ao para a Ci\^encia e Tecnologia
through Grant No. UIDB/04650/2020 and J. E. C. C. acknowledges support from Funda\c{c}\~ao para a Ci\^encia e Tecnologia
through Grants No. UIDB/00618/2020 and UID/CTM/04540/2019.
%%%%%%%%%%%%%%%%%%%%%%%%%%%%%%%%%%%%%%%%%%%%%%%%%%%%%%%%%%%%%%%%%%%%%%%%%%
\appendix

\section{Basic quantities needed for the studies of this paper}
\label{A}

The dimension of the full Hilbert space of the Hamiltonian, Eq. (\ref{H}), is obtained by the summation over the
integer values of $2S_{\tau}\geq 0$, $2S_s\geq 0$, and $2S_{\eta}\geq 0$ of the product of the numbers of irreducible
representations of each of the three global symmetries in $[SU (2)\times SU(2)\times U(1)]/Z_2^2$ as,
\begin{equation}
\sum_{2S_{\tau}=0}^{L} 
\sum_{2S_{s}=0}^{L_s}\,
\sum_{2S_{\eta}=0}^{L_{\eta}} C_0^1\cdot
d_{\tau} (S_{\tau})\times{\cal{N}}(S_{s},L_s)\times{\cal{N}}(S_{\eta},L_{\eta}) = 4^L 
\label{FullDimension}
\end{equation}
where,
\begin{eqnarray}
L_s & = & L - 2S_{\tau}\hspace{0.20cm}{\rm and}\hspace{0.20cm}L_{\eta} = 2S_{\tau}
\nonumber \\
C_{0}^1 & = & \prod_{\alpha = s,\eta}\vert\cos\left({\pi\over 2}(2S_{\alpha}+2S_{\tau})\right)\vert = 0,1 \, .
\label{C01LL}
\end{eqnarray}
Here the coefficient $C_{0}^1 = 0,1$ selects in Eq. (\ref{FullDimension}) the irreducible representations of $SU (2)\times SU(2)\times U(1)$
associated with $[SU (2)\times SU(2)\times U(1)]/Z_2^2$.

For each fixed value of $S_{\tau}$, the dimension $d_{\tau} (S_{\tau})$ in Eq. (\ref{FullDimension}) that gives
the number of irreducible representations of the $\tau$-translational $U(1)$ symmetry refers to
the occupancy configurations that both for the $\tau$-effective lattice and corresponding $\tau$-band reads,
\begin{equation}
d_{\tau} (S_{\tau}) = {L\choose 2S_{\tau}} \, .
\label{dtau}
\end{equation}

For each fixed value of $S_{\tau}$ and $\alpha$-spin $S_{\alpha}$,
the dimension ${\cal{N}} (S_{\alpha},L_{\alpha})$ in Eq. (\ref{FullDimension}) that gives
the number of irreducible representations of the $\alpha = s,\eta$ $\alpha$-spin $SU (2)$ symmetry is
given by,
\begin{eqnarray}
{\cal{N}} (S_{\alpha},L_{\alpha}) & = & (2S_{\alpha} +1)\,{\cal{N}}_{\rm singlet} (S_{\alpha},L_{\alpha})
\hspace{0.20cm}{\rm where}
\nonumber \\
{\cal{N}}_{\rm singlet} (S_{\alpha},L_{\alpha}) & = & {L_{\alpha}\choose L_{\alpha}/2-S_{\alpha}}-{L_{\alpha}\choose L_{\alpha}/2-S_{\alpha}-1}
\nonumber \\
& = & \sum_{\{M_{\alpha n}\}}\, \prod_{n =1}^{\infty}\,{L_{\alpha n}\choose M_{\alpha n}} \, .
\label{NNsinglet}
\end{eqnarray}
Here the factor $(2S_{\alpha} +1)$ refers to the number of states of each $\alpha$-spin tower, Eq. (\ref{state}),
and $L_{\alpha n} = 2S_{\alpha} + \sum_{n'=1}^{\infty}2(n'-n)\,M_{\alpha n'} + M_{\alpha n}$, Eq. (\ref{Nnh}),
is both the number $j = 1,...,L_{\alpha n}$ of sites of the $\alpha n$-squeezed effective lattice and of
discrete momentum momentum values $q_j$ of the $\alpha n$-band.

The (i) first and (ii) second equivalent expressions of ${\cal{N}}_{\rm singlet} (S_{\alpha},L_{\alpha})$ 
in Eq. (\ref{NNsinglet}) refer to (i) the number of $\alpha$-spin $SU(2)$ symmetry algebra 
configurations and (ii) the number of occupancy configurations of
both the $\alpha n$-squeezed-effective lattices and corresponding $\alpha n$-bands whose summation
$\sum_{\{M_{\alpha n}\}}$ refers to all sets $\{M_{\alpha n}\}$ of numbers $M_{\alpha n}$ that
obey the sum-rule, $\sum_{n=1}^{\infty}2n\,M_{\alpha n} = L_{\alpha} - 2S_{\alpha}$. The equality of both expressions 
is exact \cite{Carmelo_10} and confirms that all $\alpha$-spin singlet pairs of the paired physical $\alpha$-spins $1/2$ are contained in 
$\alpha n$-pairs. 

The coupled Bethe-ansatz equations needed to derive a general expression for the $\alpha$-spin current expectation values 
$\langle \hat{J}_{\alpha}^z\rangle$, Eq. (\ref{rel-currents-gen}), are in functional form given by,
\begin{eqnarray}
q_{\tau} (k) & = & k + {1\over\pi}\sum_{\alpha =s,\eta}\sum_{n=1}^{\infty}
\int_{-\infty}^{\infty} d\Lambda\, {\bar{M}}_{\alpha n} (\Lambda)
\nonumber \\
& \times & 2\pi\sigma_{\alpha n} (\Lambda)\arctan \Bigl({\sin k - \Lambda\over nu}\Bigr) 
\nonumber \\
{\rm with} && q^{\Delta} = q_{\tau} (0) \, ,
\label{Bat}
\end{eqnarray}
where $q^{\Delta}$ is the shift $\tau$-band momentum appearing in Eq. (\ref{qqq}) and,
\begin{eqnarray}
&& q_{sn} (\Lambda) = {1\over \pi}\int_{-\pi}^{\pi} dk\,{\bar{N}}_{\tau} (k)\,
2\pi\rho (k)\arctan\Bigl({\Lambda - \sin k\over n u}\Bigr)
\nonumber \\
&& - {1\over\pi}\int_{-\infty}^{\infty} d\Lambda'\,{\bar{M}}_{s n} (\Lambda')\,
2\pi\sigma_{sn} (\Lambda')
\nonumber \\
&& \times\Bigl\{\arctan\Bigl({\Lambda - \Lambda'\over 2nu}\Bigr) + 2\sum_{l=1}^{n-1}
\arctan\Bigl({\Lambda - \Lambda'\over 2lu}\Bigr)\Bigr\}
\nonumber \\
&& - {1\over \pi}\sum_{n'=1(n'\neq n)}^{\infty}\int_{-\infty}^{\infty} d\Lambda'\,{\bar{M}}_{s n'} (\Lambda')\,
2\pi\sigma_{sn'} (\Lambda')
\nonumber \\
&& \times \Bigl\{\arctan\Bigl({\Lambda - \Lambda'\over \vert n-n'\vert u}\Bigr) +
\arctan\Bigl({\Lambda - \Lambda'\over (n+n')u}\Bigr)
\nonumber \\
&& + \sum_{l=1}^{{n+n'-\vert n-n'\vert\over 2}-1}2\arctan\Bigl({\Lambda - \Lambda'\over (\vert n-n'\vert + 2l)u}\Bigr)\Bigr\} \, ,
\label{Basn}
\end{eqnarray}
\begin{eqnarray}
&& q_{\eta n} (\Lambda) = \sum_{\iota =\pm 1}\arcsin (\Lambda - i\,\iota\,n\,u)
\nonumber \\
&& - {1\over \pi}\int_{-\pi}^{\pi} dk\,{\bar{N}}_{\tau} (k)\,
2\pi\rho (k)\arctan\Bigl({\Lambda - \sin k\over n u}\Bigr)
\nonumber \\
&& - {1\over \pi}\int_{-\infty}^{\infty} d\Lambda'\,{\bar{M}}_{\eta n} (\Lambda')\,
2\pi\sigma_{\eta n} (\Lambda')
\nonumber \\
&& \times \Bigl\{\arctan\Bigl({\Lambda - \Lambda'\over 2nu}\Bigr) + 2\sum_{l=1}^{n-1}
\arctan\Bigl({\Lambda - \Lambda'\over 2lu}\Bigr)\Bigr\}
\nonumber \\
&& - {1\over\pi}\sum_{n'=1(n'\neq n)}^{\infty}\int_{-\infty}^{\infty} d\Lambda'\,{\bar{M}}_{\eta n'} (\Lambda')\,
2\pi\sigma_{\eta n'} (\Lambda')
\nonumber \\
&& \times \Bigl\{\arctan\Bigl({\Lambda - \Lambda'\over \vert n-n'\vert u}\Bigr) +
\arctan\Bigl({\Lambda - \Lambda'\over (n+n')u}\Bigr)
\nonumber \\
&& + 2\sum_{l=1}^{{n+n'-\vert n-n'\vert\over 2}-1}\arctan\Bigl({\Lambda - \Lambda'\over (\vert n-n'\vert + 2l)u}\Bigr)\Bigr\} \, .
\label{BAetan}
\end{eqnarray}
Here the distributions ${\bar{N}}_{\tau} (k)$, ${\bar{M}}_{s n} (\Lambda)$, and 
${\bar{M}}_{\eta n} (\Lambda)$ where $n = 1,...,\infty$ obey the relations 
given in Eq. (\ref{NNrela}) and the functions $2\pi\rho (k)$, $2\pi\sigma_{sn} (\Lambda)$,
and $2\pi\sigma_{\eta n} (\Lambda)$ are such that,
\begin{eqnarray}
2\pi\rho (k) & = & {d q_{\tau} (k)\over dk} \hspace{0.20cm}{\rm and}\hspace{0.20cm}
2\pi\sigma_{\alpha n} (\Lambda) = {d q_{\alpha n} (\Lambda)\over d\Lambda}
\hspace{0.20cm}{\rm where}
\nonumber \\
\alpha & = & s, \eta \hspace{0.20cm}{\rm and}\hspace{0.20cm}n = 1,...,\infty \, .
\label{rhosigma}
\end{eqnarray}

These functions are thus defined by the following coupled integral equations,
\begin{eqnarray}
2\pi\rho (k) & = & 1 + {\cos k\over\pi u}\sum_{\alpha =s,\eta}\sum_{n=1}^{\infty}n
\int_{-\infty}^{\infty} d\Lambda\, {\bar{M}}_{\alpha n} (\Lambda)\,2\pi\sigma_{\alpha n} (\Lambda)
\nonumber \\
& \times & {1\over 1 + \left({\sin k - \Lambda\over nu}\right)^2} \, ,
\label{Grho}
\end{eqnarray}
\begin{eqnarray}
&& 2\pi\sigma_{sn} (\Lambda) = {1\over \pi n u}\int_{-\pi}^{\pi} dk\,{\bar{N}}_{\tau} (k)\,
{2\pi\rho (k)\over 1 + \Bigl({\Lambda - \sin k\over n u}\Bigr)^2} 
\nonumber \\
&& - {1\over 2\pi u}\int_{-\infty}^{\infty} d\Lambda'\,{\bar{M}}_{s n} (\Lambda')\,2\pi\sigma_{sn} (\Lambda')
\nonumber \\
&& \times \Bigl\{{1\over n}\,{1\over 1 + \Bigl({\Lambda - \Lambda'\over 2nu}\Bigr)^2} + \sum_{l=1}^{n-1}{2\over l}\,{1\over 
1 + \Bigl({\Lambda - \Lambda'\over 2lu}\Bigr)^2}\Bigr\}
\nonumber \\
&& - {1\over 2\pi u}\sum_{n'=1(n'\neq n)}^{\infty}\int_{-\infty}^{\infty} d\Lambda'\,{\bar{M}}_{s n'} (\Lambda')\,
2\pi\sigma_{sn'} (\Lambda')
\nonumber \\
&& \times \Bigl\{{2\over \vert n-n'\vert}{1\over 1 + \Bigl({\Lambda - \Lambda'\over \vert n-n'\vert u}\Bigr)^2} +
{2\over n+n'}\,{1\over 1 + \Bigl({\Lambda - \Lambda'\over (n+n')u}\Bigr)^2} 
\nonumber \\
&& + \sum_{l=1}^{{n+n'-\vert n-n'\vert\over 2}-1}{4\over \vert n-n'\vert + 2l}\,
{1\over 1 + \Bigl({\Lambda - \Lambda'\over (\vert n-n'\vert + 2l)u}\Bigr)^2}\Bigr\} \, ,
\nonumber \\
\label{Gsigmasn}
\end{eqnarray}
\begin{eqnarray}
&& 2\pi\sigma_{\eta n} (\Lambda) = \sum_{\iota =\pm 1}{1\over\sqrt{1 - (\Lambda - i\,\iota\,n\,u)^2}}
\nonumber \\
&& - {1\over \pi n u}\int_{-\pi}^{\pi} dk\,{\bar{N}}_{\tau} (k)\,
{2\pi\rho (k)\over 1 + \Bigl({\Lambda - \sin k\over n u}\Bigr)^2} 
\nonumber \\
&& - {1\over 2\pi u}\int_{-\infty}^{\infty} d\Lambda'\,{\bar{M}}_{\eta n} (\Lambda')\,
2\pi\sigma_{\eta n} (\Lambda')
\nonumber \\
&& \times \Big\{{1\over n}\,{1\over 1 + \Bigl({\Lambda - \Lambda'\over 2nu}\Bigr)^2} + \sum_{l=1}^{n-1}{2\over l}\,{1\over 
1 + \Bigl({\Lambda - \Lambda'\over 2lu}\Bigr)^2}\Bigr\}
\nonumber \\
&& - {1\over 2\pi u}\sum_{n'=1(n'\neq n)}^{\infty}\int_{-\infty}^{\infty} d\Lambda'\,{\bar{M}}_{\eta n'} (\Lambda')\,
2\pi\sigma_{\eta n'} (\Lambda')
\nonumber \\
&& \times \Bigl\{{2\over \vert n-n'\vert}\,{1\over 1 + \Bigl({\Lambda - \Lambda'\over \vert n-n'\vert u}\Bigr)^2} +
{2\over n+n'}\,{1\over 1 + \Bigl({\Lambda - \Lambda'\over (n+n')u}\Bigr)^2} 
\nonumber \\
&& + \sum_{l=1}^{{n+n'-\vert n-n'\vert\over 2}-1}{4\over \vert n-n'\vert + 2l}\,
{1\over 1 + \Bigl({\Lambda - \Lambda'\over (\vert n-n'\vert + 2l)u}\Bigr)^2}\Bigr\} \, ,
\nonumber \\
\label{Gsigmaetan}
\end{eqnarray}
where $n=1,...,\infty$.

They obey the sum rules,
\begin{eqnarray}
{1\over 2\pi}\int_{-\pi}^{\pi} dk\,{\bar{N}}_{\tau} (k)\,2\pi\rho (k) & = & {N_{\tau}\over L}
\nonumber \\
{1\over 2\pi}\int_{-\infty}^{\infty} d\Lambda\,{\bar{M}}_{\alpha n} (\Lambda)\,
2\pi\sigma_{\alpha n} (\Lambda) & = & {M_{\alpha n}\over L} \, ,
\label{sumrulesNN}
\end{eqnarray}
where $\alpha = s,\eta$ and $n = 1,...,\infty$.

The derivation of the $\alpha$-spin expectation value $\langle \hat{J}_{\alpha}^z\rangle$, Eq. (\ref{rel-currents-gen}),
involves accounting for any $S_{\alpha}>0$ energy eigenstate for the interplay of the Bethe-ansatz equations
with the general expression for the $\alpha$-spin HWS's current expectation value
$\langle \hat{J}_{\alpha}^z (HWS)\rangle$, Eq. (\ref{currentHWS}). 
To reach that goal, the $\tau$-band momentum-rapidity variable $k$ and 
$\alpha n$-band rapidity variables $\Lambda$ in the Bethe-ansatz equations, Eqs. (\ref{Bat})-(\ref{BAetan}), are to be replaced 
by those given in Eq. (\ref{kLambdaPhi}). 

For the spin current expectation values $\langle \hat{J}_{s}^z\rangle$ 
that leads after some careful algebra to the following expression,
\begin{eqnarray}
&& \langle \hat{J}_{s}^z\rangle = {L_{s,+1/2} - L_{s,-1/2}\over L_{s,+1/2} + L_{s,-1/2}}\,{t L\over\pi}\int_{-\pi}^{\pi}dk {\bar{N}}_{\tau} (k) \sin k
\nonumber \\
&& + {2t\ L\over \pi^2 u}\sum_{n=1}^{\infty} {N_{sn,+1/2} - N_{sn,-1/2}\over N_{sn,+1/2} + N_{sn,-1/2}} \times 
\nonumber \\
 && \int_{-\infty}^{\infty}d\Lambda {\bar{M}}_{sn}^h (\Lambda)\int_{-\pi}^{\pi}dk 
 {\bar{N}}_{\tau} (k) {2\pi\rho (k) \sin k\over 1 + \left({\Lambda - \sin k\over nu}\right)^2} \, .
 \label{Jsz}
 \end{eqnarray}
 That of the charge current expectation values $\langle \hat{J}_{\eta}^z\rangle$ is found to read,
\begin{eqnarray}
 && \langle \hat{J}_{\eta}^z\rangle = - {L_{\eta,+1/2} - L_{\eta,-1/2}\over L_{\eta,+1/2} + L_{\eta,-1/2}}\,
 {t L\over\pi}\int_{-\pi}^{\pi}dk {\bar{N}}_{\tau}^h (k) \sin k
 \nonumber \\
 && + {4t L\over \pi}\sum_{n=1}^{\infty} {N_{\eta n,+1/2} - N_{\eta n,-1/2}\over N_{\eta n,+1/2} + N_{\eta n,-1/2}} \times 
 \nonumber \\
 && \int_{-\infty}^{\infty}d\Lambda
 {\bar{M}}_{\eta n}^h (\Lambda)\Bigl(n\,{\rm Re}\left\{{(\Lambda - i n u)\over\sqrt{1 - (\Lambda - i n u)^2}}\right\}
 \nonumber \\
 && + {1\over 2\pi u} \int_{-\pi}^{\pi}dk {\bar{N}}_{\tau} (k) {2\pi\rho (k) \sin k\over 1 + \left({\Lambda - \sin k\over nu}\right)^2}\Bigr) \, .
 \label{Jetaz}
 \end{eqnarray}
In these expressions $L_{\alpha,\pm 1/2}$ and $N_{\alpha n,\pm 1/2}$ are for $\alpha = s,\eta$ and $n = 1,...,\infty$ the number 
of physical $\alpha$-spins of projection $\pm 1/2$, Eq. (\ref{MM}), and the number of those which the $\alpha n$-pairs ``see'' 
and interchange position with, Eq. (\ref{Nnh}), respectively, ${\bar{N}}_{\tau} (k)$ , ${\bar{N}}_{\tau}^h (k)$, and 
${\bar{M}}_{\alpha n}^h (\Lambda)$ where $\alpha = s,\eta$ and $n = 1,...,\infty$
are defined by Eq. (\ref{NNrela}), and $2\pi\rho (k)$ by Eqs. (\ref{Grho})-(\ref{Gsigmaetan}).

The interplay between the Bethe-ansatz equations and the $\alpha$-spin current expectation values
expressions given in Eqs. (\ref{currentHWS}) and (\ref{rel-currents-gen}) used to reach the
general expressions, Eqs. (\ref{Jsz}) and (\ref{Jetaz}), accounts for all coupling processes to
spin and charge discussed in Sec. \ref{SECIIIA} that are behind the expressions for the momentum difference 
$(P_{\Phi} - P)$ given in Eqs. (\ref{PeffUalpha}) and  (\ref{PeffUalphaNnonHWS}).
Such coupling processes lead to the $\alpha = s,\eta$ pre-factors $(L_{\alpha,+1/2} - L_{\alpha,-1/2})/(L_{\alpha,+1/2} + L_{\alpha,-1/2})$
and $(N_{\alpha n,+1/2} - N_{\alpha n,-1/2})/(N_{\alpha n,+1/2} + N_{\alpha n,-1/2})$ in
Eqs. (\ref{Jsz}) and (\ref{Jetaz}).

The calculations to reach Eqs. (\ref{Jsz}) and (\ref{Jetaz}) combine those
to obtain Eq. (\ref{rel-currents-gen}) and the above mentioned procedures 
involving the Bethe-ansatz equations, Eqs. (\ref{Bat})-(\ref{BAetan}).
They are relatively easy for small values of $N_{\alpha,- 1/2}$ and 
${\cal{N}}_{\alpha n,+1/2} = {\cal{N}}_{\alpha n,- 1/2}$ in
$L_{\alpha,\pm 1/2} = N_{\alpha,\pm 1/2} + {\cal{N}}_{\alpha,\pm 1/2}$, Eq. (\ref{MM}),
and $N_{\alpha n,\pm 1/2} = N_{\alpha,\pm 1/2} + {\cal{N}}_{\alpha n,\pm 1/2}$, Eq. (\ref{Nnh}),
where $\alpha = s$ and $\alpha = \eta$.
The calculations become quite lengthy as these values increase, but remain straightforward and controlled.


\begin{references}
\bibitem{Gutzwiller_63}
	M. Gutzwiller, 
	{\it Effect of correlation on the ferromagnetism of transition metals},
	Phys. Rev. Lett. {\bf 10}, 159 (1963).
\bibitem{Hubbard_63}
	J. Hubbard, 
	{\it Electron correlations in narrow energy bands},
	Proc. R. Soc. London A {\bf 276}, 238 (1963); 
	J. Hubbard, 
	{\it Electron correlations in narrow energy bands. II.
	The degenerate band case},
	Proc. R. Soc. London A {\bf 277}, 237 (1964).
\bibitem{Lieb_68}
        E. H. Lieb, F. Y. Wu, 
        {\it Absence of Mott transition in an exact solution of the short-range one-band model in one dimension},
        Phys. Rev. Lett. {\bf 20}, 1445 (1968).
\bibitem{Lieb_03}
        E. H. Lieb and F. Y. Wu, 
        {\it The one-dimensional Hubbard model: A reminiscence},
        Physica A {\bf 321}, 1 (2003).
\bibitem{Takahashi_72}
        M. Takahashi, 
        {\it One-dimensional Hubbard model at finite temperature},
        Progr. Theor. Phys {\bf 47}, 69 (1972). 
\bibitem{Martins_97}
        P. B. Ramos and M. J. Martins,  
        {\it Algebraic Bethe ansatz approach for the one-dimensional Hubbard model},
        J. Phys. A: Math. Gen. {\bf 30}, L195 (1997).            
\bibitem{Martins_98}
        M. J. Martins and P. B. Ramos, 
        {\it The quantum inverse scattering method for Hubbard-like models},
        Nucl. Phys. B {\bf 522}, 413 (1998).   
\bibitem{Essler_05} 
	F. H. L. Essler, H. Frahm, F. G\"ohmann, A. Kl\"umper, and V. E. Korepin, 
	{\em The one-dimensional Hubbard model} (Cambridge University Press, Cambridge, UK, 2005).  
\bibitem{Ostlund_91}
	S. \"Ostlund and E. Mele, 
	{\it Local canonical transformations of fermions},
	Phys. Rev. B {\bf 44}, 12413 (1991).
\bibitem{Carmelo_10}
	J. M. P. Carmelo, S. \"Ostlund, and M. J. Sampaio,
	{\it Global $SO(3)\times SO(3)\times U(1)$ symmetry of the Hubbard model on bipartite lattices},
	Ann. Phys. {\bf 325}, 1550 (2010).
\bibitem{Carmelo_24} 
	J. M. P. Carmelo and P. D. Sacramento, 
	{\it Temperature dependence of charge transport in the half-filled one-dimensional Hubbard model},
         Phys. Rev. B {\bf 110}, L201108 (2024).    		
\bibitem{Ilievski_18} 
	E. Ilievski, J. De Nardis, M. Medenjak, and T. Prosen, 
	{\it Superdiffusion in one-dimensional quantum lattice models},
	Phys. Rev. Lett. {\bf 121}, 230602 (2018).
\bibitem{Fava_20} 
	M. Fava , B. Ware, S. Gopalakrishnan, R. Vasseur, and S. A. Parameswaran,
	{\it Spin crossovers and superdiffusion in the one-dimensional Hubbard model},
	Phys. Rev. B {\bf 102}, 115121 (2020).
\bibitem{Moca_23} 
	C. P. Moca, M. A. Werne, A. Valli, T. Prosen, and G. Zar\'and,
	{\it Kardar-Parisi-Zhang scaling in the Hubbard model},
	Phys. Rev. B {\bf 108}, 235139 (2023).
\bibitem{Pereira_12}
        R. G. Pereira, K. Penc, S. R. White, P. D. Sacramento, and J. M. P. Carmelo,
        {\it Charge dynamics in half-filled Hubbard chains with finite on-site interaction},
        Phys. Rev. B {\bf 85}, 165132 (2012).	
\bibitem{Nocera_18} 
	A. Nocera, F. H. L. Essler, and A. E. Feiguin,
	{\it Finite temperature dynamics of the Mott insulating Hubbard chain},
	Phys. Rev. B {\bf 97}, 045146 (2018).	
\bibitem{Ilievski_17} 
	E. Ilievski and J. De Nardis, 
	{\it Ballistic transport in the one-dimensional Hubbard model: The hydrodynamic approach},
	Phys. Rev. B {\bf 96}, 081118(R) (2017).
\bibitem{Carmelo_18}
	J. M. P. Carmelo, S. Nemati, and T. Prosen,
	{\it Absence of ballistic charge transport in the half-filled 1D Hubbard model},
	Nucl. Phys. B {\bf 930}, 418 (2018).
\bibitem{Kardar_86}	
	M. Kardar, G. Parisi, and Y.-C. Zhang, 
	{\it Dynamic scaling of growing interfaces}, 
	Phys. Rev. Lett. {\bf 56}, 889 (1986).
\bibitem{Krug_97}	
	J. Krug, 
	{\it Origins of scale invariance in growth processes}, 
	Adv. Phys. {\bf 46}, 139 (1997).
\bibitem{Kriecherbauer_10}
	T. Kriecherbauer and J. Krug, 
	{\it A pedestrian's view on interacting particle systems, KPZ universality and random matrices},
	J. Phys. A: Math. Theor. {\bf 43}, 403001 (2010).
\bibitem{Corwin_12}
	I. Corwin, 
	{\it The Kardar-Parisi-Zhang equation and universality class}, 
	Random Matrices: Theory Applic. {\bf 01}, 1130001 (2012).	
\bibitem{Ljubotina_19}	
	M. Ljubotina, M. \v{Z}nidari\v{c}, and T. Prosen, 
	{\it Kardar-Parisi-Zhang physics in the quantum Heisenberg magnet},
	Phys. Rev. Lett. {\bf 122}, 210602 (2019).
\bibitem{Ljubotina_17}
	M. Ljubotina, M. \v{Z}nidari\v{c}, and T. Prosen, 
	{\it Spin diffusion from an inhomogeneous quench in an integrable system},
	Nat. Commun. {\bf 8}, 16117 (2017).     	                    
\bibitem{Carmelo_17A}
	J. M. P. Carmelo and T. \v{C}ade\v{z},
	{\it One-electron singular spectral features of the 1D Hubbard model at finite magnetic field},
	Nucl. Phys. B {\bf 914}, 461 (2017).
\bibitem{Ogata_90} 
        M. Ogata and H. Shiba, 
        {\it Bethe-ansatz wave function, momentum distribution, and spin correlation in the one-dimensional strongly correlated Hubbard model},
        Phys. Rev. B {\bf 41}, 2326 (1990).
\bibitem{Penc_97}  
         K. Penc, K. Hallberg, F. Mila, and H. Shiba, 
         {\it Spectral functions of the one-dimensional Hubbard model in the $U\rightarrow + \infty$ limit: How to use the factorized wave function},
         Phys. Rev. B {\bf 55}, 15475 (1997).
\bibitem{Kruis_04}	
	H. V. Kruis, I. P. McCulloch, Z. Nussinov, and J. Zaanen, 
	{\it Geometry and the hidden order of Luttinger liquids: The universality of squeezed space},
	Phys. Rev. B {\bf 70}, 075109 (2004).	
\bibitem{Carmelo_15}
	J. M. P. Carmelo, T. Prosen, and D. K. Campbell,
	{\it Vanishing spin stiffness in the spin-12 Heisenberg chain for any nonzero temperature},
	Phys. Rev. B {\bf 92}, 165133 (2015).
\bibitem{Carmelo_17}
	J. M. P. Carmelo and T. Prosen, 
	{\it Absence of high-temperature ballistic transport in the spin-$1/2$ $XXX$ chain within the grand-canonical ensemble},
	Nucl. Phys. B {\bf 914}, 62-98 (2017).	
\bibitem{Carmelo_20} 
	J. M. P. Carmelo, T. \v{C}ade\v{z}, and P. D. Sacramento, 
	{\it Bethe strings in the dynamical structure factor of the spin-$1/2$ Heisenberg $XXX$ chain}
        Nucl. Phys. B {\bf 960}, 115175 (2020).	
\bibitem{Essler_91}
	F. H. L. Essler, V. E. Korepin, and K. Schoutens,
	{Complete solution of the one-dimensional Hubbard model},
	Phys. Rev. Lett. {\bf 67}, 3848 (1991).        
\bibitem{Woynarovich_82}
	F. Woynarovich, 
	{\it Excitations with complex wavenumbers in a Hubbard chain: I. States with one pair of complex wavenumbers},
	J. Phys. C, Solid State Phys. {\bf 15}, 85 (1982).
\bibitem{Woynarovich_82A}
	F. Woynarovich, 
	{\it Excitations with complex wavenumbers in a Hubbard chain: II. States with several pairs of complex wavenumbers},
	J. Phys. C, Solid State Phys. {\bf 15}, 97 (1982).
\bibitem{Shastry_90}
	B. Shastry and B. Sutherland, 
	{Twisted boundary conditions and effective mass in Heisenberg-Ising and Hubbard rings},
	Phys. Rev. Lett. {\bf 65}, 243 (1990).
\bibitem{Mukerjee_08}
	 S. Mukerjee and B. S. Shastry, 
	 {\it Signatures of diffusion and ballistic transport in the stiffness, dynamical correlation functions, and statistics of one-dimensional systems},
	 Phys. Rev. B {\bf 77}, 245131 (2008).       
\bibitem{Medenjak_17}
	M. Medenjak, C. Karrasch, and T. Prosen,
	{\it Lower bounding diffusion constant by the curvature of Drude weight},
	Phys. Rev. Lett. {\bf 119}, 080602 (2017).	
\bibitem{Damle_05}	
	K. Damle and S. Sachdev,
	{\it Universal relaxational dynamics of gapped one-dimensional models in the quantum Sine-Gordon universality class},
	Phys. Rev. Lett. {\bf 95}, 187201 (2005).	
\bibitem{Carmelo_21} 
	J. M. P. Carmelo, \v{C}ade\v{z}, and P. D. Sacramento,
	{\it One-particle spectral functions of the one-dimensional Fermionic Hubbard model with one fermion 
	per site at zero and finite magnetic fields},
	Phys. Rev. B {\bf 103}, 195129 (2021).	
\bibitem{Jin_19} 	
	Y. Jin, O. Tsyplyatyev, M. Moreno, A. Anthore, W. K. Tan, J. P. Griffiths, I. Farrer, D. A. Ritchie,
	L. I. Glazman, A.J. Schofield, and  C. J. B. Ford,
	{\it Momentum-dependent power law measured in an interacting quantum wire beyond the Luttinger limit},
	Nat. Comm. {\bf 10}, 2821 (2019).
\bibitem{Vianez_22} 
	P. M. T. Vianez, Y. Q. Jin, M. Moreno, A. S. Anirban, A. Anthore, W. K. Tan, J. P. Griffiths, I. Farrer, D. A. Ritchie, 
	A. J. Schofield, O. Tsyplyatyev, and C. J. B. Ford,
	{\it Observing separate spin and charge Fermi seas in a strongly correlated one-dimensional conductor},
	Sci. Adv. {\bf 8}, eabm2781 (2022).
\bibitem{Tsyplyatyev_22}
	O. Tsyplyatye,
	{\it Splitting of the Fermi point of strongly interacting electrons in one dimension: A nonlinear effect of spin-charge separation}
	Phys. Rev. B {\bf 105}, L121112 (2022).
\bibitem{Carmelo_19}	
	J. M. P. Carmelo, T. \v{C}ade\v{z}, D. K. Campbell, M. Sing, and R. Claessen,
	{\it Effects of finite-range interactions on the one-electron spectral properties of TTF-TCNQ},
	Phys. Rev. B {\bf 100}, 245202 (2019).
\bibitem{Carmelo_19A}
	J. M. P. Carmelo, T. \v{C}ade\v{z}, Y. Ohtsubo, S.-i. Kimura, and D. K. Campbell,
	{\it Effects of finite-range interactions on the one-electron spectral properties of one-dimensional metals: 
	Application to Bi/InSb(001)},
	Phys. Rev. B {\bf 100}, 035105 (2019).
\bibitem{Carmelo_12} 	
	J. M. P. Carmelo, M. A. N. Ara\'ujo, S. R. White, and M. J. Sampaio,
	{\it Hubbard-model description of the high-energy spin-weight distribution in La$_2$CuO$_4$},
	Phys. Rev. B {\bf 86}, 064520 (2012).	
\end{references}
\end{document}